\documentclass[usenatbib]{mn2e}
\usepackage{lscape}
\usepackage{graphicx}
\usepackage{amssymb}
\usepackage{amsmath}
\usepackage{url}
\usepackage{aas_macros}

\newcommand{\LiI}{\mbox{Li\,{\footnotesize I}}}       
\newcommand{\NaI}{\mbox{Na\,{\footnotesize I}}}       
\newcommand{\KI}{\mbox{K\,{\footnotesize I}}}       
\newcommand{\msun}{\mbox{$\mathrm{M_{\odot}}$}}		
\newcommand{\Uwfc}{\mbox{$U_{_{\mathrm{WFC}}}$}}             
\newcommand{\gwfc}{\mbox{$g_{_{\mathrm{WFC}}}$}}             
\newcommand{\rwfc}{\mbox{$r_{_{\mathrm{WFC}}}$}}             
\newcommand{\iwfc}{\mbox{$i_{_{\mathrm{WFC}}}$}}             
\newcommand{\Zwfc}{\mbox{$Z_{_{\mathrm{WFC}}}$}}             
\newcommand{\giwfc}{\mbox{$(g-i)_{_{\mathrm{WFC}}}$}}       
\newcommand{\Ugwfc}{\mbox{$(U-g)_{_{\mathrm{WFC}}}$}}       
\newcommand{\riwfc}{\mbox{$(r-i)_{_{\mathrm{WFC}}}$}}       
\newcommand{\iZwfc}{\mbox{$(i-Z)_{_{\mathrm{WFC}}}$}}       

\usepackage{multirow}

\title[Revising formation timescales]{Pre-main-sequence
  isochrones -- II. Revising star and planet formation timescales}
\author[C.~P.~M.~Bell et al.]{Cameron~P.~M.~Bell$^{1}$\thanks{E-mail:
  bell@astro.ex.ac.uk (CPMB)}, Tim Naylor$^{1}$,
N.~J.~Mayne$^{1}$, R.~D.~Jeffries$^{2}$ and S.~P.~Littlefair$^{3}$\\
$^{1}$ School of Physics, University of Exeter, Exeter EX4 4QL\\
$^{2}$ Astrophysics Group, Research Institute for the Environment,
Physical Sciences and Applied Mathematics, Keele University,\\
Staffordshire ST5 5BG\\
$^{3}$ Department of Physics and Astronomy, University of Sheffield,
Sheffield S3 7RH}

\begin{document}

\date{Accepted ?, Received ?; in original form ?}

\pagerange{\pageref{firstpage}--\pageref{lastpage}} \pubyear{2013}

\maketitle

\label{firstpage}

\begin{abstract}
We have derived ages for 13 young ($< 30\, \rm{Myr}$)
star-forming regions and find they are up to a factor two older than
the ages typically adopted in the literature. This result has
wide-ranging implications, including that circumstellar discs survive
longer ($\simeq 10-12\, \rm{Myr}$) and that the average Class\,I
lifetime is greater ($\simeq 1\, \rm{Myr}$) than currently believed.

For each star-forming region we derived two ages from colour-magnitude
diagrams. First we fitted models of the evolution between the zero-age
main-sequence and terminal-age main-sequence to derive a homogeneous
set of main-sequence ages, distances and reddenings with statistically
meaningful uncertainties. Our second age for each star-forming region was derived by
fitting pre-main-sequence stars to new semi-empirical model
isochrones. For the first time (for a set of clusters younger than
$50\, \rm{Myr}$)
we find broad agreement between these two ages, and since these are
derived from two distinct mass regimes that rely on different aspects
of stellar physics, it gives us confidence in the new age scale. This
agreement is largely due to our adoption of empirical
colour-$T_{\rm{eff}}$ relations and bolometric corrections for
pre-main-sequence stars cooler than $4000\, \rm{K}$.

The revised ages for the star-forming regions in our sample are -- $\sim
2\, \rm{Myr}$ for NGC\,6611 (Eagle Nebula; M\,16), IC\,5146 (Cocoon
Nebula), NGC\,6530 (Lagoon Nebula; M\,8), and NGC\,2244 (Rosette
Nebula); $\sim 6\, \rm{Myr}$ for $\sigma$\,Ori, Cep\,OB3b, and
IC\,348; $\simeq 10\, \rm{Myr}$ for
$\lambda$\,Ori (Collinder\,69); $\simeq 11\, \rm{Myr}$ for NGC\,2169;
$\simeq 12\, \rm{Myr}$ for NGC\,2362; $\simeq 13\,
\rm{Myr}$ for NGC\,7160; $\simeq 14\, \rm{Myr}$ for $\chi$\,Per
(NGC\,884); and $\simeq 20\, \rm{Myr}$ for NGC\,1960 (M\,36).

\end{abstract}

\begin{keywords}
  stars: evolution -- stars: formation -- stars: pre-main-sequence --
  stars: fundamental parameters --
  techniques: photometric -- open clusters and associations: general
  -- Hertzsprung-Russell and colour-magnitude diagrams
\end{keywords}

\section{Introduction}
\label{introduction}

Robust and precise ages for young stars are a requirement for the
advancement of our understanding of star and planet formation and
evolution. These ages provide timescales with which to constrain the
physical processes driving, for example, disc dissipation and
planet formation (e.g. core accretion versus gravitational collapse;
\citealp{Pollack84,Boss97}). Pre-main-sequence (pre-MS) ages are based on the gravitational
contraction of young stellar objects (YSOs), which become increasingly faint as
they contract towards the main-sequence (MS).
It has been well documented
that pre-MS ages are imprecise and contradictory being heavily model-
(and even colour-) dependent (e.g. \citealp{Naylor02,Hartmann03}) with the
derived ages differing by factors of up to 2 -- 3
(e.g. \citealp{White99,Dahm05}; \citealp*{Hillenbrand08}).

Having highlighted the problems with the pre-MS evolutionary models,
we have attempted, through a series of papers, to provide consistent
pre-MS ages. In \cite{Mayne07}, pre-MS stars in a series
of young star-forming regions (SFRs) were used to create a set
of empirical isochrones. When plotted in absolute magnitude-intrinsic
colour space these isochrones
placed the clusters on a relative age ladder independent of theoretical assumptions
(fiducial ages were assigned to a subset of these
SFRs). The main problem with this work is that it suffered from a lack
of self-consistent distance measurements, instead relying on
heterogeneously derived literature estimates. To rectify this 
\cite{Mayne08} used MS
model isochrones to fit the young MS population of many
of the clusters studied in \cite{Mayne07} and derived self-consistent distances
and extinctions. These
revised distances, led to a subtle modification of the \cite{Mayne07}
pre-MS age scale. Although self-consistent, the
\cite{Mayne07} and \cite{Mayne08} pre-MS age scale only provided relative
ages. Whilst the low-mass stars in a young SFR are still on the
pre-MS, the most massive objects have reached the MS, so in an
effort to create an absolute age scale, 
\cite{Naylor09} used MS stars
between the zero-age main-sequence
(ZAMS) and terminal-age main-sequence (TAMS),
in conjunction with the $\tau^{2}$ fitting statistic, to derive
absolute ages. These MS ages were found to be a factor 1.5 -- 2
times older than the pre-MS ages based on the
\cite{Mayne07} and \cite{Mayne08} pre-MS age scale.

In contrast to the high level of model dependency in pre-MS
evolutionary models, MS models have been shown to provide high levels
of consistency (in both age and distance) between models from
different groups (e.g. \citealp{Mayne08,Naylor09}). It would seem,
therefore, obvious to use MS ages, but these are imprecise compared
with the pre-MS ages for two reasons. First, that the rate of change
of the slope of the isochrone for a given age increment is subtle
(i.e. $\Delta (B-V)/\sigma_{(B-V)}$ is small, despite the precision in
$B-V$). Second, in young SFRs only the most massive stars have evolved
sufficiently to lie on the MS and therefore such ages are subject to
large uncertainties caused by small number statistics. In terms of
statistical and systematic uncertainties, pre-MS ages have small
statistical but large systematic uncertainties, whereas for the MS
ages the opposite is true. The approach we take in this paper,
therefore, is to use the more precise pre-MS ages, but only after we
have demonstrated why they differ from the MS ages, and brought them
into broad agreement with the MS ages. Whilst previous studies have
found either tentative evidence for agreement between the age scales
for older stellar populations
($> 50\, \rm{Myr}$; e.g. \citealp{Lyra06}) or increasing disparity at younger ages ($<
20\, \rm{Myr}$; e.g. \citealp{Piskunov04,Naylor09}), the recent study by
\cite*{Pecaut12} has demonstrated consistent pre-MS and MS ages for the
Upper Sco subgroup of Scorpius-Centaurus based on a detailed analysis
of B-, A-, F- and G-stars in the Hertzsprung-Russell (H-R) diagram.

To test whether further agreement between MS and pre-MS ages
can be found for young ($< 30\, \rm{Myr}$) pre-MS clusters,
we must address the main sources of uncertainty when attempting to
fit pre-MS stellar populations with model isochrones. The primary
issues affecting pre-MS isochrone fitting in
CMDs are; i) photometric
calibration of the data for what are very red stars, ii)
transformation of the model isochrones 
from the theoretical H-R to the CMD plane,
iii) incorporating the colour and gravity
dependence of the interstellar extinction, and iv) the colour excesses
in photometric data arising from the combined
effects of stellar activity and circumstellar disc material.

The first two of these issues were explicitly addressed in
\cite{Bell12} (hereafter referred to as Paper~1). We demonstrated that the use of MS standards to
transform photometric observations of young, very red pre-MS objects
to a standard photometric system can introduce significant
errors in the position of the pre-MS locus in CMD space. This
is due to differences in the spectrum of a MS
and pre-MS star of the same colour and can introduce an error of up to
a factor of 2 in age for young ($\lesssim 10\, \rm{Myr}$) clusters 
\citep{Bell12}. Hence, instead of the normal
practice of transforming both photometric observations and theoretical
models into a standard photometric system, it is crucial for precise
photometric studies of pre-MS objects to leave the observations in
their natural photometric system. Furthermore, Paper~1 tested several sets of pre-MS
isochrones against a series of well-calibrated CMDs of the Pleiades over a
contiguous wavelength range of $0.4-2.5\, \rm{\mu m}$ and we showed that
in optical colours no pre-MS model followed the observed Pleiades
sequence at $T_{\rm{eff}} \lesssim 4000\, \rm{K}$. This discrepancy
was quantified in individual
photometric bandpasses and showed that the model isochrones
overestimated the flux by
a factor of 2 at $0.5\, \rm{\mu m}$, with this difference
decreasing as a function of increasing wavelength, becoming negligible
at $2.2\, \rm{\mu m}$. This can also introduce an error up to a factor
of 2 in age for clusters younger than $10\, \rm{Myr}$ \citep{Bell12}.

The third issue to consider is the colour dependence of the
interstellar reddening. Reddening estimates for a given cluster can be derived from
fitting the higher mass stars with a model isochrone in the $U-B, B-V$
colour-colour diagram assuming a given reddening vector (that often
neglects the colour and gravity dependency). This
measured value is then applied to stars of all masses assuming a
similar non-dependence on colour and gravity. This can inaccurately
modify the shape of the pre-MS locus, especially if
the target stars are highly reddened. Thus, to derive the reddening
from high-mass stars in a $U-B, B-V$ colour-colour diagram requires a
more sophisticated approach with reddening vectors that incorporate
the colour and gravity dependence. To then apply this measured value
to lower mass stars, the only self-consistent and homogenous method
(assuming that spectra of each star are unavailable) is to create
reddening and extinction grids based on atmospheric models, the
appropriate system responses and a reliable representation of the
interstellar extinction law. Neglecting these colour-dependent terms
can introduce an error of $\simeq 0.1\, \rm{mag}$ in the derived
true distance modulus \citep{Mayne08}, thereby translating into an
error of $\simeq 20-30$ per cent in the derived pre-MS age. If the
extinction towards a given cluster is spatially variable, then this
makes the process of deriving ages and distances more complicated. 
In some regions this may be due to patchy foreground extinction, but
for many young SFRs ($\lesssim 10\,
\rm{Myr}$) remnant gas and dust from the star formation process makes
this issue more problematic still.

The final issue is which photometric bandpasses to adopt for
constructing CMDs that minimise the colour excesses arising from
processes including chromospheric activity, accretion, and
circumstellar disc material. Accretion predominantly affects the flux
shortward of the $B$-band \citep{Gullbring98}, whereas circumstellar
disc material becomes significant for the near-infrared (near-IR) $JHK$
bandpasses \citep{Lada92}. Lastly, \cite{Stauffer03} showed that
chromospheric activity in low-mass stars can result in an increased scatter in the $B-V$
and $U-B$ colours, whereas the $V-I$ colour remains unaffected. Hence,
as a good compromise, we adopt the optical $g$- and $i$-bandpasses for
this study, and investigate the pre-MS populations of our sample of
SFRs using the $g, g-i$ CMD.

In this study we address, and account for, each of these issues to
create a revised age
scale for a sample of well-studied SFRs, and in doing so, largely
remove the discrepancy between MS and pre-MS ages. In
Section~\ref{the_data} we describe the
data collection, reduction and photometric calibration, as well as the
various youth diagnostics used to identify pre-MS objects. The stellar
interior models, atmospheric models and the effects of reddening and
extinction are discussed in Section~\ref{the_models}, with the newly
derived colour- and gravity-dependent reddening vectors
explained. Section~\ref{fundamental_parameters_from_ms_stars}
describes the maximum-likelihood fitting technique, the model CMD, and
the fitting of MS photometric data to derive reddenings, distances and
ages. The creation of the semi-empirical pre-MS isochrones and the
reddening and extinction grids are explained in
Section~\ref{creating_recalibrated_semi-empirical_pre-ms_isochrones}. Section~\ref{fitting_pre-ms_ages}
describes the fitting of the pre-MS photometric data, with the
literature sources used to identify pre-MS stars for our sample of
SFRs given in Appendix~\ref{literature_memberships} and a revised
version of the maximum-likelihood fitting technique designed to deal
with possible non-member contamination in the CMD discussed in
Appendix~\ref{non-member_contamination_model}. The results from the MS
and pre-MS data are brought together in
Section~\ref{comparing_the_ms_and_pre-ms_ages} and the implications of
these discussed in Section~\ref{implications}. Finally, our conclusions
are given in Section~\ref{conclusions}. For readers interested in the
final age, distance and reddening for each SFR, these results are
shown in Table~\ref{adopted_parameters} in
Section~\ref{final_assigned_pre-ms_ages}.

\section{The Data}
\label{the_data}

The observations presented were obtained
using the 2.5-m Isaac Newton
Telescope (INT) on La Palma at the same time as our Pleiades
observations described in Paper~1. We refer
the reader to Paper~1 for details of our observational techniques,
photometric calibration, and data reduction. As in Paper~1, the
astrometric calibration was provided using objects in the Two-Micron
All-Sky Survey (2MASS, \citealp{Cutri03}), with a rms of approximately
$0.2\, \rm{arcsec}$ for the fit of pixel positions as a function of
RA and Dec.

 It is important for what
follows that our final catalogues are in the natural photometric
system of the Wide-Field Camera (WFC) on the INT (hereafter INT-WFC)
calibrated to the absolute AB photometric system \citep{Oke83}. In
Paper~1 we adopted a clipping radius of $2 \times$ the full-width at
half maximum (FWHM) of the
seeing, however, in these more crowded fields this was reduced to $0.8
\times \rm{FWHM}$ (for a full discussion see King et al. in preparation).
Table~\ref{tab:observations} shows the central coordinates for each
field-of-view and limiting magnitude in the INT-WFC
$(UgriZ)_{_{\rm{WFC}}}$ bandpasses for our sample of SFRs.

\begin{table*}
\caption[]{The central coordinates for each field-of-view and limiting
magnitude in the INT-WFC $(UgriZ)_{_{\rm{WFC}}}$ bandpasses (for a
signal-to-noise of 10) for our
sample of SFRs. Notes are as follows. (1) No $\Uwfc$-band photometric
data was taken. (2) No usable data.}
\centering
\begin{tabular}{c c c c}
\hline
SFR&$\alpha$ (J2000.0)&$\delta$ (J2000.0)&Limiting magnitude in
$(UgriZ)_{_{\rm{WFC}}}$ bandpasses\\
\hline
Cep\,OB3b&$22^{\rm{h}}\, 55^{\rm{m}}\, 43.3^{\rm{s}}$&$+62^{\circ}\,
40'\, 13.6''$&21.4, 23.4, 23.5, 23.2, 22.3\\
$\chi$\,Per&$02^{\rm{h}}\, 22^{\rm{m}}\, 48.7^{\rm{s}}$&$+57^{\circ}\,
07'\, 30.0''$&21.4, 22.9, 22.9, 22.7, 21.9\\
IC\,348$^{(1)}$&$03^{\rm{h}}\, 44^{\rm{m}}\, 30.0^{\rm{s}}$&$+31^{\circ}\,
59'\, 59.9''$&-- , 22.0, 22.0, 22.1, 21.4\\
IC\,5146&$21^{\rm{h}}\, 53^{\rm{m}}\, 24.0^{\rm{s}}$&$+47^{\circ}\,
15'\, 36.0''$&21.4, 22.8, 22.9, 22.5, 21.6\\
$\lambda$\,Ori\,A&$05^{\rm{h}}\, 36^{\rm{m}}\, 25.0^{\rm{s}}$&$+09^{\circ}\,
38'\, 25.8''$&20.5, 22.6, 22.6, 22.6, 21.7\\
$\lambda$\,Ori\,B&$05^{\rm{h}}\, 34^{\rm{m}}\, 58.5^{\rm{s}}$&$+09^{\circ}\,
38'\, 25.8''$&20.6, 22.6, 22.5, 22.4, 21.6\\
$\lambda$\,Ori\,C&$05^{\rm{h}}\, 36^{\rm{m}}\, 25.0^{\rm{s}}$&$+10^{\circ}\,
02'\, 25.6''$&20.6, 22.8, 22.6, 22.6, 21.7\\
$\lambda$\,Ori\,D&$05^{\rm{h}}\, 34^{\rm{m}}\, 48.6^{\rm{s}}$&$+10^{\circ}\,
02'\, 25.6''$&20.8, 22.9, 22.6, 22.9, 21.9\\
 NGC\,1960&$05^{\rm{h}}\, 36^{\rm{m}}\, 18.0^{\rm{s}}$&$+34^{\circ}\,
08'\, 24.1''$&19.0, 22.8, 22.8, 22.6, 21.8\\
NGC\,2169&$06^{\rm{h}}\, 08^{\rm{m}}\, 24.0^{\rm{s}}$&$+13^{\circ}\,
57'\, 54.0''$&21.7, 22.9, 22.8, 22.5, 21.8\\
NGC\,2244&$06^{\rm{h}}\, 31^{\rm{m}}\, 55.5^{\rm{s}}$&$+04^{\circ}\,
56'\, 34.3''$&21.4, 22.8, 22.7, 22.4, 21.5\\
NGC\,2362&$07^{\rm{h}}\, 18^{\rm{m}}\, 36.5^{\rm{s}}$&$-24^{\circ}\,
57'\, 00.0''$&21.2, 22.7, 22.5, 22.1, 21.2\\
NGC\,6530&$18^{\rm{h}}\, 04^{\rm{m}}\, 05.0^{\rm{s}}$&$-24^{\circ}\,
22'\, 00.0''$&20.3, 21.5, 21.5, 21.5, 20.8\\
NGC\,6611&$18^{\rm{h}}\, 18^{\rm{m}}\, 48.0^{\rm{s}}$&$-13^{\circ}\,
47'\, 00.0''$&20.5, 23.0, 22.4, 21.9, 21.0\\
NGC\,7160&$21^{\rm{h}}\, 53^{\rm{m}}\, 40.0^{\rm{s}}$&$+62^{\circ}\,
36'\, 12.0''$&19.0, 22.2, 22.3, 22.6, 22.0\\
$\sigma$\,Ori\,A&$05^{\rm{h}}\, 40^{\rm{m}}\, 14.2^{\rm{s}}$&$-02^{\circ}\,
20'\, 18.0''$&19.5, 22.8, 22.8, 22.6, 21.5\\
$\sigma$\,Ori\,B$^{(2)}$&$05^{\rm{h}}\, 40^{\rm{m}}\, 13.1^{\rm{s}}$&$-02^{\circ}\,
51'\, 47.7''$&-- , 23.0, 22.8, 22.8, 21.6\\
$\sigma$\,Ori\,C&$05^{\rm{h}}\, 38^{\rm{m}}\, 07.7^{\rm{s}}$&$-02^{\circ}\,
20'\, 18.0''$&18.3, 22.8, 22.8, 22.5, 21.6\\
$\sigma$\,Ori\,D$^{(1)}$&$05^{\rm{h}}\, 38^{\rm{m}}\, 07.7^{\rm{s}}$&$-02^{\circ}\,
51'\, 50.9''$&-- , 22.9, 22.7, 22.6, 21.4\\
\hline
\end{tabular}
\label{tab:observations}
\end{table*}

For the field-of-view centred on IC\,348, in all but the central CCD
there were too few stars to create the profile correction necessary to
correct the optimal photometry. Hence for stars
in these regions aperture photometry was performed with the radius of
the aperture matching that used in the case of the profile correction
(15 pixels). The resultant fluxes were then processed in an identical
manner to that in the standard reduction. The only
difference is that the non-stellar flag `N' (see
\citealp{Burningham03}) is now based on the
ratio of the flux measured through an aperture of 7 pixels. To
create the final optical photometric catalogue, all stars flagged
`H' (poor profile correction) in the optimally extracted catalogue
were replaced with measurements based on the aperture photometric
reduction.

For Cep\,OB3b, $\chi$\,Per, IC\,5146, and $\lambda$\,Ori\,A
observations were taken on two separate nights for which there is a
photometric solution. The two sets of observations were reduced
separately and a normalisation procedure performed on the two optical
photometric catalogues, so that the zero-point is the average of the
two nights. As discussed in \cite{Naylor02}, the calculated zero-point
shift is a good indicator of the internal consistency of the
photometry, as well as the accuracy with which the profile corrections
were performed. For regions where the normalisation procedure was
used an accuracy of 2 -- 3 per cent was found across all colours.

\begin{table*}
\caption[]{The photometric catalogues presented in this study. The
  members catalogues use the data where stars variable from
  night-to-night have been replaced by the values from one night
  as described in Paper~1. Sources
  used in the identification of pre-MS members are as follows (see also
  Appendix~\ref{literature_memberships}). (1) \protect\cite{Naylor99};
  (2) \protect\cite{Getman06}; (3) \protect\cite{Pozzo03};
  (4) \protect\cite*{Ogura02}; (5) \protect\cite{Littlefair10}; (6)
  \protect\cite{Herbig98}; (7) \protect\cite{Luhman03}; (8)
  \protect\cite{Luhman05a}; (9) \protect\cite*{Luhman05b}; (10)
  \protect\cite{Preibisch02}; (11) Second ROSAT PSPC
  catalogue;
  (12) \protect\cite*{Cohen04}; (13)
  \protect\cite{Littlefair05}; (14) \protect\cite{Herbig02}; (15)
  \protect\cite{Harvey08}; (16) \protect\cite{Dolan01}; (17)
  \protect\cite{Barrado04a}; (18) \protect\cite{Sacco08}; (19)
  \protect\cite{Barrado07}; (20) \protect\cite{Barrado11}; (21)
  \protect\cite{Jeffries13}; (22) \protect\cite{Jeffries07b};
  (23) \protect\cite{Wang08}; (24) \protect\cite{Balog07}; (25)
  \protect\cite{Dahm05}; (26) \protect\cite{Dahm07}; (27)
  \protect\cite{Damiani06b}; (28) \protect\cite{Damiani04}; (29)
  \protect\cite{Prisinzano07}; (30) \protect\cite{Henderson12}; (31)
  \protect\cite{Guarcello07}; (32) \protect\cite{Guarcello09}; (33)
  \protect\cite{Sicilia04,Sicilia05}; (34) \protect\cite{Sicilia06};
  (35) \protect\cite{Kenyon05}; (36) \protect\cite{Burningham05b}; (37)
  \protect\cite{Sacco08}; (38)
  \protect\cite*{Sanz-Forcada04}; (39) \protect\cite{Cody10}.}
\begin{tabular}{c c c c}
\hline
SFR&Table number&Description&Source\\
\hline
Cep\,OB3b&9&Full catalogue&\\
&10&Members&1+2 (X-ray); 3 (Spectroscopic); 4 (H$\alpha$); 5 (Periodic
variable)\\
$\chi$\,Per&11&Full catalogue&\\
&12&Positionally isolated&\\
IC\,348&13&Full catalogue&\\
&14&Members&6 (H$\alpha$); 7+8+9 (Spectroscopic); 10+11 (X-ray); 12+13
(Periodic variable)\\
IC\,5146&15&Full catalogue&\\
&16&Members&14 (H$\alpha$); 15 (IR excess)\\
$\lambda$\,Ori&17&Full catalogue&\\
&18&Members&16+17+18 (Spectroscopic); 19 (IR-excess); 20 (X-ray)\\
NGC\,1960&19&Full catalogue&\\
&20&Members&21 (Spectroscopic)\\
NGC\,2169&21&Full catalogue&\\
&22&Members&22 (Spectroscopic)\\
NGC\,2244&23&Full catalogue&\\
&24&Members&23 (X-ray); 24 (IR excess)\\
NGC\,2362&25&Full catalogue&\\
&26&Members&25 (Spectroscopic); 26 (IR excess); 27 (X-ray)\\
NGC\,6530&27&Full catalogue&\\
&28&Members&28 (X-ray); 29 (Spectroscopic); 30 (Periodic variable)\\
NGC\,6611&29&Full catalogue\\
&30&Members&31 (X-ray); 32 (IR excess)\\
NGC\,7160&31&Full catalogue&\\
&32&Members&33 (Spectroscopic and extinction); 34 (IR excess)\\
$\sigma$\,Ori&33&Full catalogue&\\
&34&Members&35+36+37 (Spectroscopic); 38 (X-ray); 39 (Periodic variable)\\
\hline
\end{tabular}
\label{tab:photo_catalogues}
\end{table*}

For the identification of pre-MS members we have used literature
sources that use specific youth indicators to discriminate
between them and older field
stars. Appendix~\ref{literature_memberships} discusses the adopted
literature sources for each SFR, a summary of which is presented in
Table~\ref{tab:photo_catalogues}. For each SFR we have typically used
a range of membership criteria (e.g. X-ray, spectroscopic features,
periodic variability, and IR excess) and these will each have an
associated inherent bias.

To derive consistent pre-MS ages from CMDs
of such populations it is vital that these biases be understood, and
if possible minimised. In brief, IR excesses predominantly identify
objects with circumstellar discs, X-rays are more likely to select
stars that are not actively accreting, H$\alpha$ is biased towards
stars that are actively accreting, and periodic variability
preferentially selects weak-lined T-Tauri stars (although classical
T-Tauri stars can be identified if the temporal density of
observations is sufficiently high;
e.g. \citealp{Littlefair05}). Members that have been selected via
non-spectroscopic methods are more likely to suffer from
foreground and background contamination. 
Spectroscopic indicators are probably
the only unbiased diagnostic, however herein lies a subtle bias which
can be introduced through a pre-selection of candidate targets. Full
spectroscopic coverage of all stars in a given field-of-view is
unfeasible and so a subset of stars is chosen, generally based on
their position in CMD space. This therefore introduces an inherent
bias into the observed subset of stars, however as demonstrated by the
studies of \cite{Kenyon05} and \cite{Burningham05b}, even adopting a
relatively conservative photometric selection does not exclude a
statistically significant number of members, and hence the observed
pre-MS locus is largely unbiased by such pre-selection.

All of the catalogues presented in this study are freely available
from the Cluster Collaboration
home page\footnote[1]{\url{http://www.astro.ex.ac.uk/people/timn/Catalogues/}}
and the CDS archive. Table~\ref{tab:photo_catalogues} shows the full
list of photometric
catalogues, with a sample of the full photometric catalogue for
$\lambda$\,Ori shown in Table~\ref{tab:lamori_photometry}.
Figs.~\ref{fig:ngc2244_diff_iso} -- \ref{fig:chiper_pms} showing CMDs
in the INT-WFC photometric
system, as well as the members catalogues (see
Table~\ref{tab:photo_catalogues}), use data where
stars variable from night-to-night have been replaced by the values
from one night as described in Paper~1. For the data plotted in
Figs.~\ref{fig:ngc2244_diff_iso} -- \ref{fig:chiper_pms} we retain
only those objects with uncertainties in both
colour and magnitude of $\leq 0.1\, \rm{mag}$,
however the full photometric catalogues include all identified
sources. 

\begin{table*}
\caption{A sample of the $\lambda$\,Ori photometric catalogue with colours and magnitudes in the
  INT-WFC photometric system. Due to space restrictions, we only show
  the $\gwfc$ and $\giwfc$ colours and
  magnitudes as a representation of its content. The full photometric
  catalogue (available online) also presents photometry in the
  $\Ugwfc$, $\riwfc$, and
  $\iZwfc$ colours as well as the $\Uwfc$, $\rwfc$, $\iwfc$, and
  $\Zwfc$ magnitudes. Columns list unique
  identifiers for each star in the catalogue: ID, R.A. and
  Dec. (J2000.0), CCD pixel coordinates of the star, and for each of
  $\gwfc$ and $\giwfc$ there is a magnitude,
  uncertainty in the magnitude and a flag (OO represents a `clean
  detection'; see the Cluster Collaboration homepage for a full
  description of the flags.)}
\begin{tabular}{c c c c c c c c c c c c}
\hline
Field&ID&R.A. (J2000.0)&Dec. (J2000.0)&x&y&$g_{_{\rm{WFC}}}$&$\sigma_{
g_{_{\rm{WFC}}}}$&Flag&$(g-i)_{_{\rm{WFC}}}$&$\sigma_{
(g-i)_{_{\rm{WFC}}}}$&Flag\\
\hline
50.02&4&05 35 12.784&+09 36
47.51&973.790&1426.059&8.148&0.010&OS&0.309&0.014&SS\\
54.03&70&05 35 13.200&+10 14
24.79&991.198&929.162&8.221&0.010&OS&0.205&0.014&SS\\
\hline
\end{tabular}
\label{tab:lamori_photometry}
\end{table*}

\subsection{Main-sequence literature data}
\label{literature_data}

\begin{table}
\caption[]{Literature sources for the $UBV$ photometric data.}
\centering
\begin{tabular}{c c}
\hline
SFR&Source\\
\hline
Cep\,OB3b&\cite*{Blaauw59}\\
$\chi$\,Per&\cite{Johnson55}\\
IC\,348&\cite*{Harris54}\\
IC\,5146&\cite{Walker59}\\
$\lambda$\,Ori&\cite{Murdin77}\\
NGC\,1960&\cite{Johnson53}\\
NGC\,2169&\cite{Hoag61}\\
NGC\,2244&\cite{Johnson62}\\
NGC\,2362&\cite{Johnson53}\\
NGC\,6530&\cite{Walker57}\\
NGC\,6611&\cite{Walker61}\\
NGC\,7160&\cite{Hoag61}\\
\multirow{2}{*}{$\sigma$\,Ori}&\cite*{Hardie64}\\
&\cite{Greenstein58,Guetter79}\\
\hline
\end{tabular}
\label{tab:ms_literature}
\end{table}

As the WFC saturates at magnitudes $\gwfc \lesssim 10\, \rm{mag}$ we were
required to include literature data so that we could derive ages,
distances and reddenings from the MS populations. To
determine statistically meaningful parameters from these
objects we require a significant mass range
to; i) display measurable evolution between the ZAMS and TAMS and ii)
derive robust reddening and distance
measurements which can then be applied to the low-mass regime when fitting
the pre-MS data. $UBV$ photometry provides us with the necessary information to
derive all of these quantities. Reddenings can be
calculated from fitting data in the $U-B, B-V$ colour-colour
diagram and distances can be derived through the traditional
MS fitting technique, utilising stars that have yet to turn
off the ZAMS and begin their evolution towards the TAMS. Additionally, the
upper region of the $V, B-V$ CMD is age
sensitive, tracing out the evolution between the ZAMS and TAMS.
Table~\ref{tab:ms_literature} details the
literature sources for the MS photometry for our
sample of SFRs.

To maintain consistency within our $UBV$ photometric data, we used,
whenever possible, the $UBV$ photoelectric
photometry of Johnson and collaborators. These pioneering works defined
and characterised the $UBV$ photometric system, and provide us with the
levels of calibration and consistency required.
Note that robust uncertainties are cited in many of the sources,
however as the uncertainty in colour is always smaller than for the
observed magnitude, we therefore treat the uncertainties as
uncorrelated.

\section{The Models}
\label{the_models}

To derive parameters, such as age and distance, from fitting
photometric data in CMDs we require model isochrones, which must
be calibrated to the same photometric system as that of the data.
Model isochrones are generated using stellar interior models that predict
the bolometric luminosity ($L_{\rm{bol}}$), effective temperature
($T_{\rm{eff}}$) and surface gravity (log$\,g$) for a given mass. These
quantities are then transformed into colours and magnitudes to compare to
measured values. Colours are calculated using a colour-$T_{\rm{eff}}$
relation and magnitudes via bolometric corrections (BCs) to $L_{\rm{bol}}$. Both
relations can be derived by folding atmospheric models through the appropriate
photometric filter responses and we describe the details of doing so
in Appendix~B of Paper~1. Here we describe our choice of interior and
atmospheric models.

\subsection{Pre-main-sequence models}
\label{pre-ms_models}

The pre-MS interior models used in this study are the same as those
adopted in Paper~1, namely those of \cite{Baraffe98} with a
solar-calibrated mixing-length parameter $\alpha=1.9$,
\cite{D'Antona97} and \cite{Dotter08} models (hereafter BCAH98 $\alpha=1.9$, DAM97 and
DCJ08 respectively). Note that the interior models of
\cite{Baraffe98} with a general mixing-length parameter $\alpha=1.0$
and \cite*{Siess00} are not used in this analysis as both models
systematically overestimate stellar
luminosities for masses $\gtrsim 0.6\, \msun$ and fail to match the
observed Pleiades MS (see Paper~1).

\subsection{Main-sequence models}
\label{main-sequence_models}

We adopted the Geneva interior models of \cite{Lejeune01}, specifically
the basic `c' grid \citep{Schaller92}. The rate of change
of $L_{\rm{bol}}$ and $T_{\rm{eff}}$ with time changes discontinuously at the
TAMS, and as the grid of interior models is much coarser than that
required, intermediate age models were created by temporally
interpolating between existing calculations. For this an
interpolation routine provided by the Geneva group was used to create
a grid with spacing $\Delta \mathrm{log(age)}=0.02$.

It should be noted that the \cite{Schaller92} MS models do not
consider the pre-MS evolutionary phase, but only that from the ZAMS
onwards. The ages derived for our sample of SFRs
are driven by the most massive stars i.e. those that have evolved
significantly from the ZAMS, which, as the SFR ages, decrease in
mass. For our oldest SFR (NGC\,1960) the age is based on stars with
spectral types of B{\footnotesize II}--{\footnotesize III} (see \citealp{Johnson53}), which,
from the evolutionary models of \cite{Siess00}, have masses of $\gtrsim
7\, \msun$ and take $< 0.5\, \rm{Myr}$
to arrive on the ZAMS. Hence, although the \cite{Schaller92} models do
not include pre-MS evolution, the timescales involved are
significantly shorter than the uncertainties on the derived ages
themselves and therefore can be safely neglected.

\subsection{Atmospheric models}
\label{atmospheric_models}

The atmospheric models used are the same as those in Paper~1 and
consist of \textsc{phoenix} BT-Settl models
\citep*{Allard11}\footnote[2]{\url{http://phoenix.ens-lyon.fr/Grids/BT-Settl/}}
for $400 \leq T_{\rm{eff}} \leq 7800\, \rm{K}$ and the Kurucz
\textsc{atlas9}/ODFnew models
\citep{Castelli04}\footnote[3]{\url{ftp://ftp.stsci.edu/cdbs/grid/ck04models}}
for $8000 \leq T_{\rm{eff}} \leq 50\,
000\, \rm{K}$. Despite differences in the underlying physical
assumptions and adopted parameters, we found that at the transitional
$T_{\rm{eff}}=8000\, \rm{K}$ all derived colours from both sets of
models agreed to within $0.02\, \rm{mag}$.

To transform the MS interior models into the Johnson $UBV$
photometric system, we derived BCs using Eqn.~B2 of
Paper~1 and the
standard $UBV$ responses of \cite{Bessell90}. To account for the
fact that the $UBV$ photoelectric system is based on an energy
integration method, Eqn.~B2 of Paper~1 was modified so that the
factor of $\lambda$ in the integrands on the third term were
removed, such that,

\begin{eqnarray}
\label{bc_final}
  BC_{_{R_{\lambda}}} & = & M_{\rm{bol},\odot} -
  2.5\, \mathrm{log}\left(\frac{4\pi(10\mathrm{pc})^{2}F_{\rm{bol}}}{L_{\odot}}\right) \\
  & + &~2.5\, \mathrm{log}\left(\frac{\int_{\lambda}
      F_{\lambda}10^{-0.4 A_{\lambda}}R_{\lambda}\, d\lambda}{\int_{\lambda}
    f^\circ_{\lambda}R_{\lambda}\, d\lambda}\right) -
   m^\circ_{_{R_{\lambda}}}, \nonumber
\end{eqnarray}

\noindent where $F_{\rm{bol}} = \sigma T^{4}_{\rm{eff}}$ is
the total flux emergent at the stellar surface and all other symbols
retain their original definitions (see Paper~1). For the solar values we used
$M_{\rm{bol}, \odot} = 4.74$ and $L_{\odot} = 3.855 \times 10^{33}\,
\rm{erg\, s^{-1}}$ \citep*{Bessell98}. To define $f^\circ_{\lambda}$
and $m^\circ_{\lambda}$ we required a Vega reference spectrum and for
this we used the CALSPEC
alpha\_lyr\_stis\_005\footnote[4]{\url{http://www.stsci.edu/hst/observatory/cdbs/calspec.html}}
spectrum with $V=0.03$ and all colours equal to zero.
We hereafter refer to the transformed MS
isochrones as the Geneva-Bessell isochrones.

In the MS interior models, some of the most luminous stars have
associated log$\,g$ values that lie just below the range provided by
the atmospheric models. In such cases, it was necessary to extrapolate
the models by setting the colour equal to that of the nearest
log$\,g$. As the log$\,g$ dependence is very small at such
$T_{\rm{eff}}$, this extrapolation affects the model isochrone at $\ll
0.01\, \rm{mag}$ level.

Note that the commonly referred to
colour-$T_{\rm{eff}}$ relation can also be represented simply in terms
of BCs as it is the difference in the BCs at a specific $T_{\rm{eff}}$
that explicitly defines the colour of the star i.e.

\begin{equation}
B-V(T_{\mathrm{eff}})=BC_{V}(T_{\mathrm{eff}})-BC_{B}(T_{\mathrm{eff}}), \nonumber
\end{equation}

\noindent and so from now on the transformation from H-R to CMD space will be
discussed in terms of the BC-$T_{\rm{eff}}$ relation.

\subsection{Reddening vectors in the Bessell system}
\label{reddening_vectors_in_the_bessell_system}

In broadband photometry, the effective wavelength of a given bandpass
moves depending on the incident stellar flux distribution, and hence the colour
(and gravity) of the star (see \citealp{Bessell98}). This variation
results in a non-linear reddening vector in the
$U-B, B-V$ colour-colour diagram, which further depends, albeit to a lesser
extent, on the intrinsic reddening of the observed source
\citep{Hiltner56,Wildey63}. 

\begin{figure}
\centering
\includegraphics[width=\columnwidth]{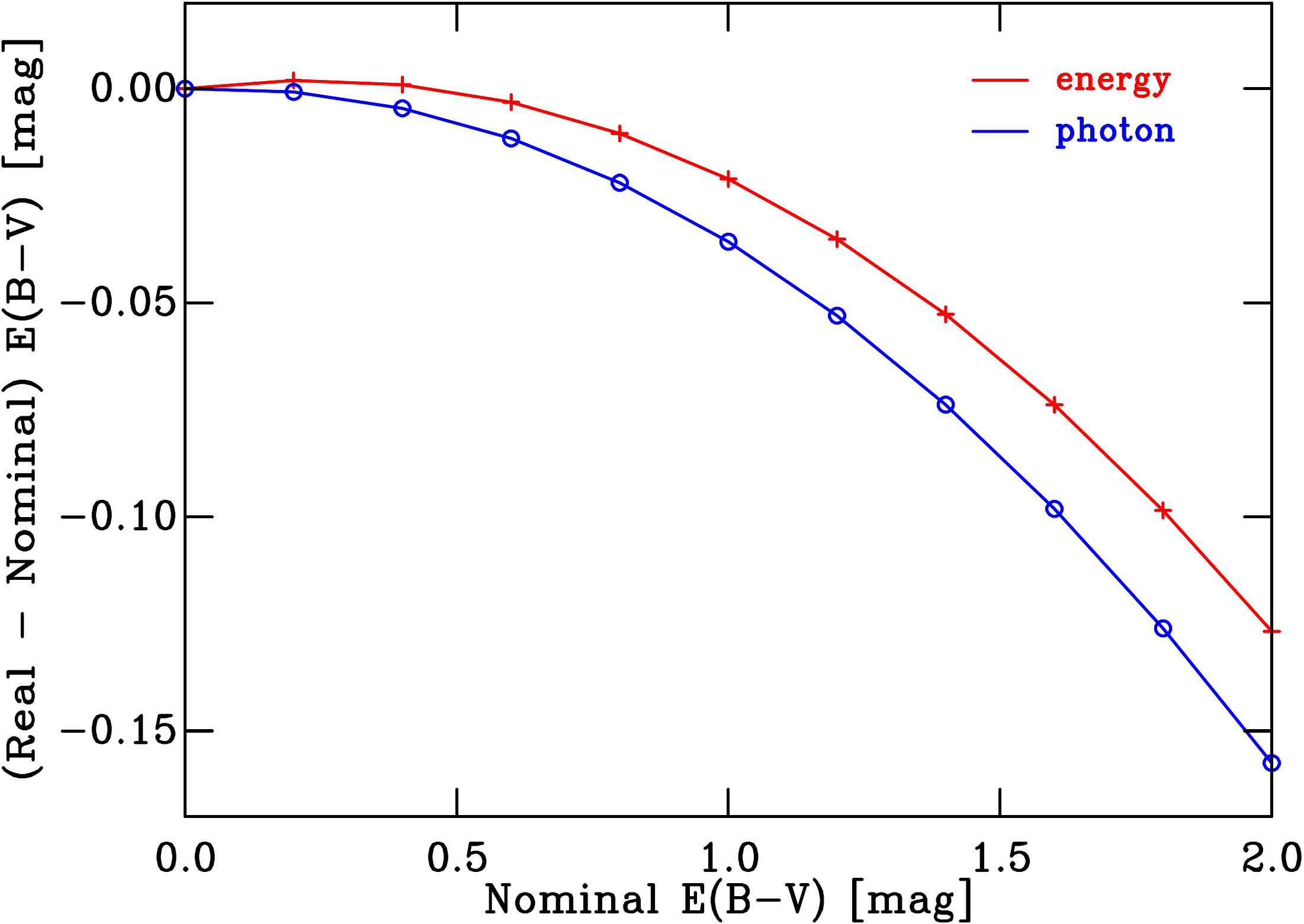}
\caption[]{Shows the difference between the nominal $E(B-V)$ as
  predicted by the parameterised extinction law of \protect\cite{Cardelli89}
  and that derived from the \textsc{atlas9}/ODFnew atmospheric models
  at an intrinsic colour of $(B-V)_\circ = -0.15$ using energy
  integration (red) and photon counting (blue) flux measurements as a
  function of nominal $E(B-V)$.}
\label{fig:ebmv}
\end{figure}

Whilst canonical linear reddening laws (such as the standard $E(U-B)=0.72
\times E(B-V)$; \citealp{Johnson53}) can be used to de-redden stars
photometrically, our derived age is heavily dependent upon the
subtle evolution between the ZAMS and TAMS, and therefore, for the
level of precision required in this study it is important to
understand how stars of varying colour move as a function of reddening
in the $U-B, B-V$ colour-colour diagram.
We can quantitatively model these
reddening vectors by reddening the atmospheric models according to the
parameterised extinction law of \cite*{Cardelli89} and folding them
through the standard $UBV$ responses of \cite{Bessell90}.
Note that we only use atmospheric models with
intrinsic colours $(B-V)_\circ \leq 0.05$ to derive the reddening vectors. The
parameterised extinction law can be visualised in terms of
a nominal $E(B-V)$ (hereafter
$E(B-V)_{\rm{nom}}$) which represents the amount of intervening
interstellar material between an observer and the object, but does not
necessarily correspond precisely to the observationally measured $E(B-V)$ value
since this value will vary with the colour of the star.

To determine the relationship between the nominal and measured $E(B-V)$
values we reddened the atmospheric models using the
\cite{Cardelli89} extinction law in steps of 0.2 from
$E(B-V)_{\rm{nom}}=0.0-2.0$. For the value of the total-to-selective
extinction ratio we adopted $R_{V}=3.2$ as an appropriate value
for the diffuse interstellar medium (ISM).
Measured $E(B-V)$ values were inferred from
the reddened atmospheric models at an intrinsic colour
$(B-V)_\circ=-0.15$. Fig.~\ref{fig:ebmv} shows the difference
between the nominal and measured $E(B-V)$ as a function of
$E(B-V)_{\rm{nom}}$ and highlights the non-linear relation between the two
quantities.

As discussed in \cite{Bessell90}, the Johnson $B-V$ colour is best
reproduced through the use of the standard $B$ and $V$ bandpasses, however,
the $U-B$ colour is best represented using the modified
$U_{\rm{x}}$ and $B_{\rm{x}}$ responses (which account for
increased atmospheric absorption; see
\citealp{Bessell90}). Therefore, for both the calculated colours and
reddening vectors we have used the $U_{\rm{x}}$ and $B_{\rm{x}}$
responses for the $U-B$ colour. The reddening vectors are,

\begin{eqnarray}
\label{av_ebmv}
\frac{A_{V}}{E(B-V)} & = & 3.264 + \{0.088 \times E(B-V)\} \\
& + & \{0.018 \times E(B-V)^{2}\} + \{0.450 \times (B-V)_\circ\} \nonumber \\
& + & \{0.333 \times (B-V)_\circ^{2}\}, \nonumber
\end{eqnarray}

\noindent and

\begin{eqnarray}
\label{eumb_ebmv}
\frac{E(U-B)}{E(B-V)} & = & 0.687 + \{0.061 \times E(B-V)\} \\
& + & \{0.013 \times E(B-V)^{2}\} - \{0.064 \times (B-V)_\circ\}, \nonumber
\end{eqnarray}

\noindent where the dependence on both the colour of the star and the
intrinsic reddening of the source are explicitly incorporated. These
reddening vectors are accurate to within $0.01\, \rm{mag}$, however this degrades
to $\simeq 0.03\, \rm{mag}$ at the limiting $E(B-V) \sim 1.9$.

The value of $R_{V}$ is typically assumed to be a
function of the line-of-sight towards a given object. Whereas in most
cases it is appropriate to assume that $R_{V} \simeq 3.2$, there is
substantial evidence that this is not always the case and that for
very young SFRs the reddening law may be significantly
different. The effect of this on our results is discussed further in
Section~\ref{anomalous_extinction} for our sample of SFRs.

\section{Fitting the young main-sequence}
\label{fundamental_parameters_from_ms_stars}

We used the $\tau^{2}$ fitting statistic\footnote[5]{The $\tau^{2}$
  fitting statistic
  software and implementation details are available at
  \url{http://www.astro.ex.ac.uk/timn/tau-squared}} of \cite{Naylor06}
and \cite{Naylor09}
to derive ages, distances and reddenings from the MS
populations of our sample of SFRs.
The $\tau^{2}$ fitting statistic can be viewed as a generalisation of the
$\chi^{2}$ statistic where both the model isochrone and the data
are two-dimensional distributions (uncertainties in both colour and
magnitude for the photometric data and a widening of the
isochrone due to an intrinsic binary fraction). The best-fit model is
found by minimising $\tau^{2}$.
We used the Geneva-Bessell models to fit the
MS stars, with reddenings calculated using the $U-B, B-V$
colour-colour diagram, and ages and distances derived simultaneously
using the $V, B-V$ CMD.

The shaded region in Fig.~\ref{fig:ngc1960_ebmv}, and all subsequent
figures showing photometric data fitted using $\tau^{2}$, represents
the two-dimensional model distribution. In this context, the model
distributions are probability density distributions (the term $\rho$
in Eqn.~1 of \citealp{Naylor06}) in CMD space obtained by applying
binarity to a model isochrone at a given age. In all
figures showing fitted data, for which a mean reddening was derived
(see Section~\ref{uniform_reddening}), the models have been adjusted
to apparent colour-magnitude space using the measured $E(B-V)$ and the
reddening vector given in
Section~\ref{reddening_vectors_in_the_bessell_system}. For regions
where the reddening was found to be spatially variable (see
Section~\ref{variable_reddenning}) the age and distance were
calculated by fitting the photometric data in the de-reddened
intrinsic colour-magnitude space.

\subsection{The model CMD}
\label{model_cmd}

We first created the probability distribution function to fit to the
photometric data.
The model CMD is created using a Monte Carlo method to simulate
$10^{6}$ stars over a given
mass range. An important part of our fitting method is that we include
binaries, though as shown in \cite{Naylor06} for low-mass stars and
\cite{Naylor10} for high-mass stars the precise form of the mass-ratio
distribution and binary fraction used is not crucial. For masses of
$14.4\, \msun$ and greater we followed the formalism of
\cite{Naylor10} and assumed an O-star binary fraction of 75 per
cent. Of these, 25 percent are evenly distributed over $0.95 < q
\leq 1.0$, 75 per cent are evenly distributed over $0.2 \leq q \leq
0.95$, and a lower restriction of $q \geq 0.2$ is adopted. For masses
below $14.4\, \msun$ we adopted a binary fraction of 50 per cent and a
uniform secondary distribution ranging from zero to the mass of the
primary.

The mass function adopted does not have a significant impact on the
best-fitting parameters (see \citep{Naylor09}), and so for each star
a mass is drawn from a
power law distribution ($dN/dM \propto M^{-1.35}$) which results in a
roughly even distribution of stars as a function of magnitude. If the
star happens to be a binary the companion mass is
assigned as described above. The interior model then
assigns an $L_{\rm{bol}}$, $T_{\rm{eff}}$ and log$\,g$ which can be
converted into CMD space using the appropriate BC-$T_{\rm{eff}}$
relation. Binary companions that lie below the lower mass limit of the
interior models are assumed to have a flux equal to zero and thus
provide no contribution to the overall luminosity of the binary
system.

\subsection{Extinction fitting}
\label{extinction_fitting}

\begin{figure}
\centering
\includegraphics[width=\columnwidth]{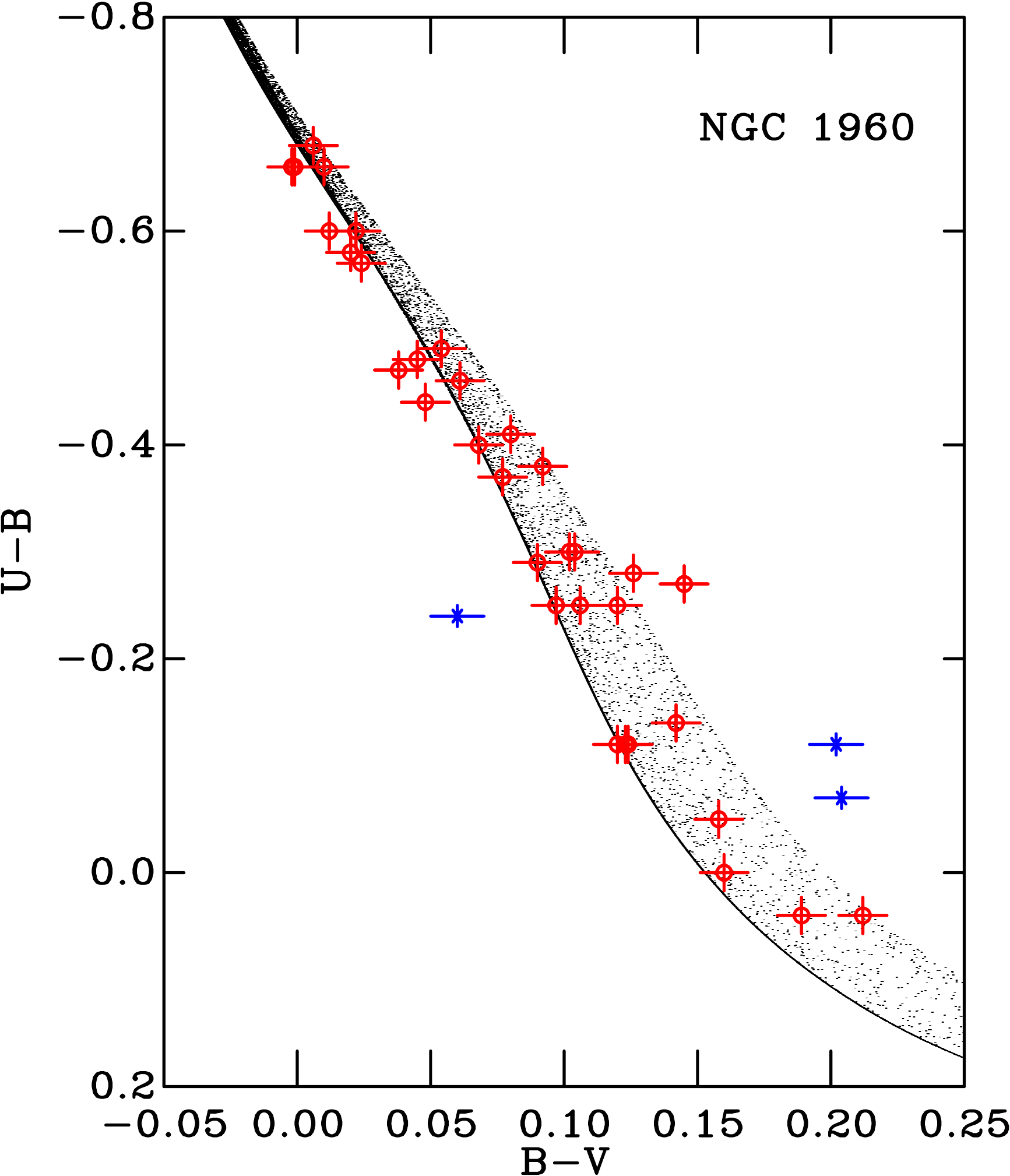}
\caption[]{The best-fitting $U-B, B-V$ colour-colour diagram for
  NGC\,1960 with a measured uniform reddening of
  $E(B-V)=0.20$. Circles represent the data of
  \protect\cite{Johnson53}, with the associated uncertainties shown as
  the bars. Asterisks represent stars that were clipped before
  deriving the best-fit (see text).}
\label{fig:ngc1960_ebmv}
\end{figure}

Having established the importance of the reddening (see
Section~\ref{reddening_vectors_in_the_bessell_system}) we must first
derive this, as changing this value will subsequently modify the
age and distance derived by fitting the $V, B-V$ CMD. For this, we
followed the two-step method as described in \cite{Mayne08}.

\subsubsection{Uniform reddening}
\label{uniform_reddening}

For each SFR a mean reddening was determined by fitting a $U-B, B-V$ model 
isochrone (which included an intrinsic binary fraction) using the reddening vector
given in Eqn.~\ref{eumb_ebmv}.
We used $\tau^2$ to find the best fitting extinction.
An example of such a fit is shown in Fig.~\ref{fig:ngc1960_ebmv} for NGC\,1960. 
Only stars defined as members by \cite{Johnson53} and those bluewards of
$B-V=0.25$ were used. We further removed three additional stars (Boden\,50, 86 and 110)
due to a combination of their positions in the $U-B, B-V$
colour-colour diagram and high $\tau^{2}$ values (see
Fig.~\ref{fig:ngc1960_ebmv}).
For the remaining stars the best-fitting $E(B-V)$ was 0.20. 

A second, less successful example of fitting the $U-B, B-V$ colour-colour 
diagram assuming a uniform reddening is shown in the left panel of 
Fig.~\ref{fig:lamori_red_dered} for the $\lambda$\,Ori association. 
Only stars identified as members by \cite{Murdin77} have been used. 
The best-fiting $E(B-V)$ is 0.14, but as is clear from Fig.~\ref{fig:lamori_red_dered}
the data show an unacceptable scatter about the fitted sequence.
This conclusion is supported by the associated $\Pr(\tau^{2})$ which
is $\simeq 10^{-5}$ (as compared to 0.2 for NGC\,1960).
$\Pr(\tau^{2})$ is exactly analogous to $\Pr(\chi^{2})$, giving
the probability that a dataset resulting from observing a SFR
whose parameters were those of the best-fit, would yield a value of
$\tau^{2}$ which was greater than that actually obtained. 
In this case it is indicating that the model is an unacceptable fit to the data, implying the
reddening is spatially variable across the $\lambda$\,Ori
association. 

Our adopted procedure, therefore, is to fit the $U-B, B-V$ data for all the SFRs with a 
uniform extinction model.
If $\Pr(\tau^{2})>0.05$ the best fitting extinction is adopted.
In all such cases the uncertainties in $E(B-V)$ were less than $\pm\, 0.01\, \rm{mag}$,
and so are negligible in our analysis. 
If the fit is poor (i.e. $\Pr(\tau^{2}) < 0.05$), we de-redden the stars individually using 
the method described below (see also the right panel of Fig.~\ref{fig:lamori_red_dered}). 

\subsubsection{Variable reddening}
\label{variable_reddenning}

\begin{figure}
\centering
\includegraphics[width=\columnwidth]{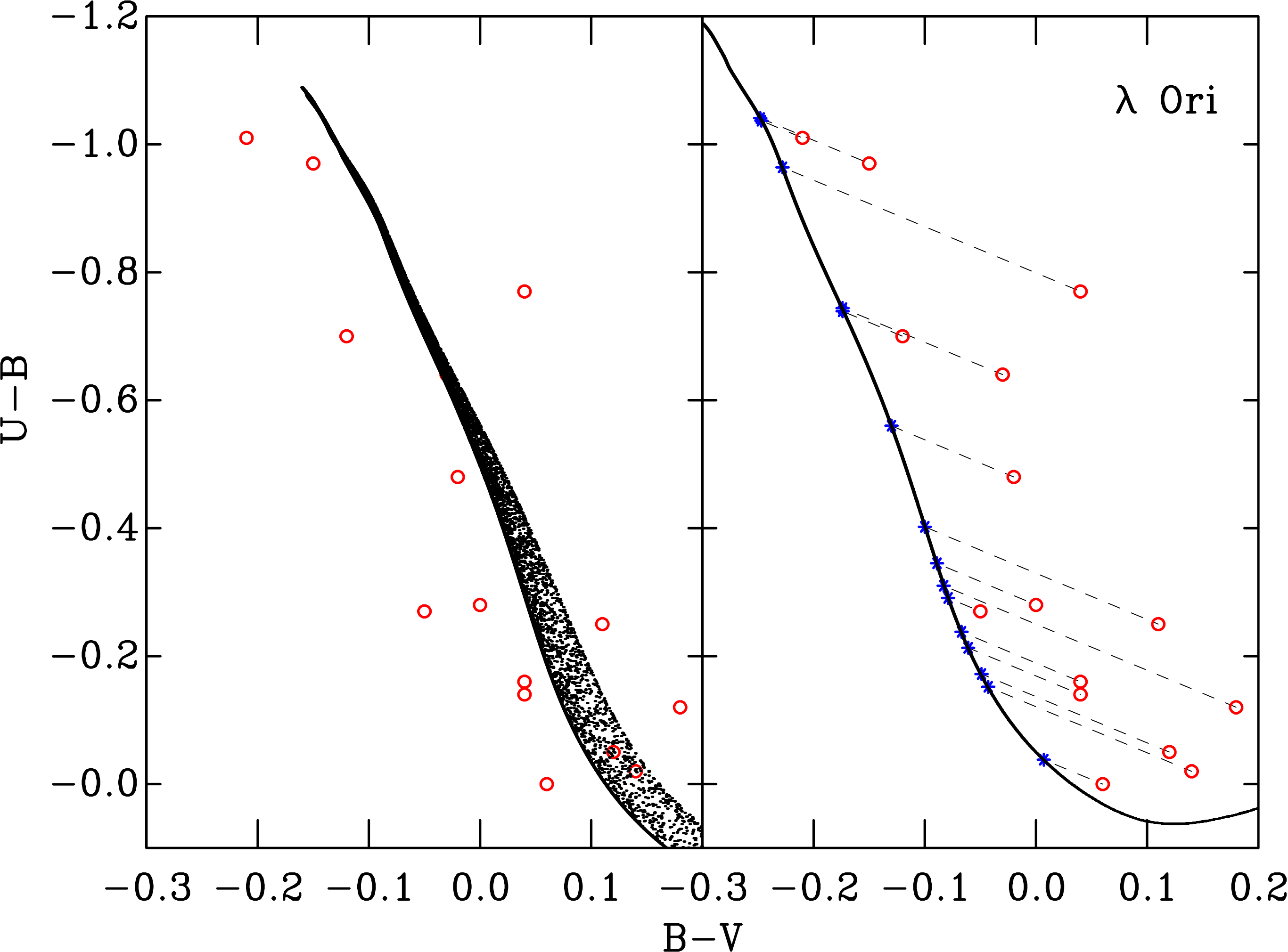}
\caption[]{Demonstration of the two methods used for calculating reddenings
  adopted in this paper.
  \textbf{Left panel:} The best-fitting $U-B, B-V$ colour-colour diagram for
  $\lambda$\,Ori with a measured uniform reddening of
  $E(B-V)=0.14$. Circles represent the data of \cite{Murdin77}.
  \textbf{Right panel:} The photometric data has been de-reddened according to the revised
  Q-method, with the asterisks corresponding to the de-reddened colours.
  Individual reddening trajectories are plotted as the dashed lines. The solid
  line is a $1\, \rm{Myr}$ Geneva-Bessell model isochrone.}
\label{fig:lamori_red_dered}
\end{figure}

To determine individual stellar reddenings we have
used a revised Q-method.
The Q-method is used to photometrically de-redden stars individually
using the $U-B, B-V$ colour-colour diagram. Whilst we carry out this
process numerically, \cite{Johnson53} parameterised the intersection of
a linear MS and linear reddening vectors to create an
extinction independent colour, Q. As noted in \cite{Mayne08}, this can result in
errors in the extinction $A_{V}$ of up to $0.1\, \rm{mag}$ due to
assuming a linear MS with
an additional error of $0.08\, \rm{mag}$ through the use of
colour independent reddening vectors. We therefore fitted a straight
line to the
Geneva-Bessell model isochrone and incorporated the colour-
and extinction-dependent reddening vector shown in Eqn.~\ref{eumb_ebmv}
to give

\begin{eqnarray}
Q & = & \{0.064 \times (B-V)^{2}\} \\
& - & (B-V)\{ [0.013 \times E(B-V)^{2}] \nonumber \\
& + & [0.061 \times E(B-V)] - 3.451\} - 0.006. \nonumber
\end{eqnarray}

By replacing the original Q-method MS straight line with a line fitted
to a section of the Geneva-Bessell model isochrone, it is necessary to
assume a given age. Unlike the evolution of the MS in the $V, B-V$
CMD, the MS in the $U-B, B-V$ colour-colour diagram moves very little
with age (see \citealp{Mayne08}). As a given sequence ages, stars of
increasingly lower mass evolve away from the MS. Hence when using the
revised Q-method to de-redden stars individually, it is important to
ensure that any post-MS objects are not included as these will be
incorrectly de-reddened and therefore occupy the wrong position in both
the $(U-B)_{\circ}, (B-V)_{\circ}$ colour-colour and $V_{\circ},
(B-V)_{\circ}$ CMD.

Where possible, it is better to fit for a mean reddening using the $\tau^{2}$
technique. The revised Q-method implicitly assumes that all stars are
either single-stars or equal-mass binaries and is thus unable to account
for the scatter in CMD space as a result of binarity and
photometric uncertainties. The effects of binarity are non-negligible.
Although in colour-colour space the single-star and equal-mass binary
sequences are co-incident, not knowing whether the star should lie on that 
sequence or somewhere in the unequal-mass binary region (see the
unequal-mass binary distribution in
Fig.~\ref{fig:lamori_red_dered}) can affect the
derived extinction on the $\simeq 0.15\, \rm{mag}$ level
\citep{Mayne08}. This effect
becomes more marked as one moves towards lower masses (redder colours)
along the MS model isochrone.

\subsubsection{Anomalous line-of-sight extinction}
\label{anomalous_extinction}

There have been numerous suggestions in the literature that the reddening law towards
very young SFRs may differ from that characteristic of the normal
ISM. Such anomalous reddenings may be caused by
photoevaporation of small dust grains by nearby massive stars or grain
growth in circumstellar environments \citep{Cardelli88,VandenAncker97}, resulting
in large values for the total-to-selective extinction ratio
$R_{V}$. It has been shown explicitly that there is an
anomalous reddening law towards NGC\,6611 (e.g. \citealp{Hiltner69,Hillenbrand93}). Polarimetric
observations by \cite*{Orsatti00,Orsatti06} have shown that the size of
silicate and graphite dust grains in NGC\,6611 might be larger than
those in the typical ISM. Values of $R_{V}$ range from $3.5-4.8$, with
a typical value of $\simeq 3.75$ \citep{Hillenbrand93}.

We have therefore recalculated the reddening vectors (Eqns.~\ref{av_ebmv} and
\ref{eumb_ebmv}) adopting this typical value of
$R_{V}=3.75$, and derived the MS age, distance and reddening 
for NGC\,6611 under
this assumption (see Table~\ref{ms_fitting_results}). Note,
however, that due to possible variations in
$R_{V}$ in NGC\,6611 the distance derived in
Section~\ref{age_and_distance_fitting} based on MS fitting may (in the
most extreme cases) be up to $0.18\, \rm{mag}$ larger or $0.75\,
\rm{mag}$ smaller. The effects of this uncertainty in the derived
distance are further discussed in Section~\ref{discussion}.

\subsection{Age and distance fitting}
\label{age_and_distance_fitting}

With a calculated mean reddening for a given SFR the next step is to redden
the model isochrones so that they can be used to fit the data in the
$V, B-V$ plane for distance and age. We used
the calculated reddening vector shown in Eqn.~\ref{av_ebmv} to create
model isochrones at the appropriate reddening. For
SFRs that showed variable reddening, and were hence de-reddened using the
revised Q-method, the model isochrones were left in intrinsic
$V_\circ, (B-V)_\circ$ space.

\begin{figure}
\centering
\includegraphics[width=\columnwidth]{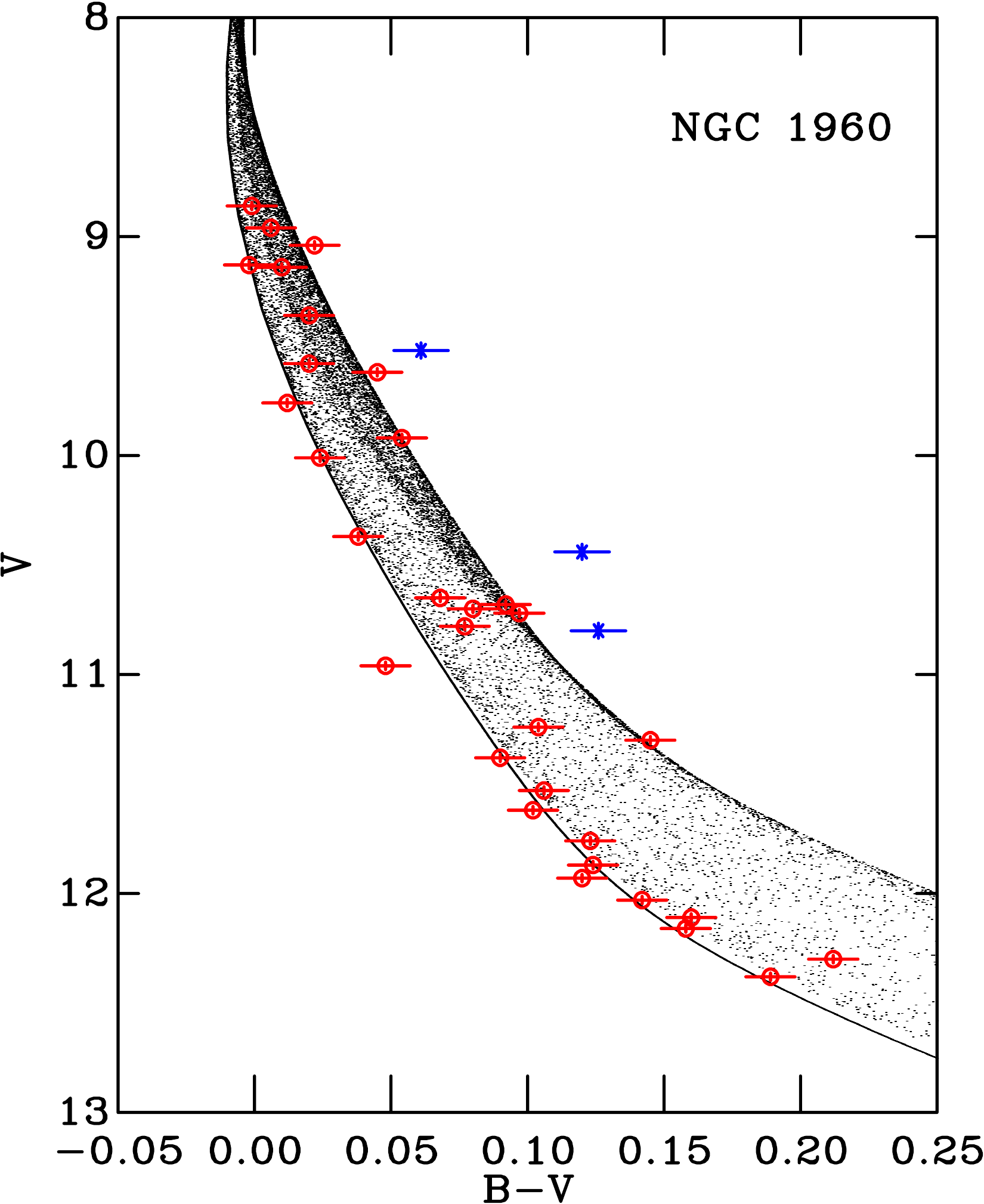}
\caption[]{The best-fitting $V, B-V$ CMD for NGC\,1960
  with a derived age of $26.3\, \rm{Myr}$ and
  distance modulus $dm=10.33$. Circles represent the data of
  \protect\cite{Johnson53}, with the associated uncertainties shown as
  the bars. Asterisks represent stars that were clipped before
  deriving the best-fit (see text).}
\label{fig:ngc1960_ms_cmd}
\end{figure}

\begin{figure}
\centering
\includegraphics[width=\columnwidth]{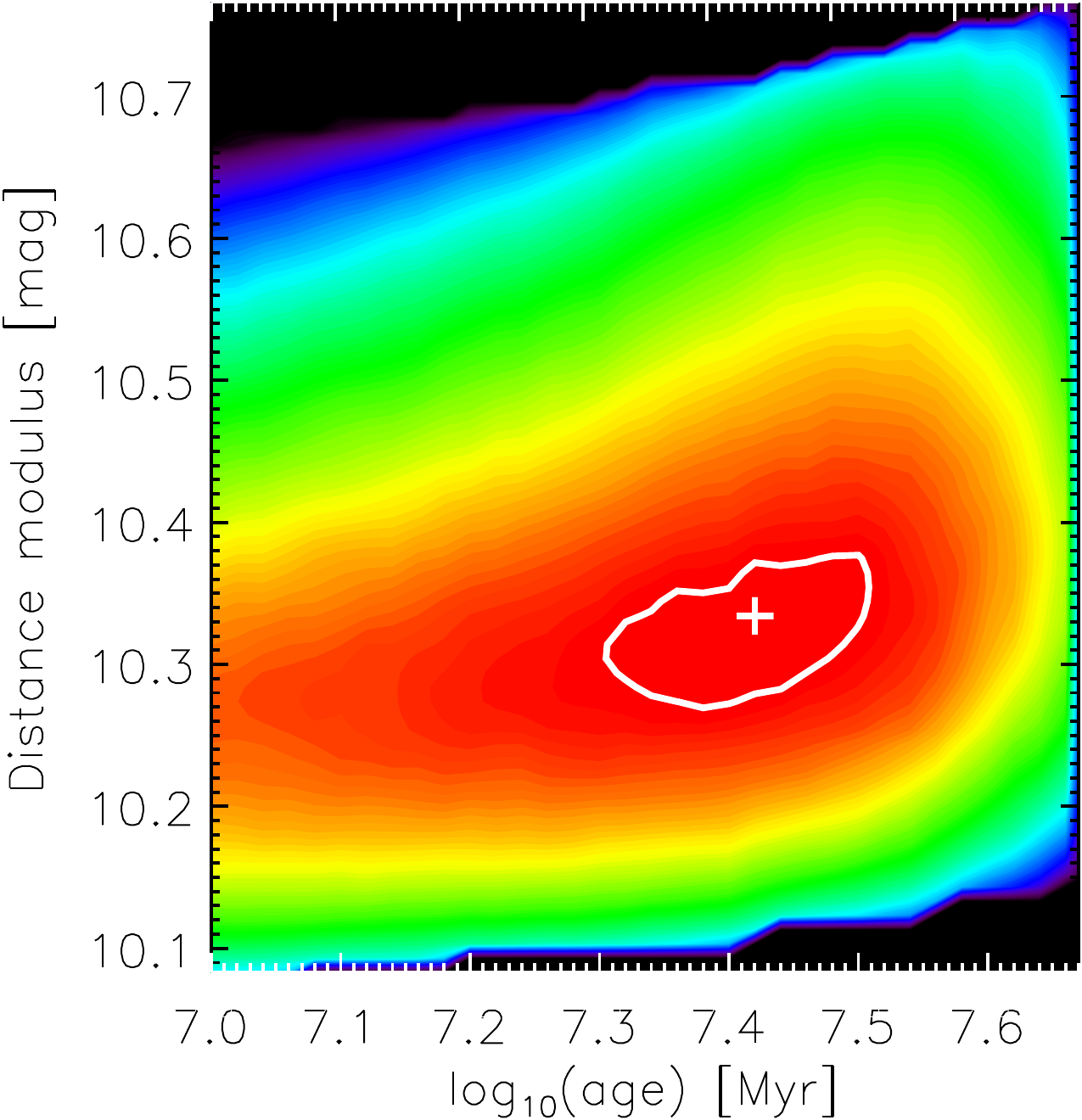}
\caption[]{The colour scale $\tau^{2}$ age-distance grid for
  NGC\,1960. The large cross denotes the lowest $\tau^{2}$ within the
  grid and hence the derived best-fitting values for the age and
  distance. The contour is at the 68 per cent level and defines the
  uncertainties in the derived age and distance.}
\label{fig:ngc1960_tau2_grid}
\end{figure}

\begin{figure*}
\centering
\includegraphics[width=\textwidth]{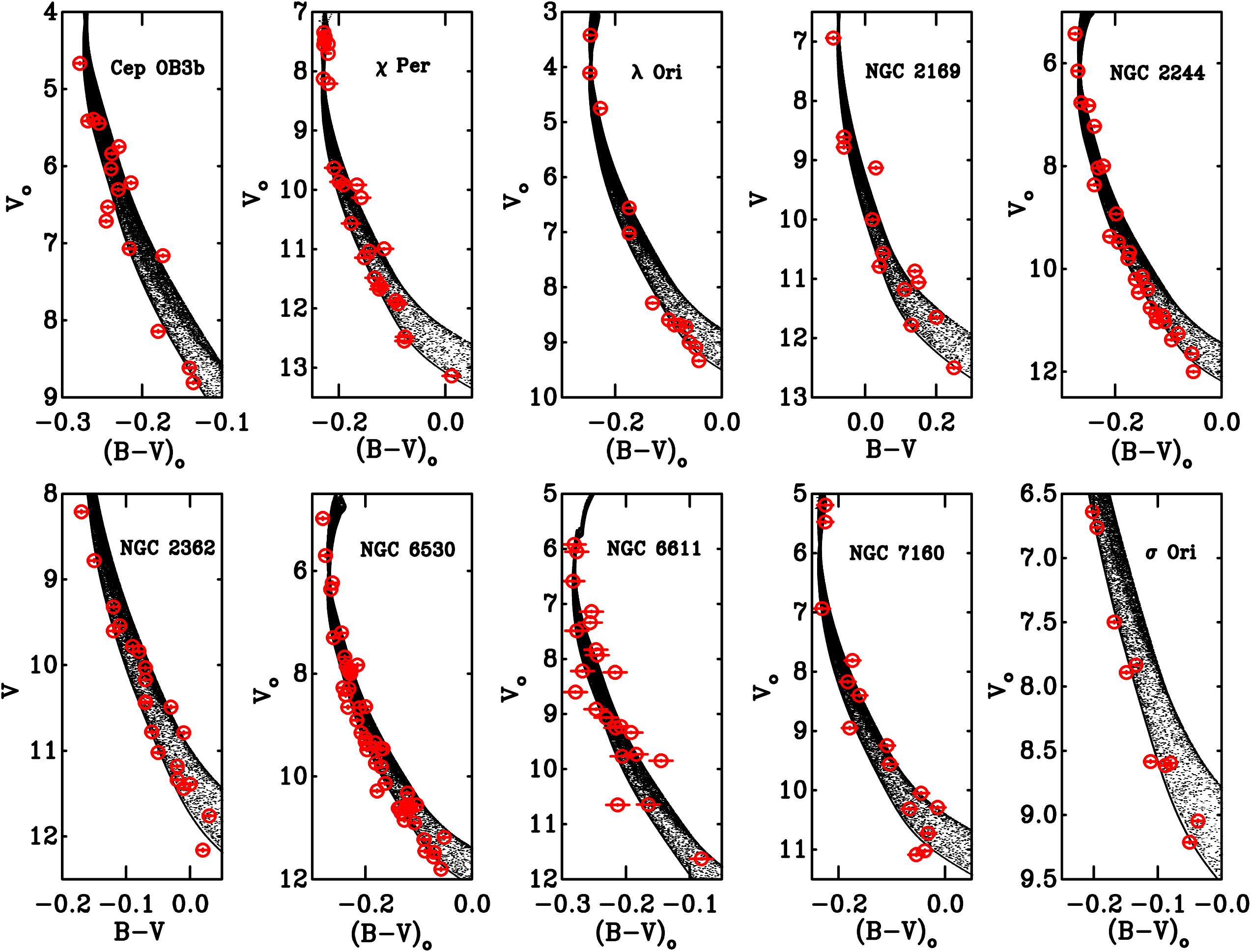}
\caption[]{The best-fitting $V, B-V$ CMDs for the remainder of the
  SFRs in our sample with the derived parameters shown in
  Table~\ref{ms_fitting_results}. Notes are as follows. (1)
  \textbf{NGC\,6611:} There are
  several stars that lie above the single-star $(U-B)_{\circ},
  (B-V)_{\circ}$ model isochrone in the colour-colour diagram and due to the
  nature of the Q-method, it was not possible to de-redden these
  sources. These stars were therefore omitted from the derived $E(B-V)$
  determination and the subsequent age-distance fit in the $V_{\circ},
  (B-V)_{\circ}$ CMD. (2) \textbf{$\sigma$\,Ori:} The $UBV$ data of
  \protect\cite{Hardie64} only provides a combined measurement for
  $\sigma$\,Ori\,ABC, where $\sigma$\,Ori\,A itself is a double-lined
  spectroscopic binary. Due to this, and the fact that our MS model
  isochrones do not include the effects of higher order multiple systems
  other than binaries, we neglect the combined measurement for
  $\sigma$\,Ori\,ABC and instead use the photometry for
  $\sigma$\,Ori\,C by \protect\cite{Greenstein58}. The fact that the
  most luminous stars have not been used results in a poorly
  constrained MS age for the SFR.}
\label{fig:ms_ages}
\end{figure*}

An example of fitting the $V, B-V$ CMD for age and distance
simultaneously is shown in Fig.~\ref{fig:ngc1960_ms_cmd} for
NGC\,1960. After fitting for $E(B-V)$, we removed three stars
(Boden\,13, 47 and 48) based on a combination of their positions in the
$V, B-V$ CMD and their associated $\tau^{2}$ values (see
Fig.~\ref{fig:ngc1960_ms_cmd}). We fitted the
remaining stars and calculated an age of $26.3^{+3.2}_{-5.2}\,
\rm{Myr}$ and a distance modulus $dm=10.33^{+0.02}_{-0.05}$ with
$\Pr(\tau^{2})=0.67$. Fig.~\ref{fig:ngc1960_tau2_grid} shows the
corresponding $\tau^{2}$ age-distance grid for NGC\,1960. The large
cross defines the lowest $\tau^{2}$ within the grid and therefore the
best-fitting values. The contour is at the 68 per cent level and
defines the uncertainties in the derived age and distance.
The best-fitting $V, B-V$ CMDs for the remainder of our sample of SFRs
are shown in Fig.~\ref{fig:ms_ages}

\begin{table*}
\caption[]{Derived ages, distances and reddenings from the MS
  populations of the SFRs. The uncertainties in the age and distance
  were calculated using the $\tau^{2}$ fitting statistic
  and represent the 68 per cent confidence
  level. Notes are as follows. (1) Individual
  reddenings derived using the revised Q-method with the median $E(B-V)$ value
  quoted and the full range in $E(B-V)$ shown in parentheses. (2)
  Parameters derived assuming the total-to-selective extinction ratio
  $R_{V}=3.75$ (see Section~\ref{anomalous_extinction}). (3)
  Unable to calculate MS age due to insufficient number of evolved stars.}
\begin{tabular}{c c c c c c c}
\hline
SFR&\multicolumn{2}{|c|}{MS Age (Myr)}&\multicolumn{2}{|c|}{Distance modulus
$dm$}&$\Pr(\tau^{2})$&$E(B-V)$\\
&Best-fit&68 per cent&Best-fit&68 per cent&&\\
\hline
NGC\,6611$^{(1,2)}$&4.6&3.9--6.0&11.38&11.08--11.44&0.17&0.71 (0.58)\\
Cep\,OB3b$^{(1)}$&6.0&3.8--6.6&8.78&8.70--8.84&0.26&0.89 (0.41)\\
NGC\,6530$^{(1)}$&6.3&5.7--7.0&10.64&10.59--10.68&0.98&0.32 (0.23)\\
NGC\,2244$^{(1)}$&6.6&5.8--7.4&10.70&10.67--10.75&0.70&0.43 (0.18)\\
$\sigma$\,Ori$^{(1)}$&8.7&4.7--13.4&8.05&7.99--8.11&0.16&0.05 (0.12)\\
$\lambda$\,Ori$^{(1)}$&10.0&8.9--11.0&8.02&7.99--8.06&0.49&0.11 (0.24)\\
NGC\,2169&12.6&10.5--17.6&9.99&9.90--10.06&0.20&0.16\\
NGC\,2362&12.6&7.9--15.3&10.60&10.57--10.66&0.31&0.07\\
NGC\,7160$^{(1)}$&12.6&10.5--13.9&9.67&9.62--9.76&0.29&0.37 (0.37)\\
$\chi$\,Per$^{(1)}$&14.5&12.8--16.7&11.80&11.77--11.86&0.49&0.52 (0.28)\\
NGC\,1960&26.3&21.1--29.5&10.33&10.28--10.35&0.67&0.20\\
IC\,348$^{(1,3)}$&--&--&6.98&6.89--7.17&--&0.69 (0.52)\\
IC\,5146$^{(1,3)}$&--&--&9.81&9.62--10.01&--&0.75 (0.61)\\
\hline
\end{tabular}
\label{ms_fitting_results}
\end{table*}

The MS
populations of both IC\,348 and IC\,5146 lack a sufficient
number of evolved stars to constrain a MS age. Individual stars were
de-reddened using a combination of the revised Q-method and spectral
types to determine the best-fit solution. A photometric parallax
distance for both SFRs was calculated in an identical manner to the
fitting routine used for the other SFRs in our sample.

The ages, distances and
reddenings for all SFRs derived using the Geneva-Bessell model isochrones
are shown in Table~\ref{ms_fitting_results}.
For SFRs where the reddening appears uniform, we show only
the mean $E(B-V)$, whereas for SFRs with variable reddening we note the
median $E(B-V)$ and the full range in $E(B-V)$ for all MS members in
parentheses.

\subsection{Discussion}
\label{discussion_ms_results}

In this section a self-consistent set of MS ages, distances and
reddenings have been derived for a sample of young ($< 30\, \rm{Myr}$)
SFRs to, in most cases, a higher level of precision than that existing
in the literature, with statistically meaningful uncertainties in the
derived ages and distances. It is instructive to place these new
derivations in context by comparing them with previous determinations
for these regions. Seven of the SFRs studied here have also been
investigated by \cite{Naylor09}. Comparing these results, the most
obvious conclusions that can be drawn are; i) the best-fit MS ages
presented in this study are, in all but one case, older than those in
\citeauthor{Naylor09}, ii) the distances derived here are consistent
with those of \citeauthor{Naylor09}, and iii) the associated
$\Pr(\tau^{2})$ values in this study are, in general, higher than
those of \citeauthor{Naylor09}, indicating that the models adopted in
this study represent a better fit to the photometric data.

The older ages derived in this study are primarily due to adopting
different atmospheric models to those of \cite{Naylor09}, and to a
lesser extent, to the use of colour- and extinction-dependent reddening
vectors. The difference is most significant in the $U$-band which
affects the position of de-reddened stars and therefore the age
required to best-fit the photometric data.
Given that both the interior models and the photometric datasets used
in this study and that of
\citeauthor{Naylor09} are the same, the fact that the associated
$\Pr(\tau^{2})$ values for the MS distance-age fits are higher in this
study suggests that the revised model atmospheres provide a better fit to the
data and consequently that the use of the updated atmospheric models
is correct. Furthermore, it suggests that the conversion from H-R to
CMD space derived in Section~\ref{atmospheric_models} represents an
improved description of the Johnson $UBV$ photometric system compared
to that of \cite{Bessell98}. Finally, it implies that the MS parameters
derived in Section~\ref{age_and_distance_fitting} are more robust than
those of \cite{Naylor09}.

\section{Semi-empirical pre-main-sequence
  isochrones}
\label{creating_recalibrated_semi-empirical_pre-ms_isochrones}

As discussed in Section~\ref{the_models}, BC-$T_{\rm{eff}}$ relations can be derived by folding
the flux distribution from model atmospheres through the filter
responses for the appropriate photometric system. However, it is well
known that such a procedure overestimates the optical flux for stars
cooler than $4000\, \rm{K}$, a result we quantified in Paper~1. This is
thought to be because the model atmospheres have an incomplete
description of the opacity in the optical. Given that the differences between
the theoretical and empirical BCs can be as much as $\simeq 0.28\,
\rm{mag}$, corresponding to
a factor 2 difference in age, it is clear we must use empirical BCs for $T_{\rm{eff}}$ lower
than $4000\, \rm{K}$.

The usual source for empirical BCs has been observations of MS stars
(e.g. \citealp{Johnson66,Schmidt82,Bessell90b,Flower96}). The problem
with such an approach for pre-MS
fitting is that empirical BCs do not have any allowance for
the difference in log$\,g$ between MS and pre-MS stars.  According to
the BCAH98 $\alpha=1.9$ models the log$\,g$ of a $0.6\, \msun$ star
increases by almost $1\, \rm{dex}$
between $3\, \rm{Myr}$ and reaching the
ZAMS. This makes a difference of order $0.05\, \rm{mag}$ to
the BC in the $g$-band predicted by the BT-Settl atmospheric models.
Hence the difference in log$\,g$ is marginally significant (20 per cent
in age), but equally importantly, if we fail to make some adjustment
for log$\,g$, there will be a discontinuity in our model isochrones at the
point that we switch from theoretical to empirical BCs,
and the size of that discontinuity will be age dependent.

Rather than using MS stars with their mix of
metallicities, we follow \cite{Stauffer98a} and \cite*{Jeffries01} and
use the Pleiades, whose metallicity is, to within the uncertainties,
solar. We derive the empirical BC at each point along the Pleiades sequence
of \cite{Bell12}, and then
express this as a correction to the theoretical BC derived for the
appropriate $T_{\rm{eff}}$ and log$\,g$ for that 
point in the sequence. 
As shown in Paper~1, the models fit the $K_{\rm{s}}$-band flux well,
and so we chose this to fix the $T_{\rm{eff}}$ at each point 
in the sequence\footnote[6]{Unlike other versions of this technique,
  this leaves us with a well defined mass scale for the younger
  clusters, tied to the Pleiades.}.
The result is a set of $T_{\rm{eff}}$-dependent corrections to the
BCs.
We then return to the theoretical BC grid in $T_{\rm{eff}}$ and
log$\,g$ and add the appropriate correction for the $T_{\rm{eff}}$ 
to each entry, irrespective of its log$\,g$.   
This yields a set of log$\,g$-dependent semi-empirical BCs. Note, that
as demonstrated in Paper~1, we only apply corrections to the
theoretical BC-$T_{\rm{eff}}$ relation up to $T_{\rm{eff}}=4000\,
\rm{K}$ as at higher $T_{\rm{eff}}$ the model isochrones match the
observed shape of the Pleiades sequence.

The assumption of log$\,g$ independence in the correction to the BC is
equivalent to assuming that the missing opacity has the same
log$\,g$ dependence as the remaining opacity.
Whilst such an assumption is far from unassailable, as the models give
a difference in BC for the appropriate change in gravity of only
$0.05\, \rm{mag}$, we need only a
very crude log$\,g$ correction to push its effects
below the level at which they matter.
Furthermore, the assumption is tested by our fitting, where at least
for stars older than $6\, \rm{Myr}$ we obtain good fits to the data (see
Section~\ref{fitting_pre-ms_ages}). 

\subsection{Reddening and extinction for pre-main-sequence stars}
\label{reddening_and_extinction_for_pre-ms_stars}

\begin{figure}
\centering
\includegraphics[width=\columnwidth]{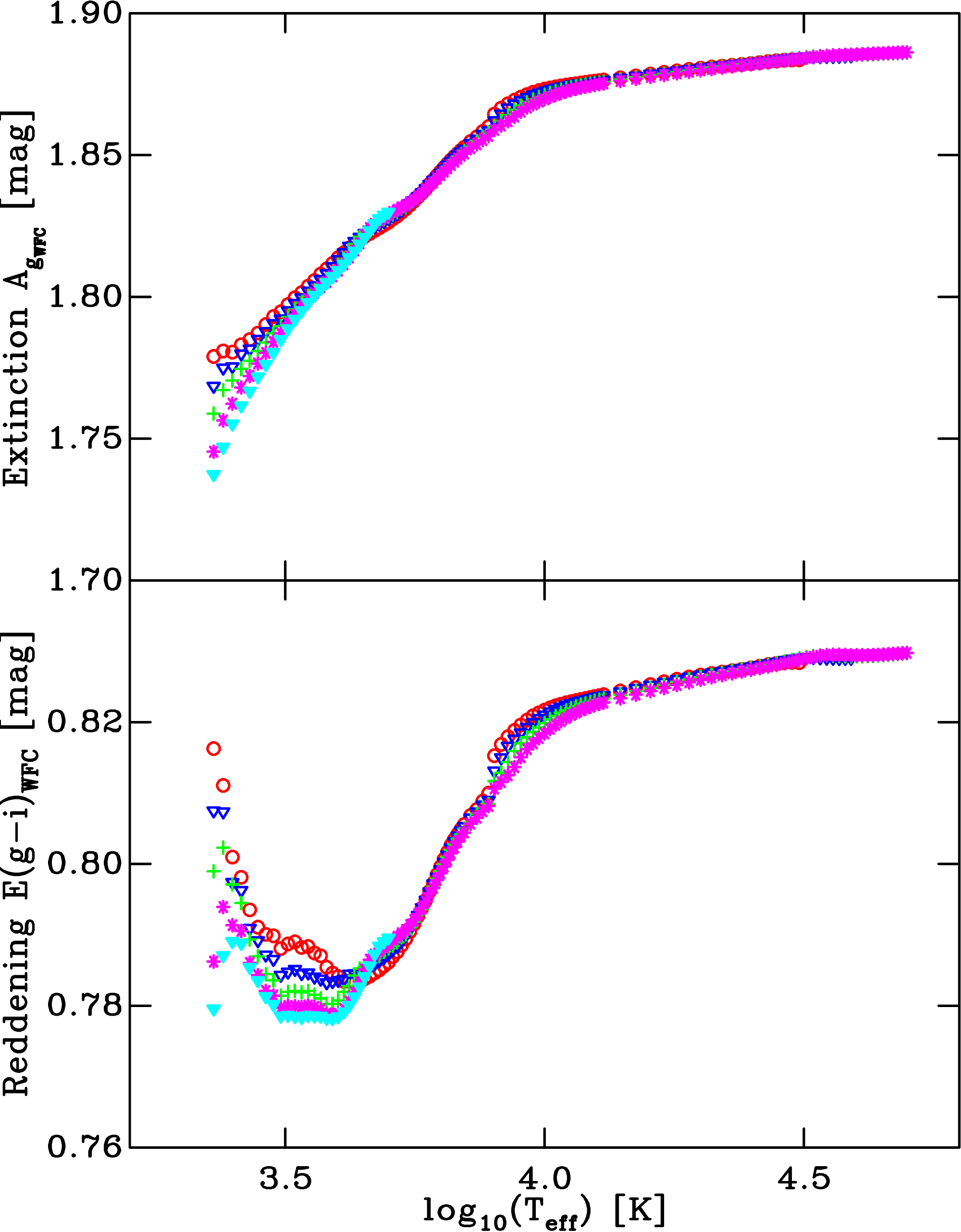}
\caption[]{The variation in extinction ($A_{\gwfc}$; top panel) and
  reddening ($E(g-i)_{_{\rm{WFC}}}$; bottom panel) as a function of
  $T_{\rm{eff}}$ between 2000 and $30\,000\, \rm{K}$ for a
  $E(B-V)_{\rm{nom}}=0.5$ calculated using the atmospheric
  models and the interstellar extinction law of
  \protect\cite{Cardelli89} with $R_{V}=3.2$. The different colour symbols represent different
  surface gravities;  log$\,g=3.5$ (red), 4.0 (blue), 4.5 (green), 5.0
  (magenta), and 5.5 (cyan).}
\label{fig:wfc_ext_red}
\end{figure}

Fig.~\ref{fig:wfc_ext_red}
shows how the calculated extinction and reddening, in the $\gwfc$-band
and $\giwfc$ colour respectively, vary as a function of $T_{\rm{eff}}$
for stars with $2000 \leq T_{\rm{eff}} \leq 30\,000\, \rm{K}$ for a
$E(B-V)_{\rm{nom}}=0.5$. For hot stars ($T_{\rm{eff}} > 10\,000\,
\rm{K}$) there is little variation, thus one can reliably model
reddening vectors for stars in this regime that apply over a large
$T_{\rm{eff}}$ (or equivalently spectral type) range. The same,
however, is not true for cool stars where a difference in the
extinction of $\simeq 0.1\, \rm{mag}$ and reddening of $\simeq 0.04\,
\rm{mag}$ is calculated between stars with $T_{\rm{eff}}=3000$ and
$T_{\rm{eff}}=10\,000\, \rm{K}$. This not only demonstrates that
applying the extinction and reddening derived from high-mass stars in
a given SFR to those in the low-mass regime is incorrect, but
furthermore that the same reddening vectors should not be used for all
spectral types.

Without spectra for a large sample of objects in a given
field-of-view, it is not possible to de-redden sources on an
individual basis and then fit the model isochrones in the
extinction-corrected CMD. Instead, the isochrone must be
appropriately reddened and then applied to the photometric data. The
only consistent and homogeneous way to redden the pre-MS model isochrones,
as a function of $T_{\rm{eff}}$, is to create extinction grids based
on atmospheric models, the photometric system responses, and a
description of the interstellar extinction law. The atmospheric
models were reddened according to the parameterised extinction law of
\cite{Cardelli89} and folded through the calculated INT-WFC system
responses (see Paper~1). The models were reddened in steps of 0.5 from
$E(B-V)_{\rm{nom}}=0.0-2.0$, with the grids comprising extinction in all
INT-WFC $(UgriZ)_{_{\rm{WFC}}}$ bandpasses as a function of $T_{\rm{eff}}$ and log$\,g$ (both
of which come from the atmospheric models). Thus to redden a pre-MS model
isochrone, the reddening derived from the more massive MS members is
used to calculate a $E(B-V)_{\rm{nom}}$ that represents the column
density of interstellar material between the Earth and the star. This
$E(B-V)_{\rm{nom}}$ is then used to interpolate within the extinction
grids for the extinction and reddening (difference in extinction
between two bandpasses) for a star of given $T_{\rm{eff}}$ and
log$\,g$ as defined by the pre-MS interior models. This process is
then repeated for each point along the model isochrone.

\section{Fitting the pre-main-sequence}
\label{fitting_pre-ms_ages}

Conceptually, there is no difference between fitting the pre-MS and MS
populations of a given SFR using the $\tau^{2}$ fitting
statistic. There are a couple of examples in the literature
(e.g. \citealp{Naylor06,Cargile10}) where $\tau^{2}$ has been
used to derive pre-MS ages by fitting the positions of probable
low-mass members using model distributions (isochrones including an
intrinsic binary fraction; see Section~\ref{model_cmd}). In these examples, the
clusters represent relatively old pre-MS populations ($\gtrsim 30\,
\rm{Myr}$) in which the pre-MS locus is well defined, and as such the
age and distance were fitted simultaneously. These studies
showed that the main contributor to the error budget in the age were
uncertainties in the derived distance. This is unsurprising given the
degeneracy between derived age and assumed distance when fitting
pre-MS model isochrones to photometric data in CMDs.

As one moves to slightly younger populations ($\simeq 10\, \rm{Myr}$), the pre-MS
locus, though still well-defined in CMD space, begins to exhibit an
enhanced luminosity spread at a given colour,
perhaps as a result of astrophysical processes in the
form of, for example, enhanced variability arising from the
inclusion of accreting objects in samples.
This spread can, however, be exaggerated by the inclusion of
non-members in the sample. Spectroscopic measurements are the only
unbiased diagnostic, although, in many cases other diagnostics (e.g.
X-ray emission, IR excess or H$\alpha$ emission) have been used to
differentiate between young SFR members and older field stars. As a
result, these methods are more likely to include contamination from
foreground and background objects than memberships based on purely
spectroscopic methods. One way of
ensuring that such non-members do not influence the derived age is to
adopt a so-called soft-clipping approach, whereby data points with
colours and magnitudes that lie several $\sigma$ away from the
observed sequence are assigned an arbitrary low probability as they
are not well described by the model distribution (e.g. \citealp{Naylor06}). In
Appendix~\ref{non-member_contamination_model} a model dealing with such
interlopers by modelling a
background population of non-member stars in conjunction with the bona fide
cluster members is introduced. The assumption of a uniform
distribution of non-members is a poor description of the physical
distribution, but as shown in
Appendix~\ref{non-member_contamination_model} it does allow the
correct best-fitting age to be derived, although it will result in
incorrect uncertainties in that age. Practically, however, our uncertainty
in the derived age is driven by the uncertainty in the
distance from our MS fitting and so we derive the uncertainty in age by
fitting at the two extremes of the distance estimate. In
Section~\ref{pre-ms_ages_derived_using_tau2} this model is thus
implemented and the $\tau^{2}$ fitting statistic is used to derive
pre-MS ages for SFRs with MS ages $\gtrsim 10\, \rm{Myr}$.

\begin{figure}
\centering
\includegraphics[width=\columnwidth]{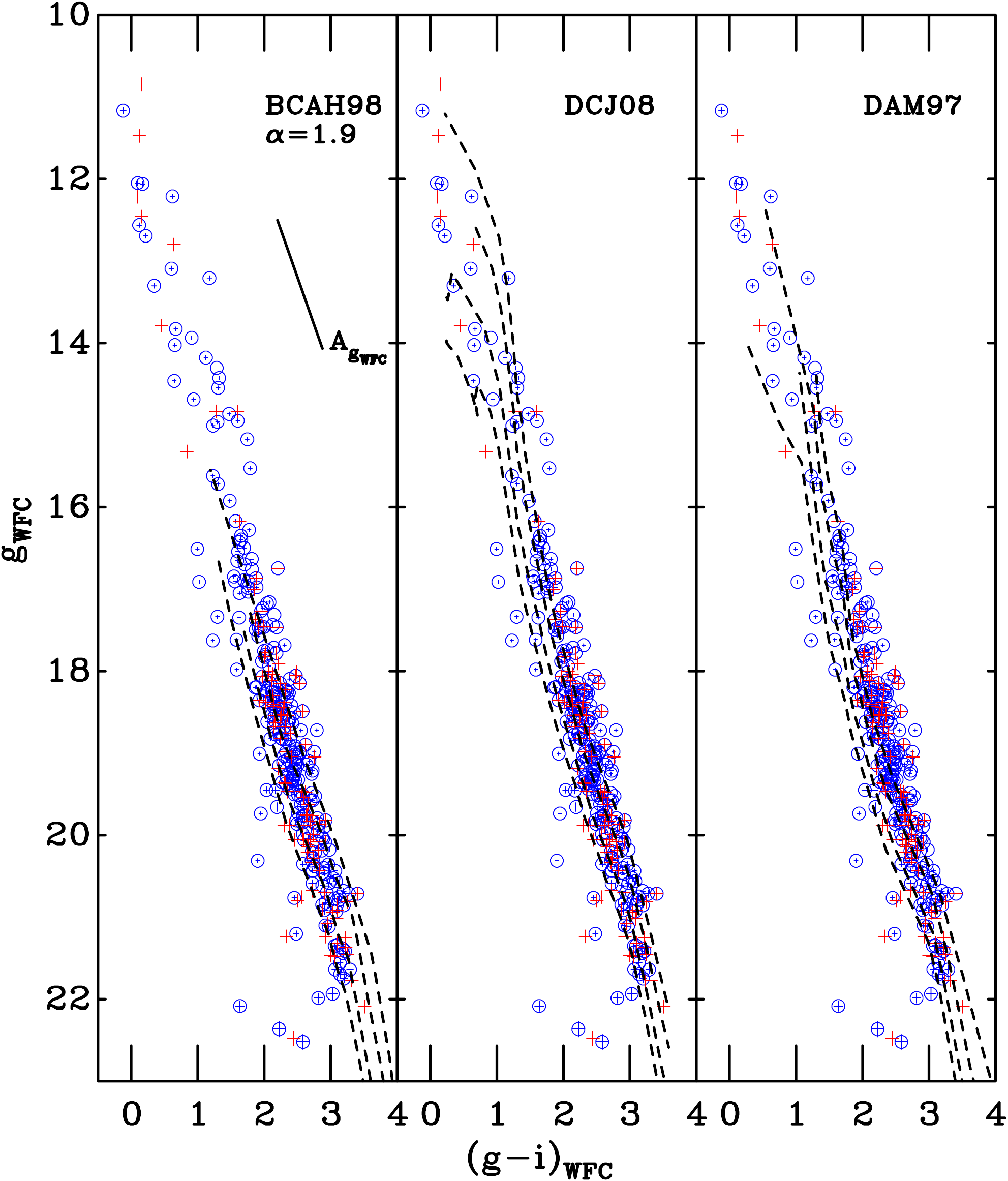}
\caption[]{Pre-MS stars selected as members of NGC\,2244 in the
  $\gwfc, \giwfc$ CMD. Circles are X-ray sources from \cite{Wang08} and crosses
  are IR excess objects from \cite{Balog07}. Semi-empirical pre-MS
  single-star model isochrones at ages of 1, 2, 5, and $10\, \rm{Myr}$
  are overlaid at
  the best-fit MS distance and reddened assuming the median value
  derived in Section~\ref{fundamental_parameters_from_ms_stars}
  according to the prescription described in
  Section~\ref{reddening_and_extinction_for_pre-ms_stars}. The
  diagonal line represents the reddening vector in the $\gwfc, \giwfc$
  plane for a star of $T_{\rm{eff}} \simeq 4500\, \rm{K}$ and log$\,g
  \simeq 4$ based on the median value for the SFR. \textbf{Left
    panel:} BCAH98 $\alpha=1.9$. \textbf{Middle panel:}
  DCJ08. \textbf{Right panel:} DAM97.}
\label{fig:ngc2244_diff_iso}
\end{figure}

\begin{figure}
\centering
\includegraphics[width=\columnwidth]{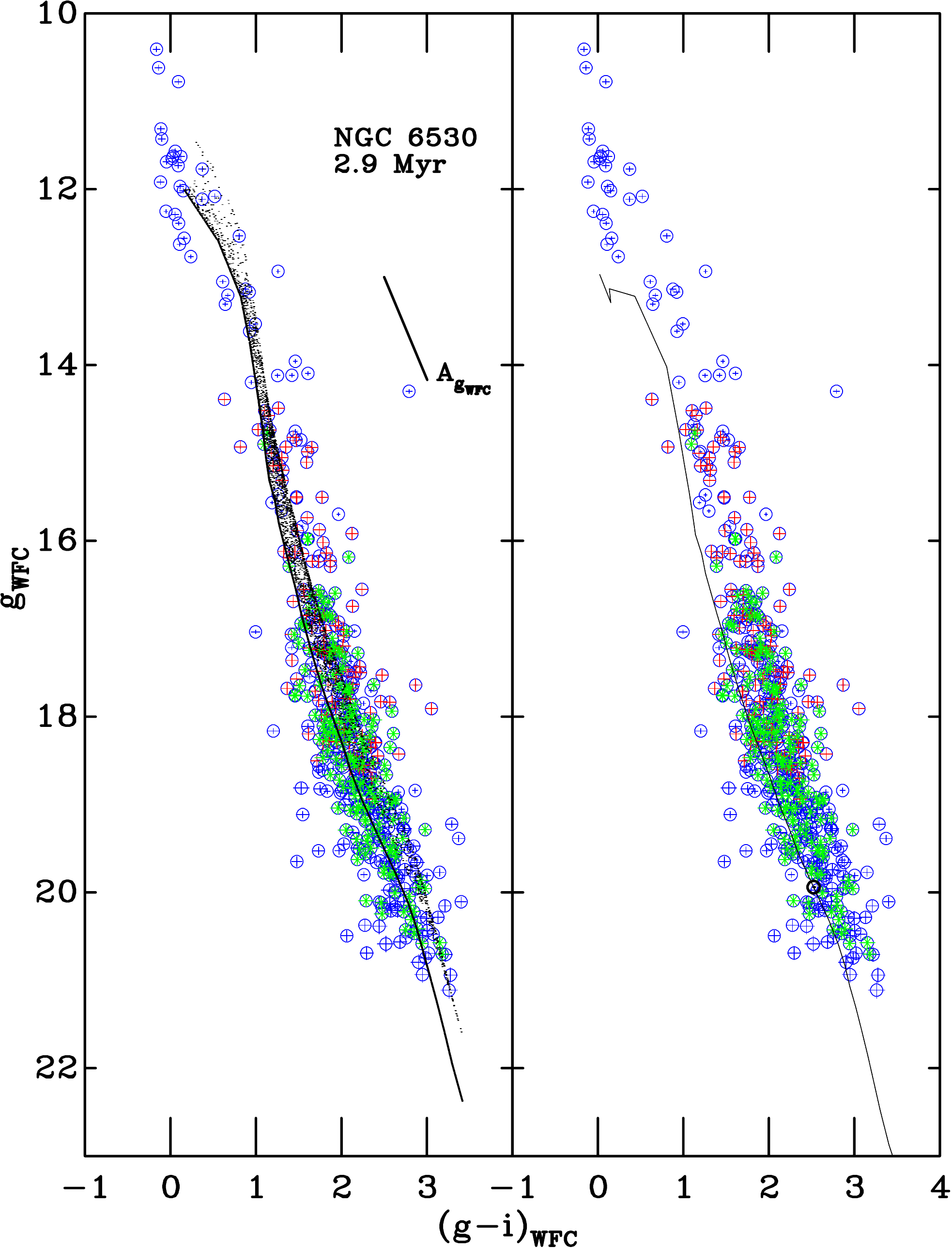}
\caption[]{Demonstration of the two methods used for assigning pre-MS
  ages adopted in this paper using pre-MS stars selected as members of
  NGC\,6530. Circles are X-ray sources from \cite{Damiani04}, the
  crosses are spectroscopic members from \cite{Prisinzano07}, and the
  asterisks are periodic variables from \cite{Henderson12}. \textbf{Left panel:} The
  best-fit semi-empirical DCJ08 pre-MS model
  distribution, as calculated using the $\tau^{2}$ fitting statistic,
  is overlaid at the best-fit MS distance and reddened assuming the
  median value derived in
  Section~\ref{fundamental_parameters_from_ms_stars} according
  to the prescription described in
  Section~\ref{reddening_and_extinction_for_pre-ms_stars}. 
  The diagonal line
  represents the reddening vector in the $\gwfc,
  \giwfc$ plane for a star with $T_{\rm{eff}} \simeq 4500\, \rm{K}$
  and log\,$g \simeq 4$ based on the median value for the SFR. \textbf{Right
  panel:} Same as for the left panel, but with a $6\, \rm{Myr}$
  semi-empirical DCJ08 pre-MS single-star model isochrone overlaid. The black circle
  marks the position of a a $0.75\, \msun$ star.}
\label{fig:tau2_nom_example}
\end{figure}

Moving to even younger ages ($< 10\, \rm{Myr}$) there are two effects
which preclude us using $\tau^2$ fitting.
First, Figs.~\ref{fig:ngc2244_diff_iso}--\ref{fig:nominal_pms_ages}
show that for some SFRs with MS ages $< 10\, \rm{Myr}$ the semi-empirical pre-MS 
isochrones do not match the shape of the observed pre-MS locus as well as in the case 
of the older, more evolved SFRs, with the models tending to cut through the pre-MS 
locus (see Fig.~\ref{fig:ngc2244_diff_iso} where this is demonstrated explicitly
for the three sets of pre-MS model isochrones using NGC\,2244).
Second, the observed
luminosity spread in CMD space, at a given colour,
becomes more pronounced (see \citealp{Hartmann01} for a discussion on the
possible sources). It is apparent from CMDs of some of our SFRs that
the observed luminosity spread can be as large as $2-3\, \rm{mag}$ at
a given colour, and so although the $\tau^{2}$ fitting statistic includes the
effects of binarity, this alone is not enough to model the observed
spread (see Fig.~\ref{fig:tau2_nom_example}).
Given that we have insufficient knowledge of what is
causing the observed luminosity spreads in these SFRs (see
Section~\ref{discussion}) as well as lacking the additional data
required to accurately model the effects of, for example, stellar
variability and accretion, we only use the
$\tau^{2}$ fitting statistic to derive absolute ages for SFRs
where the luminosity spread is commensurate with the two-dimensional
model distribution. For SFRs where the observed luminosity
spread prohibits the use of the $\tau^{2}$ fitting statistic we are
unable to derive absolute ages. In such cases, it is possible to a
create a relative age ladder of SFRs based on common positions shared
in CMD space when compared
with a pre-MS model isochrone of a given age (see
Fig.~\ref{fig:tau2_nom_example}). Therefore, in
Section~\ref{nominal_pre-ms_ages} a subset of our sample of SFRs are
assigned to such groups and nominal ages for each group discussed. 

We have investigated whether using a different technique for SFRs
younger than 7 or $8\, \rm{Myr}$ introduces a discontinuity in our age
scale using $\sigma$\,Ori as an
example. For this SFR we find that the $\tau^{2}$ fitting statistic
derives an age of $5.3\, \rm{Myr}$, whereas the nominal age is $\simeq
6\, \rm{Myr}$ (dependent upon the choice of pre-MS model isochrone;
see Table~\ref{tab:pre-ms_nominal_ages}) and therefore we are
confident that the final assigned ages are consistent across the
sample and that no discontinuities have been introduced as a result of
changing how the pre-MS age has been derived.

\subsection{Pre-main-sequence ages derived using $\tau^{2}$}
\label{pre-ms_ages_derived_using_tau2}

Using the method detailed in the previous section, 
pre-MS ages have been calculated for the SFRs with MS ages $\gtrsim
10\, \rm{Myr}$ using the $\tau^{2}$ fitting statistic in the $\gwfc,
\giwfc$ CMD. The memberships listed in
Appendix~\ref{literature_memberships} have been used to select pre-MS
member stars for each SFR.

To fit stars selected as pre-MS members, grids of model distributions
were created for each set of interior models spanning a range of ages using
the recalibrated BC-$T_{\rm{eff}}$ relation at the appropriate SFR
reddening. For SFRs where we have derived individual star-by-star
reddenings for the MS population using the revised Q-method, we adopt
the median measured reddening. Although there is a
distribution of reddenings due to variable extinction across a given
region, the reddening vector in the $\gwfc, \giwfc$ CMD, and in the
mass range we are interested in, lies almost
parallel to the observed pre-MS locus and therefore applying a fixed
reddening to the pre-MS model distributions does not significantly affect
the derived age.

\begin{figure}
\centering
\includegraphics[width=\columnwidth]{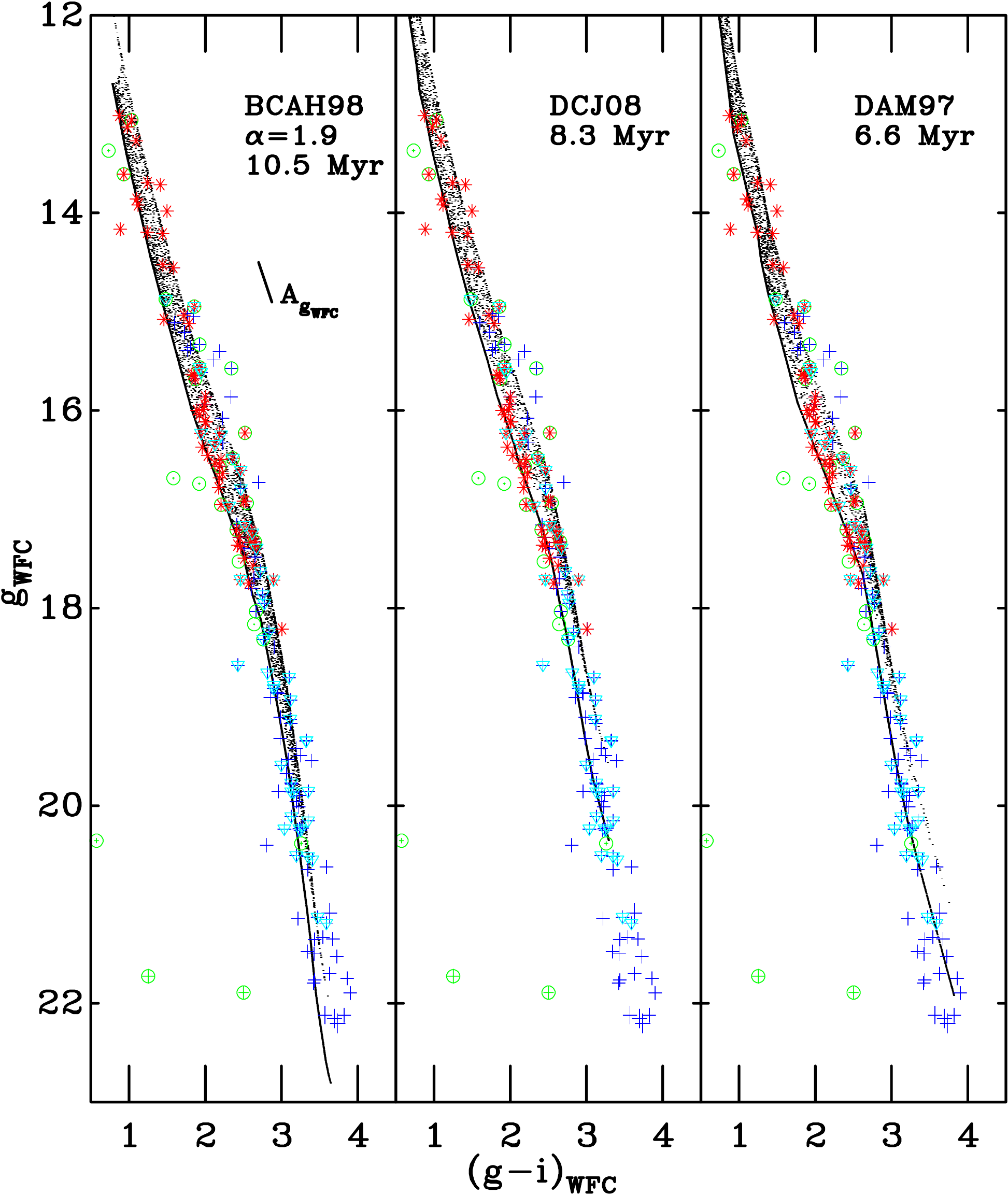}
\caption[]{Pre-MS stars selected as members of the $\lambda$\,Ori
  association in the $\gwfc, \giwfc$ CMD. Asterisks are $\LiI$ spectroscopic members from
  \protect\cite{Dolan01}, crosses are the combined H$\alpha$ spectroscopic
  and IR excess sources from \protect\cite{Barrado07}, triangles are
  $\LiI$ and H$\alpha$ spectroscopic members from \protect\cite{Sacco08},
  and circles are the X-ray sources from \protect\cite{Barrado11}.
  The best-fitting pre-MS model distributions are overlaid at the
  best-fit MS distance and reddened assuming the median value derived
  in Section~\ref{fundamental_parameters_from_ms_stars} according to
  the prescription described in
  Section~\ref{reddening_and_extinction_for_pre-ms_stars}. The diagonal line in
  the left panel represents the reddening vector in the $\gwfc,
  \giwfc$ plane for a star with $T_{\rm{eff}} \simeq 4500\, \rm{K}$
  and log\,$g \simeq 4$ based on the median value for the SFR.
  \textbf{Left panel:} BCAH98
  $\alpha=1.9$. \textbf{Middle panel:} DCJ08. \textbf{Right panel:}
  DAM97.}
\label{fig:lamori_pms_tau2}
\end{figure}

\begin{figure}
\centering
\includegraphics[width=\columnwidth]{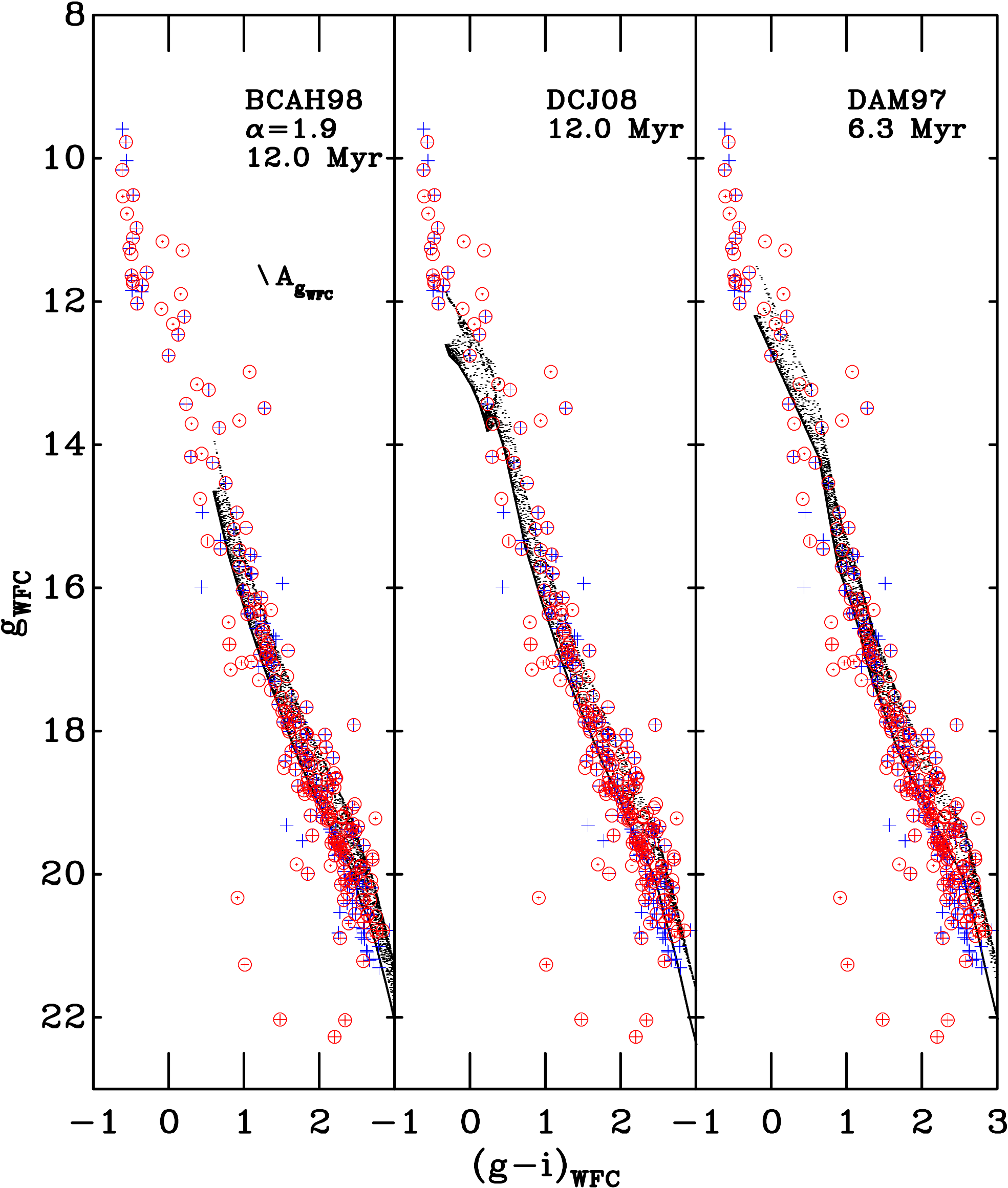}
\caption[]{Same as Fig.~\ref{fig:lamori_pms_tau2} but for
  NGC\,2362. Crosses
  are the combined $\LiI$ and H$\alpha$ spectroscopic and IR excess
  sources from \protect\cite{Dahm07} and circles are X-ray sources from
  \protect\cite{Damiani06b}.}
\label{fig:ngc2362_pms_tau2}
\end{figure}

\begin{figure}
\centering
\includegraphics[width=\columnwidth]{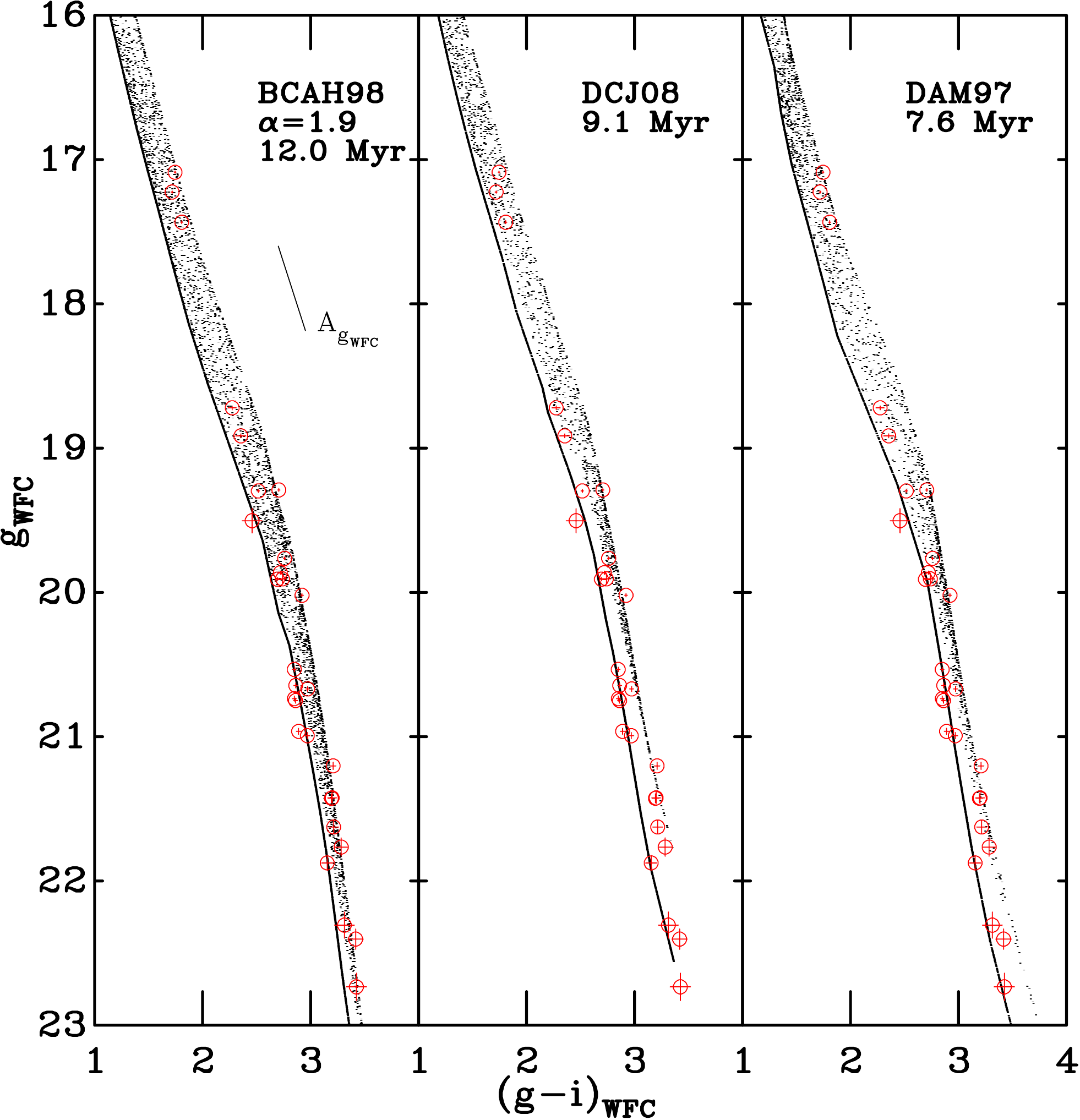}
\caption[]{Same as Fig.~\ref{fig:lamori_pms_tau2} but for
  NGC\,2169. Circles
  are $\LiI$ spectroscopic members from \protect\cite{Jeffries07b}.}
\label{fig:ngc2169_pms_tau2}
\end{figure}

\begin{figure}
\centering
\includegraphics[width=\columnwidth]{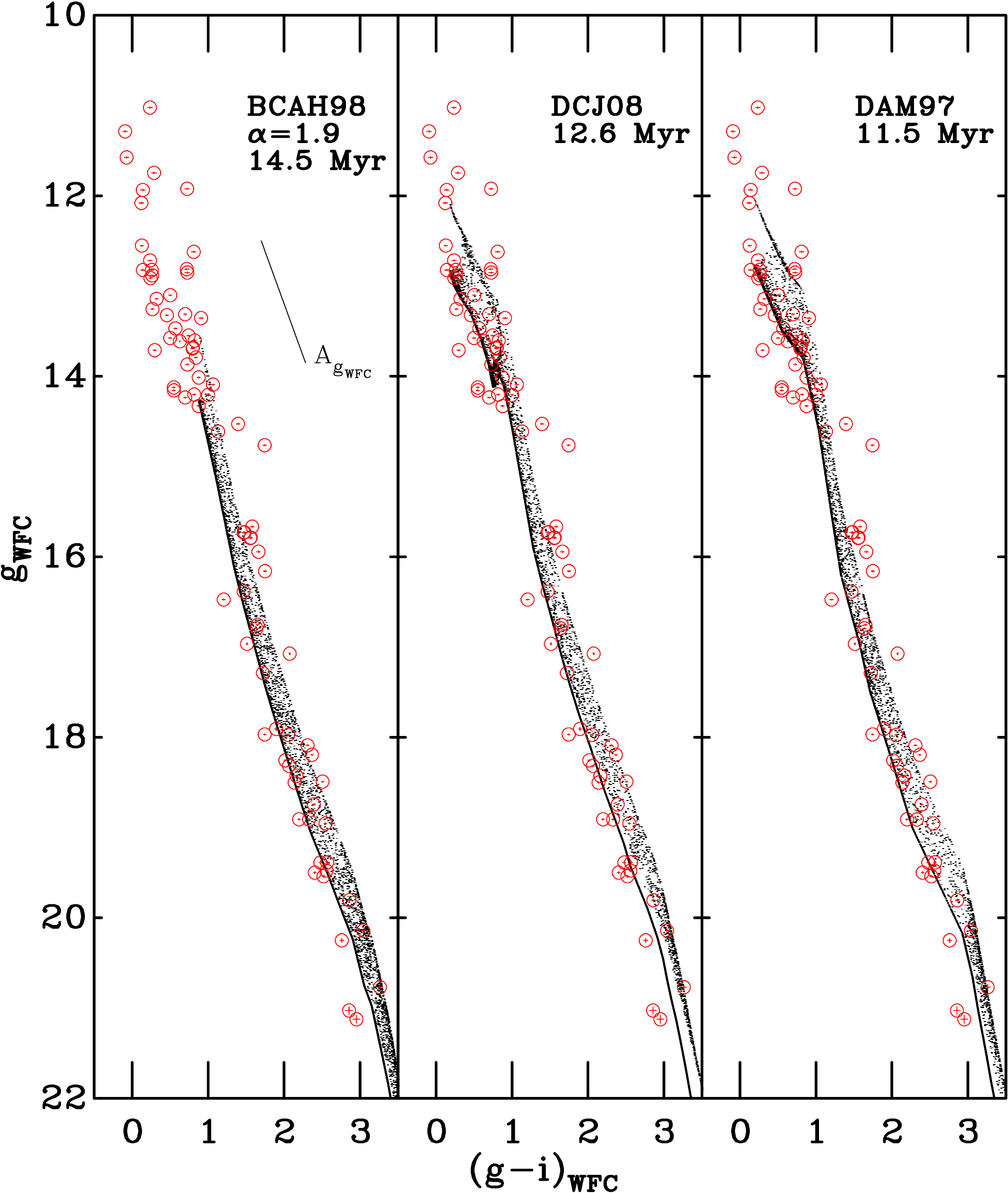}
\caption[]{Same as Fig.~\ref{fig:lamori_pms_tau2} but for
  NGC\,7160. Circles
  are $\LiI$ and H$\alpha$ spectroscopic, IR excess, and
  extinction-based members from \protect\cite{Sicilia06}.}
\label{fig:ngc7160_pms_tau2}
\end{figure}

\begin{figure}
\centering
\includegraphics[width=\columnwidth]{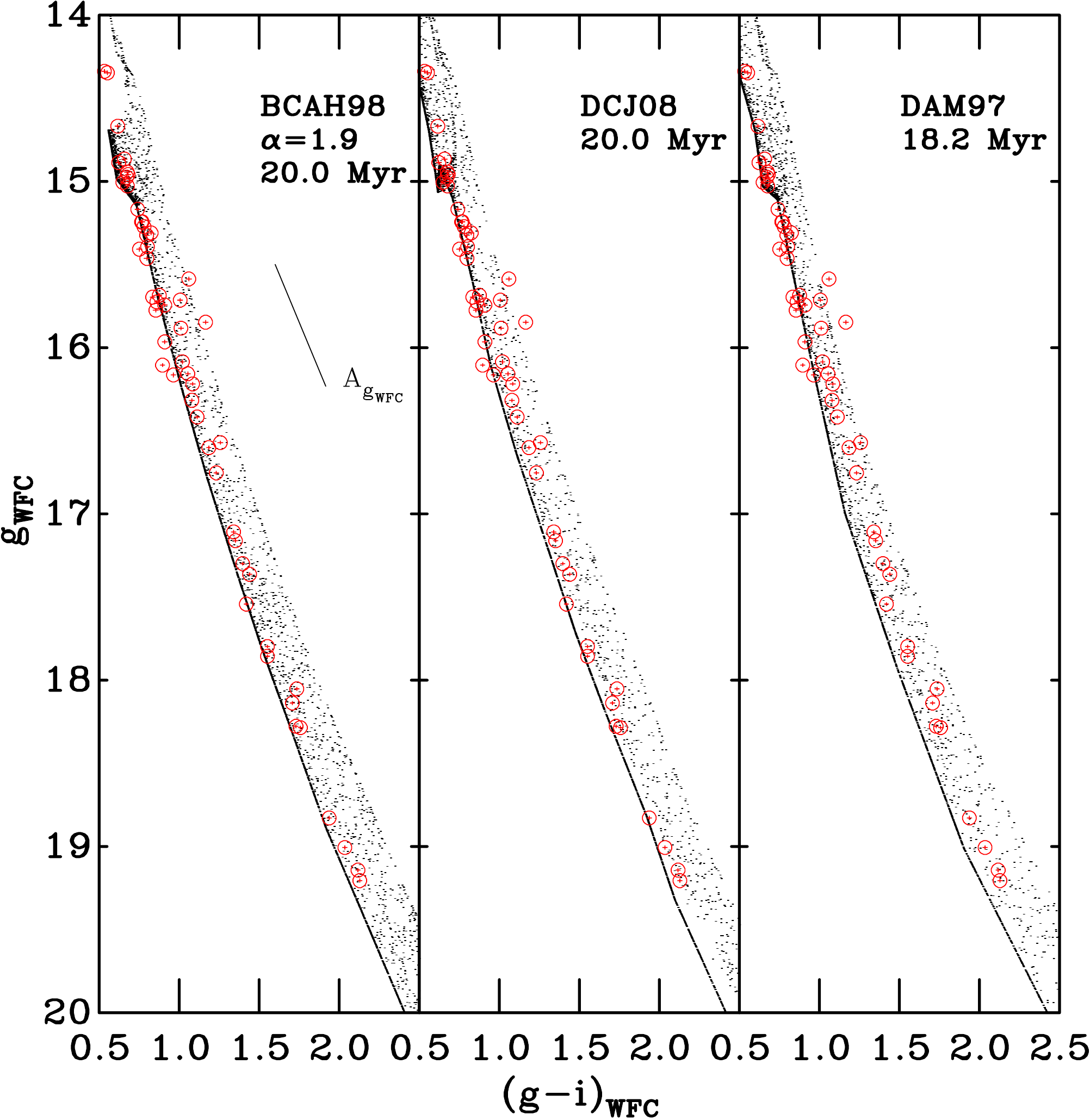}
\caption[]{Same as Fig.~\ref{fig:lamori_pms_tau2} but for
  NGC\,1960. Circles are $\LiI$ spectroscopic members from \protect\cite{Jeffries13}.}
\label{fig:ngc1960_pms_tau2}
\end{figure}

Figs.~\ref{fig:lamori_pms_tau2} -- \ref{fig:ngc1960_pms_tau2} shows
the $\gwfc, \giwfc$ CMDs of stars selected as pre-MS members of
$\lambda$\,Ori, NGC\,2362, NGC\,2169, NGC\,7160, and NGC\,1960 with the
best-fitting BCAH98 $\alpha=1.9$, DCJ08 and DAM97 model distributions (including
an intrinsic binary fraction of 50 per cent) overlaid at the best-fit
MS distance. The largest source of uncertainty in the derived pre-MS
age is attributable to the associated uncertainty in the assumed
distance, therefore the uncertainty in the pre-MS age is calculated by
deriving the corresponding age at the upper and lower distance
uncertainty limits as defined by the 68 per cent confidence contour in
the MS age-distance $\tau^{2}$ grid (see for example
Fig.~\ref{fig:ngc1960_tau2_grid}). The absolute pre-MS ages for SFRs
with MS ages $\gtrsim 10\, \rm{Myr}$ derived
using the $\tau^{2}$ fitting statistic are shown in
Table~\ref{pre-ms_fitting_results}.

\begin{table*}
\caption[]{Absolute pre-MS ages for SFRs with MS ages $\gtrsim 10\,
  \rm{Myr}$ derived using semi-empirical pre-MS model distributions and
  fitted using the $\tau^{2}$ fitting statistic. The
  best-fit pre-MS age is derived assuming the best-fit MS distance
  and the uncertainty in the pre-MS age represents the
  uncertainty in the MS distance translated into an age uncertainty.}
\begin{tabular}{c c c c c c c}
\hline
SFR&\multicolumn{6}{c}{Absolute Pre-MS Age (Myr)}\\
&\multicolumn{2}{c}{BCAH98
  $\alpha=1.9$}&\multicolumn{2}{c}{DCJ08}&\multicolumn{2}{c}{DAM97}\\
&Best-fit&68 per cent&Best-fit&68 per cent&Best-fit&68 per cent\\
\hline
$\lambda$\,Ori&10.5&10.0--11.0&8.3&7.6--8.7&6.6&6.0--7.2\\
NGC\,2169&12.0&10.5--13.2&9.1&7.9--10.5&7.6&6.3--8.3\\
NGC\,2362&12.0&10.5--12.6&12.0&9.5--12.6&6.3&5.7--6.6\\
NGC\,7160&14.5&12.6--16.6&12.6&11.0--14.5&11.5&10.5--12.5\\
NGC\,1960&20.0&19.0--20.9&20.0&19.0--20.9&18.2&17.4--19.1\\
\hline
\end{tabular}
\label{pre-ms_fitting_results}
\end{table*}

\subsection{Nominal pre-main-sequence ages}
\label{nominal_pre-ms_ages}

\subsubsection{Star-forming regions with main-sequence ages $< 10\,
  Myr$}
\label{sfrs_with_ms_ages_less_than_10}

For the reasons explained in the introduction to Section~\ref{fitting_pre-ms_ages}
we derive ages for this group of SFRs by comparing the positions of the observed sequences
to a semi-empirical pre-MS single-star model isochrones of a given age.
This could be performed by simply de-reddening the pre-MS loci using a
given reddening vector and shifting vertically using the derived MS
distance, thereby comparing the populations in the absolute
magnitude-intrinsic colour plane (e.g. \citealp{Mayne07}). However, as
was shown in Section~\ref{reddening_and_extinction_for_pre-ms_stars},
the reddening and extinction for a given object depends upon its
$T_{\rm{eff}}$ and therefore the reddening vector in the $\gwfc,
\giwfc$ plane is not a fixed vector. As we do not have the necessary
$T_{\rm{eff}}$ diagnostics for the low-mass pre-MS objects in our
sample of SFRs, we are unable to de-redden these objects
individually. Instead, we leave the sequence in the apparent
magnitude-apparent colour plane and the model isochrone is instead
reddened using the appropriate reddening and distance modulus for the
SFR. By comparing the photometric data to the models in this way SFRs
can be grouped in order of increasing age.

As the distance moduli to the SFRs cover a range of $\Delta dm \simeq
4.5$ the mass regimes probed across the sample of SFRs varies. In
addition, there are inherent lower and upper mass limits on the pre-MS
interior models, which may further be restricted due to the lowest
$T_{\rm{eff}}$ limit defined by the derived corrections (see
Section~\ref{creating_recalibrated_semi-empirical_pre-ms_isochrones}). Of
the model isochrones tested, it is clear that the BCAH98
$\alpha=1.9$ and DCJ08 models represent the best-fit to the observed
Pleiades MS for $T_{\rm{eff}} \gtrsim 4000\, \rm{K}$ (see Paper~1). Due to the
upper mass limit of $1.4\, \msun$ on the BCAH98 $\alpha=1.9$ models,
the observed pre-MS loci are compared to a semi-empirical DCJ08
single-star model isochrone at an
age of $6\, \rm{Myr}$. The reason we adopt an age of $6\, \rm{Myr}$ is
that, to within the uncertainties, the ages for SFRs with MS ages $< 10\,
\rm{Myr}$ are all consistent with $6\, \rm{Myr}$, and thus
by adopting such an age, we are still assessing whether agreement is
observed between age estimates in distinct mass regimes.

\begin{figure*}
\centering
\includegraphics[width=\textwidth]{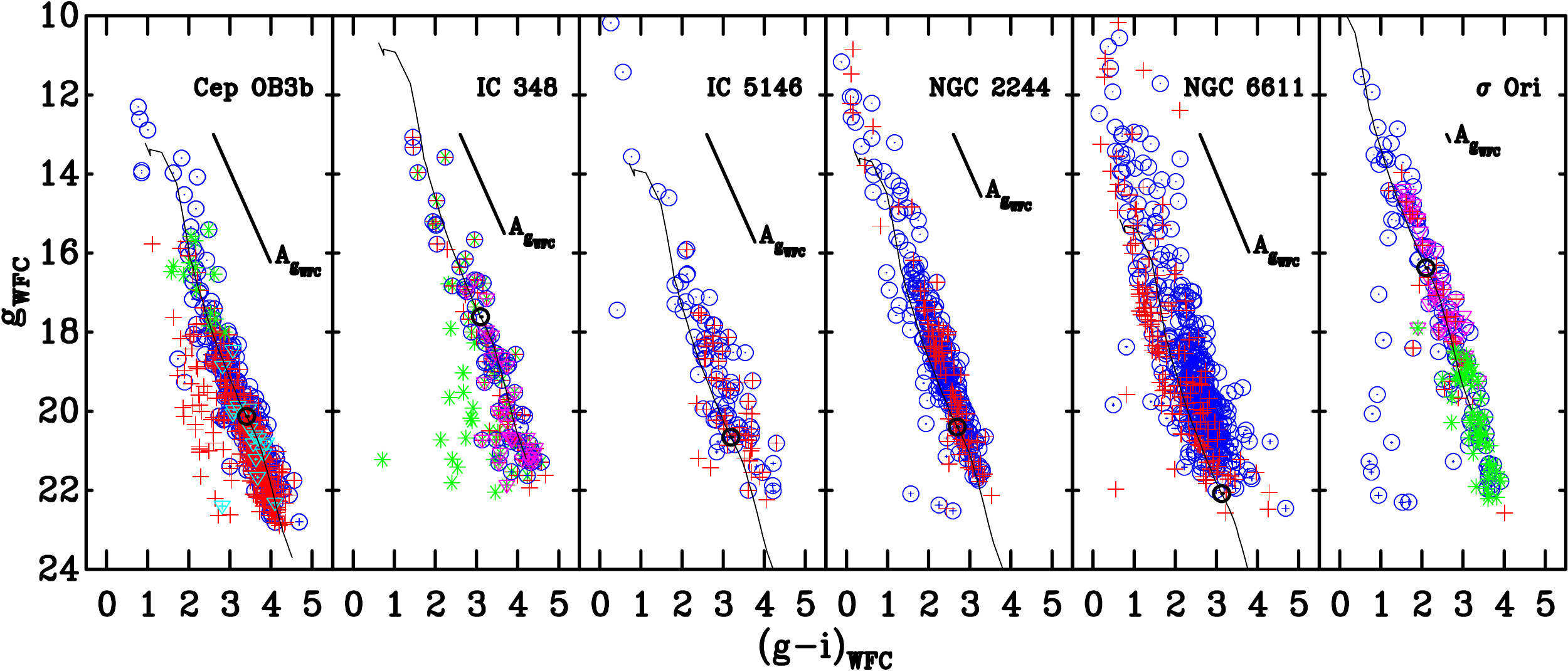}
\caption[]{Stars selected as pre-MS members for SFRs with MS ages
  $<10\, \rm{Myr}$. The $6\, \rm{Myr}$ semi-empirical DCJ08 pre-MS
  single-star model isochrone shown in each panel is overlaid at the
  best-fit MS distance and reddened assuming the median value derived
  in Section~\ref{fundamental_parameters_from_ms_stars} according to
  the prescription described in
  Section~\ref{reddening_and_extinction_for_pre-ms_stars}. The DCJ08
  model isochrone ends at a mass of $0.2\, \msun$ with the SFR data
  having different depths based on a combination of distance and mean
  extinction. The black
  circle marks the position of a $0.75\, \msun$ star. The diagonal
  line represents the reddening vector in the $\gwfc, \giwfc$ plane
  for a star with $T_{\rm{eff}} \simeq 4500\, \rm{K}$ and log$\,g
  \simeq 4$ based on the median value for the SFR.
  \textbf{Cep\,OB3b:} Crosses are periodic variables from
  \protect\cite{Littlefair10}, the
  circles are X-ray sources from \cite{Naylor99} and \cite{Getman06}, the
  asterisks are spectroscopic members from \cite{Pozzo03}, and the
  triangles are are H$\alpha$ sources from \cite{Ogura02}. \textbf{IC\,348:}
  Circles are X-ray sources from \cite{Preibisch02}, the
  crosses are the H$\alpha$, $\NaI$, and $\KI$ spectroscopic members
  from \cite{Luhman03,Luhman05a,Luhman05b}, the triangles are the
  H$\alpha$ spectroscopic members from \cite{Herbig98}, and the
  asterisks are the combined periodic variable sources from
  \cite{Cohen04} and \cite{Littlefair05}. \textbf{IC\,5146:} Circles
  are the IR excess sources of \cite{Harvey08} and the
  crosses are spectroscopic members
  from \cite{Herbig02}. \textbf{NGC\,2244:} Circles
  are X-ray sources from \cite{Wang08} and the
  crosses are IR excess objects from
  \cite{Balog07}. \textbf{NGC\,6611:} Circles
  are X-ray sources from \cite{Guarcello07} and the
  crosses are IR excess objects from
  \cite{Guarcello09}. Note that both the isochrone and the reddening
  vector have been calculated adopting the total-to-selective extinction
  ratio $R_{V}=3.75$. \textbf{$\sigma$\,Ori:} Circles are X-ray sources from
  \cite{Sanz-Forcada04}, the crosses are periodic and aperiodic
  variables from \cite{Cody10}, the asterisks are the combined $\NaI$
  and $\LiI$ spectroscopic members from \cite{Kenyon05} and
  \cite{Burningham05b}, and the triangles are the $\LiI$ and H$\alpha$
  spectroscopic members from \cite{Sacco08}.}
\label{fig:nominal_pms_ages}
\end{figure*}

Figs.~\ref{fig:tau2_nom_example} and \ref{fig:nominal_pms_ages} shows the $\gwfc,
\giwfc$ CMDs of stars selected as pre-MS members of Cep\,OB3b, IC\,348,
IC\,5146, NGC\,2244, NGC\,6530, NGC\,6611, and $\sigma$\,Ori with a $6\,
\rm{Myr}$ semi-empirical DCJ08 single-star isochrone overlaid. It is clear
that, for SFRs with MS ages $< 10\, \rm{Myr}$, these can be separated
into two distinct groups based on the comparison of the pre-MS
populations with the model isochrone. In ascending age order these two
groups comprise;

\begin{itemize}
\item NGC\,6611, IC\,5146, NGC\,6530, and NGC\,2244 -- for which the
  isochrone sits below the observed pre-MS locus,
\item Cep\,OB3b, $\sigma$\,Ori, and IC\,348 -- for which the
  isochrone sits approximately in the middle of the observed pre-MS
  locus.
\end{itemize}

\begin{table}
\caption[]{Nominal pre-MS ages for SFRs with MS ages $<10\,  \rm{Myr}$
  estimated using semi-empirical pre-MS isochrones. The ages were
  derived at a mass of $0.75\, \msun$ and assuming the best-fit MS
  distance. Notes are as follows. (1) Age derived assuming the
  total-to-selective extinction ratio $R_{V}=3.75$ (see
  Section~\ref{anomalous_extinction}).}
\centering
\begin{tabular}{c c c c}
\hline
SFR&\multicolumn{3}{c}{Nominal Pre-MS Age ($\simeq$ Myr)}\\
&BCAH98 $\alpha=1.9$&DCJ08&DAM97\\
\hline
NGC\,6611$^{(1)}$&2&1&0.5\\
IC\,5146&2&1&0.5\\
NGC\,6530&3&2&1\\
NGC\,2244&3&2&1\\
$\sigma$\,Ori&6&5&3\\
IC\,348&7&6&4\\
Cep\,OB3b&7&7&3\\
\hline
\end{tabular}
\label{tab:pre-ms_nominal_ages}
\end{table}

Due to the fact that, in some cases, the semi-empirical pre-MS isochrones do not
follow the shape of the observed pre-MS locus (see for example
Fig.~\ref{fig:ngc2244_diff_iso}), and that the mass ranges sampled are
different due to the differences in the distance between the SFRs, a
pre-MS age derived by simply laying an isochrone over the photometric
data will be biased depending on which section of the sequence is
fitted. Therefore, a more consistent approach is to estimate the
pre-MS age of a given SFR by comparing the position of a
model star of given mass (having applied the reddening and
distance modulus as for the comparison with the $6\, \rm{Myr}$
semi-empirical DCJ08 single-star pre-MS
model isochrone) with the approximate middle of the observed pre-MS
locus. Such ages are termed nominal ages and estimated adopting a
mass of $0.75\, \msun$. Table~\ref{tab:pre-ms_nominal_ages} shows the
nominal pre-MS ages for the SFRs with MS ages $< 10\, \rm{Myr}$.

\subsubsection{$\chi$\,Per}
\label{chiper}

\begin{figure}
\centering
\includegraphics[width=0.75\columnwidth]{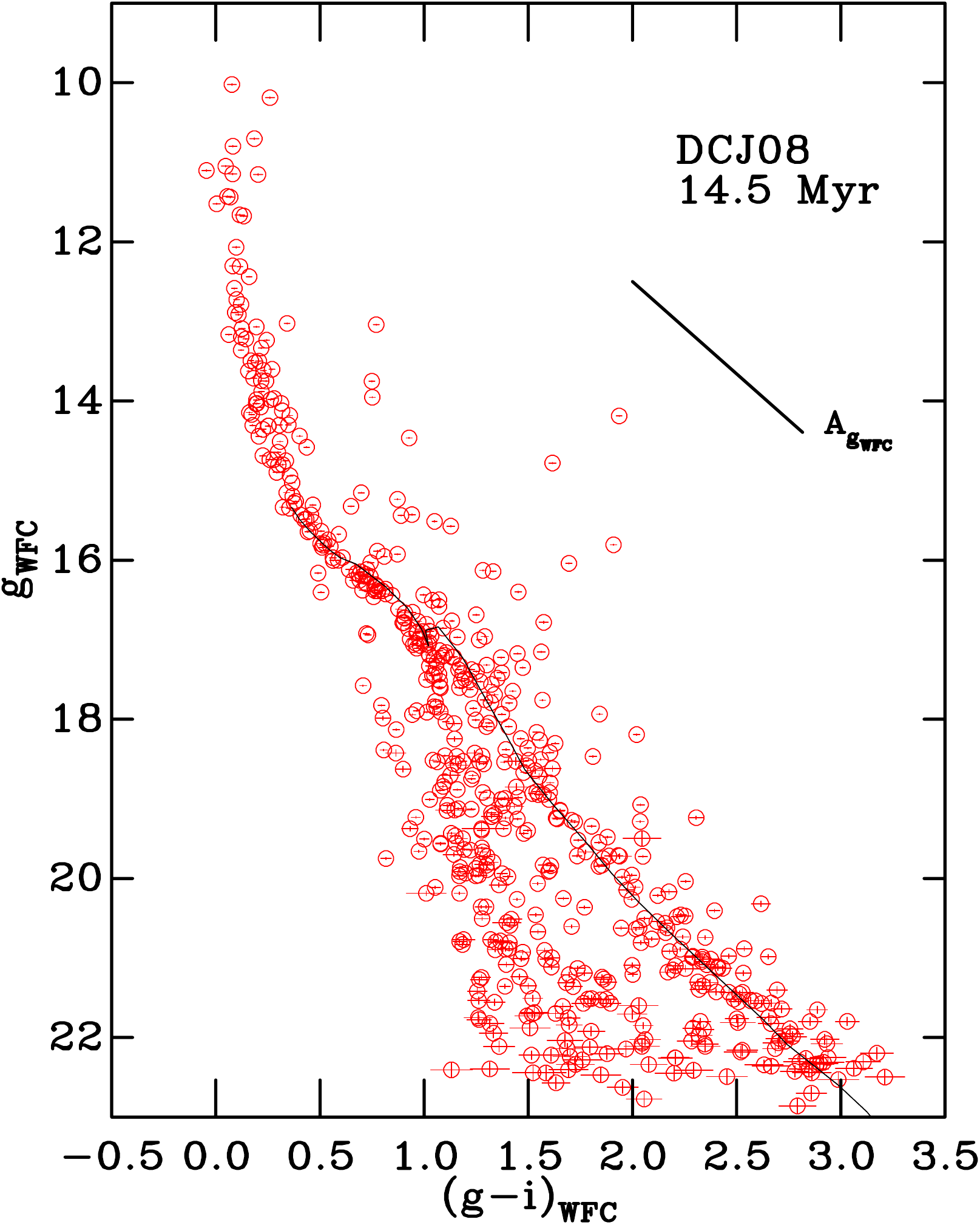}
\caption[]{Stars selected as possible members of $\chi$\,Per based on their positions
  on the sky relative to the cluster centre (see
  Appendix~\ref{chiper_members}) in the $\gwfc, \giwfc$
  CMD. The $14.5\,
  \rm{Myr}$ semi-empirical DCJ08 single-star pre-MS model isochrone is
  overlaid at the best-fit MS distance and reddened assuming the
  median value derived in
  Section~\ref{fundamental_parameters_from_ms_stars} according to the
  prescription described in
  Section~\ref{reddening_and_extinction_for_pre-ms_stars}.
  The diagonal line
  represents the reddening vector in the $\gwfc, \giwfc$ plane for a
  star with $T_{\rm{eff}} \simeq 4500\, \rm{K}$ and log$\,g \simeq 4$
  based on the median value for the SFR.}
\label{fig:chiper_pms}
\end{figure}

$\chi$\,Per could not be fitted using the simple $\tau^{2}$ model that
accounts for a non-member population as the fraction of non-members
(introduced by the selection based purely on positions on the sky
relative to the cluster centre; see Appendix~\ref{chiper_members}) was
simply too high. Given the derived
best-fit MS age of $14.5\, \rm{Myr}$, one learns nothing by comparing
the pre-MS locus with a $6\, \rm{Myr}$ pre-MS
model isochrone. Therefore, in Fig.~\ref{fig:chiper_pms} the
semi-empirical DCJ08 single-star model isochrone at an age of $14.5\,
\rm{Myr}$ is overlaid adopting the best-fit MS distance modulus
$dm=11.80$ and reddened by the median value of $E(B-V)=0.52$ according
to the prescription described in
Section~\ref{reddening_and_extinction_for_pre-ms_stars}.

From Fig.~\ref{fig:chiper_pms} it is apparent that the semi-empirical
DCJ08 single-star model isochrone matches the shape of the observed
pre-MS well across the entire colour range, as well as following the
sequence across the pre-MS-MS transition. The nominal ages
for $\chi$\,Per based on
the three semi-empirical pre-MS models are $\simeq 14\, \rm{Myr}$ for both
the BCAH98 $\alpha=1.9$ and DCJ08 isochrones, and $\simeq 12\, \rm{Myr}$ for
the DAM97 isochrones. Thus, the consistency demonstrated for the other
SFRs with MS ages $> 10\, \rm{Myr}$ is still evident for the BCAH98
$\alpha=1.9$ and DCJ08 models, whereas for the DAM97 models the
tendancy to predict a slightly younger age is again observed.

\subsection{The effects of assumptions on the pre-main-sequence
  ages}
\label{effects_of_assumptions_in_pre-ms_fitting}

When deriving ages from the pre-MS photometric data, either by using
$\tau^{2}$ fitting to the full two-dimensional CMD distribution, or by
comparing the sequence with a single-star model isochrone, we have made
two main assumptions. 
First, for SFRs where we have identified the reddening to be spatially variable, we
have adopted the median value for fitting the pre-MS. 
Second, we have assumed the same (solar) composition for all SFRs.

\subsubsection{Differential reddening}
\label{differential_reddening}

In a  $\gwfc, \giwfc$ CMD the reddening vector is roughly parallel to the pre-MS
(see Fig.~\ref{fig:nominal_pms_ages}), much as in the commonly used $V, V-I$ CMD
(e.g. \citealp{Hillenbrand97,Burningham03}). 
Thus the primary effect of variable extinction is to scatter stars along the isochrone.
To test how far mis-measured extinction might affect our results, the NGC\,7160 pre-MS data 
were fitted as described in Section~\ref{pre-ms_ages_derived_using_tau2}, but using 
reddening that was 50 per cent larger than the median value ($E(B-V)=0.37$) originally
used. The reason we adopt NGC\,7160 is that this SFR has the largest
and most variable reddening of all the SFRs with MS ages $\gtrsim 10\,
\rm{Myr}$. At the best-fit MS distance, the best-fit pre-MS age differs by only 5
per cent. 
The effect of a scatter will be considerably less than that of a global shift,
thus the fact that we have adopted the median reddening for a
given SFR, does not significantly affect the derived pre-MS ages.

\subsubsection{Composition variations}
\label{composition_variations}

In the pre-MS regime, we are unable to use sub- or super-solar metallicity
pre-MS evolutionary models to derive ages because; i) the DAM97 pre-MS
models are only available with a solar composition and ii) although the BCAH98 $\alpha=1.9$ and DCJ08
models are available for a range of metallicities, we do not have the requisite photometric
data in the INT-WFC system to calculate the empirical corrections to
the theoretical BC-$T_{\rm{eff}}$ relation as we have done using the
Pleiades for the solar metallicity case (see
Section~\ref{creating_recalibrated_semi-empirical_pre-ms_isochrones}). In
the MS regime, however, the \cite{Schaller92} models do cover a range
of metallicities and it is possible to investigate what effects
adopting a different composition would have on the derived MS
parameters.

As an example, we have investigated these effects using
NGC\,2244 as there is evidence that the metallicity for this SFR could
be sub-solar ($\rm{[Fe/H]}=-0.46$; see \citealp{Paunzen10}). The
reddening vectors and BC-$T_{\rm{eff}}$ relation were recalculated
using the $Z=0.006$ \textsc{atlas9}/ODFnew atmospheric models (the
closest to the required sub-solar composition from the available
grid). The $Z=0.008$ \cite{Schaller92} interior models (the closest to
the sub-solar composition) were used to calculate the reddening,
distance, and age as described in
Section~\ref{fundamental_parameters_from_ms_stars}. A revised distance
modulus $dm=10.34^{+0.04}_{-0.06}$ was calculated, whereas both the
age and reddening (both median and full range) were insensitive to
changes in the metallicity. Hence, if the composition of NGC\,2244 is
indeed approximately $1/3 Z_{\odot}$, the distance modulus would be
$\simeq 0.4-0.5\, \rm{mag}$ smaller. Whilst sub-solar composition
pre-MS model isochrones, at a given age, are more luminous than solar
metallicity models in CMD space, it is hard to quantify how the difference in the
derived distance modulus would translate into an age difference in
absolute terms when fitting the pre-MS population.

\section{Comparing the main-sequence and pre-main-sequence ages}
\label{comparing_the_ms_and_pre-ms_ages}

\begin{figure*}
\centering
\includegraphics[width=\textwidth]{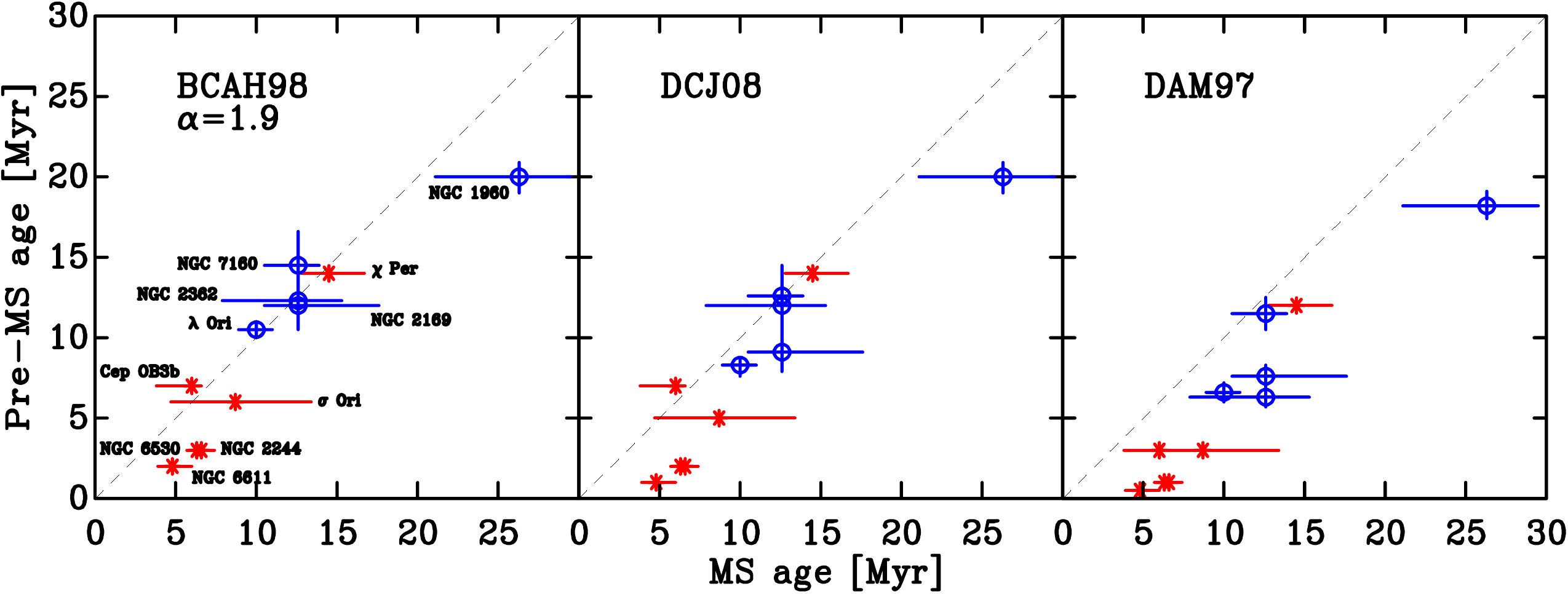}
\caption[]{The MS versus pre-MS ages
  for the SFRs in our sample. The blue circles and error bars represent the SFRs for which
  both the MS and pre-MS age were derived using the $\tau^{2}$ fitting
  statistic. The red asterisks and error bars denote SFRs for which the MS age was derived using
  the $\tau^{2}$ fitting statistic, however the pre-MS age was estimated
  by overlaying a semi-empirical pre-MS single-star model isochrone on the pre-MS
  population and is a nominal age at a mass of $0.75\, \msun$ (see
  Section~\ref{sfrs_with_ms_ages_less_than_10}).
  The uncertainties in the MS age were calculated from the
  $\tau^{2}$ age-distance fit (see Section~\ref{age_and_distance_fitting}) and
  represent the 68 per cent confidence levels. Uncertainties in the pre-MS
  age were only calculated for those SFRs where the age was derived
  using the $\tau^{2}$ fitting statistic and represent the uncertainty
  in the derived MS distance translated into an age (see
  Section~\ref{pre-ms_ages_derived_using_tau2}). Note that both
  IC\,348 and IC\,5146 are not shown in this figure as no MS ages were
  derived for either SFR (see caption of Fig.~\ref{fig:ms_ages}).
  \textbf{Left panel:} BCAH98 $\alpha=1.9$. Note that the pre-MS age
  for NGC\,2362 has been increased by $0.3\, \rm{Myr}$ to highlight the
  uncertainty on the MS age. \textbf{Middle panel:}
  DCJ08. \textbf{Right panel:} DAM97.}
\label{fig:ms_pms_age}
\end{figure*}

Having derived ages from the MS and pre-MS members for the sample of
SFRs, it is now possible to bring these two age diagnostics
together. In Sections~\ref{pre-ms_ages_derived_using_tau2} and
\ref{nominal_pre-ms_ages} it was shown
that the pre-MS ages for young SFRs are heavily model-dependent, even
after recalibrating the transformation between theoretical H-R and
observable CMD space using the observed colours of Pleiades
members. Hence to choose between the various pre-MS age scales, in
Fig.~\ref{fig:ms_pms_age} we have plotted the MS ages against the
pre-MS ages for our sample of SFRs.

The most obvious conclusion that can be drawn from
Fig.~\ref{fig:ms_pms_age} is that the DAM97 pre-MS age scale is inconsistent with the MS
age scale across almost the entire sample. That the DAM97 models tend
to predict younger pre-MS ages than other pre-MS models is not a new
finding (see for example \citealp{Dahm05}). For a given age, the
DAM97 models predict pre-MS stars that are overluminous across a mass range of
$\simeq 0.5-1.8\, \msun$ with respect to both the BCAH98 $\alpha=1.9$
and DCJ08 models, resulting in an approximate difference of a
factor of two in age through isochrone fitting for SFRs with ages
$\simeq 10-12\, \rm{Myr}$.  

The levels of agreement between the MS age scale and the
pre-MS age scales of BCAH98 $\alpha=1.9$ and DCJ08 are much higher
than for the DAM97 models. For SFRs with pre-MS ages of $\gtrsim 6\,
\rm{Myr}$ (on the BCAH98 $\alpha=1.9$ and DCJ08 scales) both models
predict pre-MS ages that are generally consistent with the MS ages
derived in Section~\ref{age_and_distance_fitting}.
However, for the very youngest
clusters ($< 6\, \rm{Myr}$) there remains a discrepancy between the two age
diagnostics, with the MS ages being approximately a factor of 2 older
than the pre-MS ages. 
The work presented in this paper has therefore removed the age discrepancy between the
MS and pre-MS for all but the very youngest SFRs, a significant improvement
over the result of \cite{Naylor09} where a factor 2 discrepancy was
still observed at ages of $10\, \rm{Myr}$. That we have found
agreement between the MS and pre-MS ages (for SFRs with ages $\gtrsim
6\, \rm{Myr}$), which are based on different mass regimes that rely on
different aspects of stellar physics, instils confidence in the
pre-MS ages derived using the BCAH98 $\alpha=1.9$
and DCJ08 model isochrones.

This agreement between pre-MS and MS ages builds upon that already
demonstrated by \cite{Pecaut12}. A detailed analysis of stars hotter
than $4000\, \rm{K}$, which should therefore be
comparable with our age scale, showed they were a factor of 2.5 less luminous
compared with predictions for a $5\, \rm{Myr}$ old population by four
sets of pre-MS evolutionary models. Deriving isochronal ages
separately for B-, A-, F-, and G-type stars, as well as the M-type
supergiant Antares,
\cite{Pecaut12} not only calculated consistent ages from the pre-MS
and post-MS populations, but also increased the age of Upper Sco by
approximately a factor of 2, calculating a revised mean age of $\simeq 11\,
\rm{Myr}$. Thus these combined works have increased the number of SFRs
with significantly revised ages to 14, thereby lending support to the claim of
\cite{Pecaut12} that similarly aged SFRs may be in need of further
investigation and possible amendment.

\subsection{Final assigned ages}
\label{final_assigned_pre-ms_ages}

Having demonstrated that the BCAH98 $\alpha=1.9$ and DCJ08
and MS age scales are generally consistent, we are
now in a position to assign the finalised age to each of the 13
SFRs in our sample. As discussed in
Section~\ref{introduction}, the MS ages are much more uncertain in a
statistical sense but on average they confirm that the pre-MS ages,
which are statistically much more precise but potentially have more
systematic error, are on a reasonable scale. Given that the
differences between the pre-MS ages derived using the BCAH98
$\alpha=1.9$ and DCJ08 models, are typically of the order of $\simeq
1-2\, \rm{Myr}$, we adopt the mean of these two ages for a given SFR.

There is still the
question of what age to assign for the youngest SFRs
(pre-MS ages $< 8\, \rm{Myr}$). Given the uncertainties in the
distance to each of the SFRs, the nominal ages given in
Table~\ref{tab:pre-ms_nominal_ages} are
consistent with two groups of clusters at ages of 2 and $6\,
\rm{Myr}$. For the $6\, \rm{Myr}$ group, the MS ages are consistent with the
pre-MS nominal ages, and so our final assigned age for this group is
$6\, \rm{Myr}$. For the very youngest group, the MS ages ($\simeq 5-6\,
\rm{Myr}$) are approximately a factor of two greater than the pre-MS
ages. For reasons we discuss in Section~\ref{discussion}
we adopt an age of $2\, \rm{Myr}$ for this group.
The resulting ages we propose
should be used for these SFRs in further studies are given in
Table~\ref{adopted_parameters}.

\begin{table*}
\caption[]{Final age, distance and reddening for the sample of
  SFRs. Uncertainties in the derived ages for SFRs older than
  $6\, \rm{Myr}$ are shown in
  Table~\ref{pre-ms_fitting_results}. Notes are as follows. (1)
  Individual reddenings derived using the revised Q-method with the
  median $E(B-V)$ value quoted (otherwise the value shown represents
  the mean uniform reddening) with the full range
  shown in Table~\ref{ms_fitting_results}. (2) Parameters derived
  assuming the total-to-selective extinction ratio $R_{V}=3.75$ (see
  Section~\ref{anomalous_extinction})}
\begin{tabular}{c c c c}
\hline
Age (Myr)&SFR&Distance modulus $dm$&$E(B-V)$\\
\hline
\multirow{4}{*}{2}&NGC\,6611 (Eagle Nebula; M\,16)$^{(1,2)}$&$11.30 \leq 11.38 \leq 11.44$&0.71\\
&IC\,5146 (Cocoon Nebula)$^{(1)}$&$9.62 \leq 9.81 \leq 10.01$&0.75\\
&NGC\,6530 (Lagoon Nebula; M\,8)$^{(1)}$&$10.59 \leq 10.64 \leq 10.68$&0.32\\
&NGC\,2244 (Rosette Nebula)$^{(1)}$&$10.67 \leq 10.70 \leq 10.75$&0.43\\
\hline
\multirow{3}{*}{6}&$\sigma$\,Ori$^{(1)}$&$7.99 \leq 8.05 \leq 8.11$&0.05\\
&Cep\,OB3b$^{(1)}$&$8.70 \leq 8.78 \leq 8.84$&0.89\\
&IC\,348$^{(1)}$&$6.89 \leq 6.98 \leq 7.17$&0.69\\
\hline
10&$\lambda$\,Ori (Collinder\,69)$^{(1)}$&$7.99 \leq 8.02 \leq
8.06$&0.11\\
11&NGC\,2169&$9.90 \leq 9.99 \leq 10.06$&0.16\\
12&NGC\,2362&$10.57 \leq 10.60 \leq 10.66$&0.07\\
13&NGC\,7160$^{(1)}$&$9.62 \leq 9.67 \leq9.76$&0.37\\
14&$\chi$\,Per (NGC\,884)$^{(1)}$&$11.77 \leq 11.80 \leq 11.86$&0.52\\
20&NGC\,1960 (M\,36)&$10.28 \leq 10.33 \leq 10.35$&0.20\\
\hline
\end{tabular}
\label{adopted_parameters}
\end{table*}

\subsection{Discussion}
\label{discussion}

On the balance of the evidence presented in
Sections~\ref{age_and_distance_fitting} and
\ref{sfrs_with_ms_ages_less_than_10}, how reasonable is the distinction between
the $2\, \rm{Myr}$ and $6\, \rm{Myr}$ groups?
Although it is statistically impossible to differentiate
between these two groups based on their MS ages (all clustered
around $6\, \rm{Myr}$) there is an obvious difference in the
luminosity of the pre-MS locus between those SFRs where
an isochrone of $6\, \rm{Myr}$ lies systematically below the observed
pre-MS locus and those SFRs where the isochrone traces the approximate
middle of the locus. Furthermore, there is a visible difference between
the magnitude of the observed luminosity spread between the $2\,
\rm{Myr}$ and $6\, \rm{Myr}$ groups. For a given colour, the spread in
the $2\, \rm{Myr}$ group
covers approximately $3\, \rm{mag}$, whereas in the $6\, \rm{Myr}$ group the
observed spread is approximately a magnitude smaller. Note also that in the
older SFRs this spread almost entirely vanishes and is explainable by the
presence of binaries and higher order multiple systems.

The main difference between the absolute pre-MS ages derived using the
$\tau^{2}$ fitting statistic and the nominal pre-MS ages is that the
former include an intrinsic binary fraction whereas the latter do
not and are solely based on comparison with a single-star model
isochrone. Thus, there is a suggestion that the pre-MS ages for both the 2
and $6\, \rm{Myr}$ groups require a correction to account for
binarity. The difference between the lower single-star and upper
equal-mass binary envelopes in a coeval model isochrone is $\simeq
0.75\, \rm{mag}$, and so if we na{\"i}vely assume the most extreme
case (i.e. 50 per cent of stars are single and 50 per cent are in
equal-mass binaries) this would necessitate that the single-star
isochronal ages be increased by a factor that translates to a shift of
$\simeq 0.38\, \rm{mag}$ fainter. For more realistic mass ratio
distributions, however, this shift would be smaller. Adopting the most
extreme case, such a shift would increase the pre-MS ages for the 2
and $6\, \rm{Myr}$ groups by an additional factor of $1.5-2$. This
would have the effect of decreasing the disparity between the MS and
pre-MS ages for the $2\, \rm{Myr}$ group, whilst causing disagreement
between the two ages for the $6\, \rm{Myr}$ group.

Such a
shift appears unlikely given that at ages of $\simeq 10\, \rm{Myr}$ we
would expect to see evolved high-mass stars in SFRs like Cep\,OB3b,
however these are not observed in the optical CMDs. Furthermore, the $2\,
\rm{Myr}$ group represents the earliest visible stages of the star
formation process and any shift would increase the age of these SFRs
to $\simeq 4-5\, \rm{Myr}$, suggesting that only after such times do
embedded protostars become optically visible.

Any
possible quantification of the shift in the ages due to binarity are
hampered by the underlying uncertainty in the causes of the observed
luminosity spread and the resulting implications for the evolution of
single- and binary-star systems (e.g. \citealp{Preibisch12}). In
addition, these young SFRs likely contain significant numbers of stars
with circumstellar discs, thus further complicating the issue due to
the effects of accreting objects observed with a range of
accretion rates and viewing angles (see \citealp{Mayne10}).

In Section~\ref{anomalous_extinction} we briefly discussed the
uncertainty associated with the distance derived from MS fitting
arising from possible variations in $R_{V}$ towards very young
SFRs. The revised ages given in
Section~\ref{final_assigned_pre-ms_ages} demonstrate that we have
derived consistent ages from both the high-mass and low-mass
populations for a range of SFRs down to ages of $\simeq 6\,
\rm{Myr}$. There is still, however, disparity between these two ages
for the very youngest SFRs (the so-called $2\, \rm{Myr}$ group) for
which the pre-MS ages are approximately factors of between 2 and 3
younger. Given the degeneracy between age and distance in deriving
ages using pre-MS evolutionary models, it would be interesting to note
whether variations in $R_{V}$ as a function of age are observed. If
this is the case, and the typical value of $R_{V}$ is larger
in the youngest SFRs, then the derived distance would be smaller than
that derived in Section~\ref{age_and_distance_fitting} by a factor of
$\Delta dm = \Delta R_{V} \times E(B-V)$. This could offer a simple
solution to the disparity of the MS and pre-MS ages for the youngest
SFRs as a decreased distance would necessitate an older age to fit a
given photometric pre-MS locus in CMD space, however, further
observational work is required in such SFRs to ascertain whether or
not this is in fact the case.

We could improve on our work if we understood the observed luminosity
spread as well as the problems associated with the evolutionary models and
physical processes that affect the associated spectral energy
distributions (SEDs) of young pre-MS stars.
The models adopted in this study all
assume that the mixing-length parameter $\alpha$ is constant for all
evolutionary stages and identical for all masses. Studies
investigating whether this is a reasonable assumption
(e.g. \citealp*{Ludwig99,Ludwig08}) have found that $\alpha$ can vary
as a function of spectral type, the effects of which would be more
pronounced at earlier evolutionary phases where the stars are fully
convective and the superadiabatic region is more extended. In addition
to inadequacies in the theoretical
models (see \citealp{Baraffe02}), there are also physical processes that affect the SEDs
associated with pre-MS stars, and these can be almost impossible to
incorporate into evolutionary codes. An obvious example is the
enhanced levels of activity observed on pre-MS stars. High levels of
activity, presumably driven by intense surface magnetic fields, can
inhibit convective flows at the stellar surface and result in
starspots covering a large fraction of the photosphere. Thus the
colours of young low-mass stars can be somewhat different than the
colours of older stars of the same mass and $T_{\rm{eff}}$
(e.g. \citealp{Stauffer03}). An additional consequence of the
inhibited convective flows
is that the radii and $T_{\rm{eff}}$ of stars with intense magnetic fields can
differ from stars of a similar mass but with a much weaker magnetic
field (e.g. \citealp*{Chabrier07}; \citealp{Yee10}). These combined
effects further complicate the transformation from H-R to CMD space
i.e. for a star of given mass, age, metallicity, and log$\,g$ there is not a
single conversion from, for example, $T_{\rm{eff}}$ to $\giwfc$, but
instead a range.

In addition to the effects discussed above, there is a fierce debate
about whether on-going or early episodes of accretion can result in a
marked alteration to the luminosity of very young ($< 10\, \rm{Myr}$)
objects, when compared to the standard non-accreting models
(e.g. \citealp*{Baraffe09,Hosokawa11}). Short-lived phases of intense accretion
during the early Class\,I phase of a YSO have been advocated as a
possible explanation for the observed luminosity spreads in CMDs of
young SFRs. This would naturally impact on the derived ages (and
masses) for young objects derived from CMDs, however, this is further
compounded by the fact that it is not the currently observed
properties that ultimately affects the position of a given object in
CMD space, but rather, the \emph{accretion history} of that specific object.
Observational evidence pertaining to effects on the luminosity
evolution of young pre-MS objects as a result of accretion history has
been reported by \cite{Littlefair11}, and
therefore it is conceivable that a considerable portion of the
observed luminosity spread in CMDs of young SFRs may be due
to such
variable accretion histories within a coeval stellar population.

\section{Implications}
\label{implications}

In Section~\ref{final_assigned_pre-ms_ages} a set of revised
ages were assigned to a range of young $(< 30\, \rm{Myr})$ SFRs. Two
areas of pre-MS evolution that are heavily dependent upon the
adopted ages are; i) the survival timescales for circumstellar discs
and ii) the evolutionary timescales of YSOs. Therefore, in this
section we discuss the implications of the revised age scale in terms
of these two aspects.

\subsection{Circumstellar disc lifetimes}
\label{implications_discs}

Circumstellar discs appear to be a ubiquitous by-product of the star
formation process and are a driving factor in the evolution of stars
and planetary systems. Mid- to far-IR \textit{Spitzer}
observations of low-mass pre-MS stars indicate that by $\simeq 5\,
\rm{Myr}$ approximately 80 per cent of primordial discs have
dissipated (e.g. \citealp{Carpenter06,Dahm07}), agreeing with estimates based
on near-IR observations (e.g. \citealp*{Haisch01};
\citealp{Hillenbrand05}). These timescales are almost exclusively based
on ages determined from pre-MS isochrone fitting to young stellar
populations and thus any revision of pre-MS SFR ages will naturally
alter the expected lifetime of circumstellar discs.

\begin{figure}
\centering
\includegraphics[width=\columnwidth]{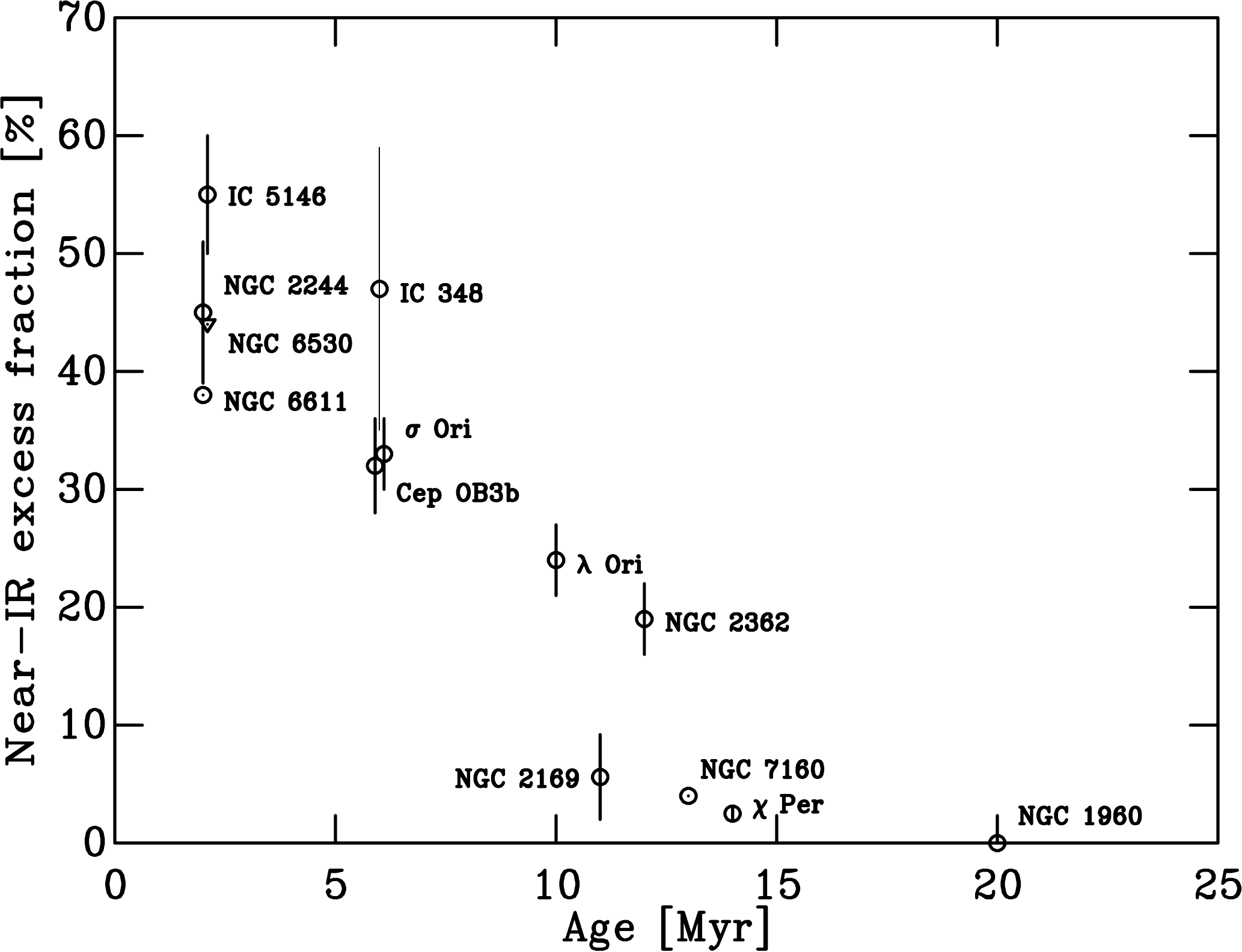}
\caption[]{Fraction of stars (typically spectral types mid-K and later) with
  near-IR excess disc emission as a function of our revised age. Circles
  represent the disc fraction for T-Tauri stars based on
  \textit{Spitzer} observations (see text for references). The
  triangle represents the disc fraction for NGC\,6530 based on
  $JHK_{\rm{s}}$ observations from \protect\cite{Prisinzano07}. Note that the ages for
  IC\,5146, NGC\,6530, $\sigma$\,Ori, and Cep\,OB3b have been shifted
  slightly to highlight the uncertainties in individually derived disc
  fractions.}
\label{fig:disc_frac_age}
\end{figure}

Fig.~\ref{fig:disc_frac_age} shows the disc frequency of late-type
stars (typically mid-K and later) with near-IR excess emission in
different SFRs as a function of our revised age. Disc fractions have,
in all but one case, been taken from studies based on \textit{Spitzer}
observations including NGC\,6611 \citep{Guarcello07}, IC\,5146
\citep{Harvey08}, NGC\,2244 \citep{Balog07}, Cep\,OB3b
\citep{Allen12}, NGC\,2169 (Hernandez et al. in preparation),
$\sigma$\,Ori \citep{Hernandez07}, IC\,348 \citep{Lada06}, $\lambda$\,Ori
\citep{Barrado07}, NGC\,2362 \citep{Dahm07}, NGC\,7160
\citep{Sicilia06}, $\chi$\,Per \citep{Currie10}, and NGC\,1960
\citep{Smith12}. Note that in cases where both \textit{Spitzer} and
\cite{Haisch01} disc fractions are available, those of
\textit{Spitzer} have been adopted as they are less likely to be
affected by sensitivity issues, particularly in the $L$-band (see
\citealp{Lyo03}).

The disc fraction associated
with NGC\,6530 comes from the study of \cite{Prisinzano07} who use
Two-Micron All-Sky Survey (2MASS) $JHK_{\rm{s}}$ colour-colour
diagrams and various Q-indices (analogous to the Johnson Q-index; see
\citealp{Damiani06a}) to determine stars with excess, which
they attribute to circumstellar material. As demonstrated by
\cite{Kenyon01}, $L$-band photometry is almost essential for
constraining the disc bearing population of a given sample of
stars. The $JHK$ bandpasses do not extend to long enough wavelengths
to unambiguously demonstrate the presence of circumstellar material,
whereas observations at wavelengths $\gtrsim 3.5\, \rm{\mu m}$ provide
a higher contrast relative to photospheric emission from the central
star. Thus the value quoted by \cite{Prisinzano07} should only be
treated as an approximation.

Fig.~\ref{fig:disc_frac_age} shows a remarkable correlation between
our revised ages for the sample of SFRs and the associated
circumstellar disc fraction, strengthening that originally identified
by \cite{Haisch01}. The caveat, however, is that despite this
correlation there are likely issues arising from; i)
differences in the mass range sampled across the SFRs and ii)
sensitivity issues for the latest spectral types. Recent
evidence suggests that circumstellar disc dispersal timescales are
dependent upon the mass of the star, with discs around lower mass stars surviving longer
(e.g. \citealp{Carpenter06,Kennedy09}). Furthermore, it can be hard to
unambiguously disentangle the disc emission from that of the stellar
photosphere, especially for M dwarfs (e.g. \citealp*{Ercolano09}).
Hence, as the spectral type
range is unlikely to be identical for all SFRs, this could mean that
such effects may modify the correlation. What is clear from
Fig.~\ref{fig:disc_frac_age} is that
circumstellar discs appear to survive longer than previously
believed, indicating a disc half-life of $\simeq 5-6\, \rm{Myr}$ with
approximately 80 per cent of stars having lost their discs on
timescales of $\simeq 10-12\, \rm{Myr}$.

An important young SFR not included in
our sample but that is commonly used to investigate
the evolution of pre-MS processes is Upper Sco. This SFR, in addition to
$\lambda$\,Ori, has been used as a benchmark for the study of
circumstellar disc lifetimes and fractions. The \textit{Spitzer}
analysis of \cite{Carpenter06} shows that Upper Sco has a disc
fraction (in the same spectral type range as adopted above) of $19 \pm
4$ per cent, in good agreement with the disc fractions of both
$\lambda$\,Ori and NGC\,2362 (see
Fig.~\ref{fig:disc_frac_age}). Furthermore, using H$\alpha$ as a proxy
for stars that are actively accreting, \cite{Barrado07} re-analysed
the data of \cite{Dolan01} and found that $\simeq 11$ per cent of
stars are actively accreting in $\lambda$\,Ori (see also
\citealp{Sacco08}), compared with $\simeq 10$ per cent in Upper Sco
\citep*{Lodieu11}.

As discussed in \cite{Ayliffe09}, there is a long-standing concern
that the timescales required by numerical models to form a gas giant
Jupiter-like planet ($\simeq 10\, \rm{Myr}$;
\citealp{Perri74,Pollack96}) may be larger than the observed
circumstellar disc lifetimes ($\simeq 5\, \rm{Myr}$;
\citealp{Haisch01}). Hence the findings presented in this study offer a
simple solution to this apparent discrepancy. If the SFRs used to
measure the dissipation timescale of circumstellar discs are on
average a factor of 2 older, then the discs survive for approximately
twice as long and thus provide the necessary reservoirs to form gas
giant planets on $\simeq 10\, \rm{Myr}$ timescales.

\subsection{Evolutionary lifetimes of young stellar objects}
\label{implications_ysos}

Having discussed the effects of the revised age scale in terms of
circumstellar disc lifetimes, this naturally leads onto the lifetimes
of different evolutionary stages of YSOs. The generally adopted method
to derive such timescales is to use the number of objects in each
class to establish relative lifetimes (e.g. \citealp*{Wilking89}). A
recent study by \cite{Evans09} proposed that if star formation is a
continuous process with a duration larger than the age of Class\,II
(classical T-Tauri stars), then relative YSO evolutionary lifetimes can be
estimated by taking the ratio of objects in each class and multiplying
by the adopted lifetime for a Class\,II object. Based on various
studies of young SFRs (see references within Section~5.3 of
\citealp{Evans09}), a lifetime of $\simeq 2\, \rm{Myr}$ was assigned
to the Class\,II evolutionary phase, with the lifetimes of the other
classes derived accordingly.

It is clear then that any revision to the pre-MS ages of young SFRs
will have an impact on the adopted Class\,II lifetime used to
calculate relative lifetimes for other YSO evolutionary phases. Whilst
an in-depth discussion on how the relative ages are affected by the
revised ages is beyond the scope of this study,
Fig.~\ref{fig:disc_frac_age} suggests that the adopted lifetime of
Class\,II objects is likely underestimated. The ages shown in
Table~\ref{adopted_parameters} are approximately a factor of 2
older than current ages for the same SFRs. Regions such as
$\lambda$\,Ori, with two independent age determinations at $\simeq 10\,
\rm{Myr}$ (see also Upper Sco; \citealp{Pecaut12}) which show that
$\simeq 20$ per cent of stars retain their optically thick
circumstellar discs (indicative of Class\,II status) and that $\simeq
10$ per cent of stars are still actively accreting, imply a longer
Class\,II lifetime somewhere in the region of $\simeq 4-5\,
\rm{Myr}$. Propagating this age to the earlier Class\,I lifetime, based
on the number of YSOs presented in \cite{Evans09} and ignoring
environmental effects, indicates an average lifetime of $\simeq 1\,
\rm{Myr}$, resulting in a lifetime that is considerably longer than
most estimates (cf. $0.25-0.67\, \rm{Myr}$; \citealp{Hatchell07}).

\section{Conclusions}
\label{conclusions}

In this study we have used the stellar populations of 13 young $(<
30\, \rm{Myr})$ SFRs to critically assess the ages derived using
pre-MS isochrones in CMDs. The stages we have gone through to achieve
this are as follows.

\begin{enumerate}
\item We have derived a self-consistent set of ages, distances and
  reddenings (with typically higher precision than previous estimates
  and statistically
  meaningful uncertainties) by fitting the MS populations of our
  sample of SFRs with MS evolutionary models. For several of these
  SFRs, this represents the first parameter derivations in the literature
  based on a robust statistical fitting technique.
\item We have created new semi-empirical pre-MS isochrones
  incorporating existing stellar interior models, an empirical
  BC-$T_{\rm{eff}}$ relation, and theoretical corrections for the
  dependence on log$\,g$. These new isochrones have been used to
  calculate ages from the pre-MS populations of our sample of SFRs. We
  find that the ages derived using these semi-empirical isochrones are
  typically a factor of 2 older than current estimates.
\item Comparing the various pre-MS age scales with the more reliable
  MS age scale, we find that the DAM97 models systematically
  underestimate the ages of young SFRs, whereas the scales
  of both the BCAH98 $\alpha=1.9$ and DCJ08 models are
  generally consistent. Thus, we suggest that in an effort to create
  an absolute pre-MS age scale, either the BCAH98
  $\alpha=1.9$ or DCJ08 models should be adopted for this purpose.
\item We still note a discrepancy between the MS and pre-MS ages for
  very young SFRs (MS age $\lesssim 6\, \rm{Myr}$) for which the
  MS ages are approximately factors of between 2 and 3 greater than the pre-MS
  ages. This mismatch in the ages could be attributed to a combination of
  inherent uncertainties in the evolutionary models (e.g. the
  treatment of convection) as well as physical processes that affect
  the colours of real stars that are not incorporated in the
  models. Furthermore, sources of systematic uncertainty
  due to possible
  variations in the value of $R_{V}$ for the youngest SFRs and the
  effects of binarity could contribute to this disparity.
\item We have furthermore investigated the effects of the revised
  pre-MS age scale in terms of both circumstellar disc lifetimes and
  YSO evolutionary timescales. We conclude that circumstellar discs
  survive approximately twice as long as currently believed and that
  this could offer a practical solution to the apparent discrepancy
  that the time required to form planets is larger than the lifetime
  of discs around stars. In addition, we find that due to the revised
  ages, the typical Class\,I lifetime is also underestimated, with a
  significantly longer revised lifetime of $\simeq 1\, \rm{Myr}$
  suggested.
\end{enumerate}

\section*{Acknowledgements}

CPMB is funded by a UK Science and Technology Facilities Council
(STFC) studentship. SPL is supported by an RCUK fellowship. The
authors would like to thank
Charles~D.~H.~Williams for maintaining the Xgrid facilities at the
University of Exeter which were used to reduce the photometric data
presented in this study. We also thank
the referee for useful comments and constructive suggestions that have
greatly improved this work.
This research has made use of data obtained at the
Isaac Newton Telescope which is operated on the island of La
Palma by the Isaac Newton Group (ING) in the Spanish
Observatorio del Roque de los
Muchachos of the Institutio de Astrofisica de Canarias.
This research has made use of archival data products from the
Two-Micron All-Sky Survey (2MASS), which is a joint project of the
University of Massachusetts and the Infrared Processing and Analysis
Center, funded by the National Aeronautics and Space Administration (NASA)
and the National Science Foundation.

\bibliographystyle{mn3e}
\bibliography{references}

\appendix

\section{Pre-main-sequence literature memberships}
\label{literature_memberships}

Many of the SFRs investigated in this study lie in or near the Galactic
plane and as such a CMD for a given field-of-view will contain a high
number of field stars. Thus we must consider the issue of identifying
and isolating bona fide pre-MS members.

Pre-MS stars exhibit several characteristics that highlight
their youth and which can be used to differentiate between them and
older field
stars. These indicators only differentiate between
different types of stellar populations and not between distinct young
populations (see \citealp{Jeffries06a}).
Where no literature memberships were available, we instead
adopted an approach based on relative positions on the sky to identify possible members, however
this has the obvious drawback of including a number of non-members
in the selection.

\subsection{Cep\,OB3b}
\label{cepob3b_members}

X-ray members are from \cite{Naylor99} and \cite{Getman06} using the ROSAT PSPC and
\textit{Chandra} ACIS instruments respectively. We used a combination of both as the
area covered by the ROSAT pointing is larger than the \textit{Chandra} field-of-view.
For stars common to both catalogues, we adopt the
positions from the \textit{Chandra} dataset due to the improved spatial resolution
and sensitivity.

Spectroscopic members are taken from \cite{Pozzo03}. They took spectra
of objects within an initial photometric cut made in the $V,
V-I_{\rm{c}}$ CMD, using a $B-V, V-I_{\rm{c}}$ colour-colour diagram
of optical counterparts to the \cite{Naylor99} ROSAT sources to
identify regions with a high fraction of probable
members. Spectroscopic follow-up measurements provided lithium
equivalent widths
($W_{\lambda}(\LiI)$), $W_{\lambda}(\rm{H\alpha})$, and radial
velocities. Memberships were assigned based on the comparison between
individual radial velocities and the group mean, in addition to strong
spectroscopic features. Additional H$\alpha$ members are from
\cite{Ogura02} which comprise a subset of a Herbig-Haro object survey
using a narrowband H$\alpha$ filter.

Periodic variables are from an $I$-band variability study by
\cite{Littlefair10}. A recent \textit{Spitzer} IRAC and MIPS study
\citep{Allen12} has demonstrated that the young stars within the
cluster are concentrated into two sub-clusters; an eastern sub-cluster
near the Cep\,B molecular cloud and a western sub-cluster near the
Cep\,F molecular cloud. The literature memberships listed above only
cover the eastern sub-cluster, whereas the variability study of
\cite{Littlefair10} covers both sub-clusters. To not bias the
selection of pre-MS members, the positions on the sky of the other
literature sources were used to define a region within which
variability members were isolated.

\subsection{$\chi$\,Per}
\label{chiper_members}

There is a lack of bona fide pre-MS member diagnostics for
$\chi$\,Per. The so-called spectroscopically confirmed members of
\cite{Currie10} are not based on the presence of spectral features
(e.g. $\LiI$ or H$\alpha$) but represent a sequence defined by a
combined $14\, \rm{Myr}$ MS and pre-MS isochrone in the de-reddened
$V$-band magnitude versus spectral type diagram. The width of this
sequence was determined by; i) the physical extent of the cluster, ii)
binarity, and iii) uncertainties in the derived spectral
types. Cross-correlation of the \cite{Currie10} members with our
optical photometric catalogue results in a poorly defined pre-MS locus
with less than 10 stars occupying the single-star sequence at
magnitudes $\gwfc > 18$.

Our photometric catalogue extends to $\gwfc \simeq 23$ and therefore
stellar positions on the sky were instead used to choose an area that
minimises foreground and background contamination (see
\citealp{Mayne07}). Stars brighter than
$\gwfc < 14$ lie bluewards of the contamination and thus represent a
subset of the $\chi$\,Per stellar population where the field star
contamination is minimal. The positions of these stars were then
overlaid on the positions of the entire photometric catalogue and
circular positional selections from $5'$ down to $1'$ in increments of $1'$ were
made around the central cluster coordinates of
$\alpha_{\rm{J2000.0}}=02^{\rm{h}}\, 22^{\rm{m}}\, 05.02^{\rm{s}},
\delta_{\rm{J2000.0}}=+57^{\circ}\, 07'\, 43.44''$
\citep{Mayne07}. The resulting sequence was visually inspected
(by-eye) and the best results (fraction of candidate member-to-field
stars) was found for a radius of $3'$. To ensure that the selection of
the cluster centre was not biased, the positions of proper motion
members from \cite{Uribe02} and likely members based on spectral types
from \cite*{Slesnick02} were also overlaid. In both cases, the central
coordinates from each source were almost identical.

\subsection{IC\,348}
\label{ic348_members}

$\rm{H\alpha}$ members are taken from \cite{Herbig98} who carried out
a wide-field grism spectrograph survey of the
region, discarding stars with $W_{\lambda}(\rm{H\alpha}) < 2\,
\rm{\AA}$. Spectral types (through comparison with dwarf spectral
standards) and extinctions were derived for 80 of these stars using
follow-up spectroscopy. Additional spectroscopic members are taken
from \cite{Luhman03,Luhman05a,Luhman05b}. Memberships were assigned
using a combination of spectral types, spectral features (H$\alpha$,
$\NaI$, and $\KI$), and positions within extinction-corrected CMDs.

X-ray members are from \cite{Preibisch02} taken with the
\textit{Chandra} ACIS detector (supplemented with additional data from the Second
ROSAT PSPC catalogue\footnote[7]{\url{http://ledas-www.star.le.ac.uk/rosat/rra/}}).

Periodic variables are from \cite{Cohen04} and \cite{Littlefair05},
both of which are based on wide-field $I_{\rm{c}}$-band surveys.

\subsection{IC\,5146}
\label{ic5146_members}

$\rm{H\alpha}$ members are taken from \cite{Herbig02} who conducted a
wide-field grism spectrograph survey to identify emission stars to a
limit of $W_{\lambda}(\rm{H\alpha}) \simeq 3\, \rm{\AA}$ and limiting
magnitude of $R_{\rm{c}}=20.5$. Only stars with
$W_{\lambda}(\rm{H\alpha}) > 5\, \rm{\AA}$ and which lie above the
Pleiades MS at a distance of $1.2\, \rm{kpc}$ as described in
\cite{Herbig02} were used.

Additional IR excess members are based on the \textit{Spitzer} IRAC and MIPS
photometry of \cite{Harvey08}. The identification of YSOs from
\textit{Spitzer} is generally based on some
combination of IR excess in addition to a brightness limit (set by the
limiting magnitude fainter than which extragalactic contamination
becomes too high to reliably distinguish between objects). A
combination of colour-colour diagrams and CMDs were used to identify
predominantly Class\,I and Class\,II objects.

\subsection{$\lambda$\,Ori}
\label{lamori_members}

$\LiI$ abundance members come from
\cite{Dolan01}, extending the previous study
of \cite{Dolan99}. An initial photometric cut for stars with $12 \leq R_{\rm{c}} \leq 16$
was made, with the probable members chosen based on their position relative to a
$30\, \rm{Myr}$ isochrone in an $R_{\rm{c}}, (R-I)_{\rm{c}}$ CMD. Follow-up
spectroscopic measurements were taken and $W_{\lambda}(\LiI)$ was used
as a youth indicator, where stars with $W_{\lambda}(\LiI) \geq 0.2\,
\rm{\AA}$ were retained as members.

Additional spectroscopic members come from \cite{Barrado04a} where optical $(RI)_{\rm{c}}$
photometry was combined with 2MASS $JHK_{\rm{s}}$ observations to initially
identify possible member candidates in CMD space. Follow-up
spectroscopic observations were taken to measure
$W_{\lambda}(\rm{H\alpha})$. Memberships were assigned to those stars
that satisfied both spectroscopic and photometric criteria. Later, 
\cite{Barrado07} used \textit{Spitzer} IRAC photometry to identify probable
low-mass members based on IR excess using IRAC colour-colour diagrams in conjunction
with optical and IR CMDs. Further members are from the spectroscopic study of \cite{Sacco08}
who measured both $W_{\lambda}(\LiI)$ and
$W_{\lambda}(\rm{H\alpha})$ in addition to radial velocities.
Final membership is based on a combination of individual candidate velocities that
are consistent with the cluster dispersion, the presence
of strong $W_{\lambda}(\LiI)$ and the presence of $\rm{H\alpha}$
emission.

X-ray members are taken from \cite{Barrado11} observed using the
XMM-\textit{Newton} EPIC detector.

\subsection{NGC\,1960}
\label{ngc1960_members}

Spectroscopic members are from an upcoming paper by
\cite{Jeffries13}. Optical $VI_{\rm{c}}$ photometry was used to
identify possible cluster members from their position in
CMD space, selecting objects with $14 < V < 18.5$. Spectroscopic
observations were to identify the presence of $\LiI$ and derive radial
velocities. Objects identified with strong $\LiI$ were cross-correlated
against the PPMXL catalogue \citep*{Roeser10} adopting the central
cluster proper motion ($\mu_{\alpha}\, \rm{cos}\,\delta=2.9 \pm 2.7\,
\rm{mas\,yr^{-1}}$ and $\mu_{\delta}=-8.0 \pm 2.5\,
\rm{mas\,yr^{-1}}$)
as derived by \cite{Sanner00}. Those that were consistent within the
uncertainties were retained as likely members.

\subsection{NGC\,2169}
\label{ngc2169_members}

Spectroscopic members are from \cite{Jeffries07b} in which optical $(RI)_{\rm{c}}$
photometry was used to highlight possible cluster members from their
position in CMD space. Follow-up
spectroscopic observations were made to identify the presence of $\LiI$ and
$\rm{H\alpha}$ in addition to measuring radial velocities. A combination
of all three diagnostics were used to assign final memberships.

\subsection{NGC\,2244}
\label{ngc2244_members}

X-ray members are from \cite{Wang08} taken with the \textit{Chandra} ACIS instrument.
Cross-correlation against existing photometric studies showed that, when plotted in
a $J-H, H-K_{\rm{s}}$ colour-colour diagram, the majority of the X-ray sources occupy
a space indicative of discless Class\,III objects. A significant fraction of sources
were also identified as having a $K_{\rm{s}}$-band excess and are thus believed to be
pre-MS stars harbouring circumstellar discs.

Additional IR excess sources are taken from \cite{Balog07} who use
\textit{Spitzer} IRAC and MIPS observations to identify the Class\,I and Class\,II
population using IR colour-colour diagnostics.

\subsection{NGC\,2362}
\label{ngc2362_members}

Spectroscopic members are taken from \cite{Dahm05}. He used a wide-field
$\rm{H}\alpha$ survey to identify probable cluster members
which lie above the ZAMS in the $V, V-I_{\rm{c}}$ CMD. Spectroscopic
follow-up on these objects was used to ascertain the presence of
strong $\LiI$ and $\rm{H}\alpha$
features. \textit{Spitzer} IRAC photometry \citep{Dahm07}
was used in conjunction with existing H$\alpha$ emission data, optical
$V(RI)_{\rm{c}}$ and near-IR $JHK$ photometry, and moderate-resolution
spectroscopy to identify the disc-bearing population.

X-ray members are from \cite{Damiani06b} taken with the
\textit{Chandra} ACIS
detector. They find that 88 per cent of the X-ray sources have
optical counterparts that are good candidate low-mass pre-MS stars
based on their position in the $V, V-I_{\rm{c}}$ CMD.

\subsection{NGC\,6530}
\label{ngc6530_members}

X-ray members are taken from the \textit{Chandra} ACIS observations of \cite{Damiani04}.
\cite{Prisinzano05} used this data, in conjunction with optical $BVI_{\rm{c}}$
photometry, to assign pre-MS status to approximately 90 per cent of
the identified X-ray sources, based on their position in the $V, V-I_{\rm{c}}$ CMD.
Spectroscopic follow-up observations on a subset of cluster members was
performed by \cite{Prisinzano07} to identify $W_{\lambda}(\LiI)$, $W_{\lambda}(\rm{H\alpha})$,
and measure radial velocities. These were combined with the X-ray catalogue to
compile a list of cluster members. Extinction-free Q-indices were used,
in conjunction with optical $BVI_{\rm{c}}$ and 2MASS $JHK_{\rm{s}}$
photometry, to identify sources with near-IR excess (the H$\alpha$
spectra were used to differentiate between CTTS and WTTS).

Periodic variables are from the wide-field, high-cadence $I_{\rm{c}}$-band variability study
of \cite{Henderson12}.

\subsection{NGC\,6611}
\label{ngc6611_members}

Members are taken from the studies of \cite{Guarcello07,Guarcello09}.
\cite{Guarcello07} compiled a multiband photometric catalogue
including measurements in the optical
$BVI_{\rm{c}}$ and near-IR 2MASS $JHK_{\rm{s}}$ measurements. These were supplemented by
\textit{Chandra} ACIS X-ray observations \citep{Linsky07}. Stars with
circumstellar discs were identified using extinction-free Q-indices
\citep{Damiani06a} using a combination of optical and near-IR colour
indices. A lower Q-index limit corresponding to photospheric emission was
computed using the MS colours of \cite{Kenyon95}. Near-IR excess was identified on
the basis that the Q-index is smaller than the photospheric limit by $3 \sigma_{\rm{Q}}$,
where $\sigma_{\rm{Q}}$ is the mean error in Q. Stars with circumstellar discs were then
cross-correlated against the X-ray catalogue of \cite{Linsky07}. X-ray sources with neither
an optical nor a 2MASS counterpart were not considered members. Furthermore, X-ray
sources that appeared to lie in an area dominated by young foreground objects in a
$V, V-I_{\rm{c}}$ CMD were also classified as non-members.
\cite{Guarcello09} extended the membership list with \textit{Spitzer} IRAC observations.
Stars with circumstellar discs were identified using a combination of
IRAC colour-colour diagrams and the extinction independent Q-indices of \cite{Guarcello07}.

\subsection{NGC\,7160}
\label{ngc7160_members}

Members are based on the studies of
\cite{Sicilia04,Sicilia05,Sicilia06}. An initial photometric cut was made in the $V, V-I_{\rm{c}}$
CMD to rule out stars near or below the \cite{Siess00} ZAMS, assuming an
average extinction for all stars. This selection was further refined
using optical $(RI)_{\rm{c}}$ variability. Spectroscopic
measurements of $W_{\lambda}(\LiI)$ and $W_{\lambda}(\rm{H\alpha})$ were
used to assign memberships to probable
candidates. The spectroscopic observations were used to calculate
spectral types and extinction for these sources. Final membership is
based on the standard deviation ($\sigma$) from the average cluster
extinction, which naturally relies on the adopted spectral typing and
intrinsic colours. Retained members are those that lie within
$1\sigma$ of the average extinction.

\subsection{$\sigma$\,Ori}
\label{sigori_members}

Spectroscopic members are taken from \cite{Kenyon05} and \cite{Burningham05b}. In
\cite{Kenyon05} a photometric cut was made to select stars lying close to a $5\, \rm{Myr}$
BCAH98 isochrone in the range $14.8 \leq I \leq 18.2$ in an $I_{\rm{c}}, (R-I)_{\rm{c}}$
CMD. Spectroscopic follow-up was used
to measure $W_{\lambda}(\LiI)$,
$W_{\lambda}(\NaI)$, and radial velocities. Membership was
assigned on the basis that the radial velocity and the gravity
sensitive $\NaI$ doublet measurements were coincident with the cluster
mean, in addition that $W_{\lambda}(\LiI) \geq 0.2\,
\rm{\AA}$. Similarly, \cite{Burningham05b} initially used a
photometric cut to select possible candidate members, however a
broader selection (in terms of colour range for a given magnitude; see
their Fig.~1) was chosen to compliment those already observed
by \cite{Kenyon05}. Membership status was again based on radial
velocities, $W_{\lambda}(\LiI)$ and
$W_{\lambda}(\NaI)$. It was found that
the broader photometric selection included no more additional members
than the previous study of \cite{Kenyon05}, suggesting that
photometric selection techniques do not exclude significant numbers of
bona fide members. Members from \cite{Kenyon05} were
only used if they satisfied all selection criteria, however only those
with a membership probability greater than 80 per cent were used from
\cite{Burningham05b}. Additional spectroscopic members are from
\cite{Sacco08} who use the criteria described in
Section~\ref{lamori_members}.

X-ray members are taken from \cite{Sanz-Forcada04} using the
XMM-\textit{Newton} EPIC detector. Positions of X-ray sources have
been cross-correlated against our optical photometric catalogue.

Periodic and aperiodic variability members are from \cite{Cody10}
based on high-precision, high-cadence $I_{\rm{c}}$-band photometric
monitoring.

As discussed in \cite{Jeffries06a}, the pre-MS members of the
$\sigma$\,Ori association are split into two kinematically distinct
sub-groups of different ages differentiated by their heliocentric
radial velocities. At declinations less than
$\delta_{\rm{J2000.0}}=-02^{\circ}\, 18'\, 00.0''$ the region
is dominated by members of Group\,2 (as described in
\citealp{Jeffries06a}). From the membership selections used above, all
but six stars lie in this region (these have subsequently been
removed).

\section{$\tau^{2}$ fitting -- A model for dealing with possible non-member contamination}
\label{non-member_contamination_model}

A CMD of a given SFR, even after isolating the pre-MS using youth
indicators, may contain some contamination from foreground
or background sources that have not been rejected from the selection
process. Therefore, a prescription is needed
to deal with a possible non-member population which may influence the
derived age when fitting for a pre-MS age using the $\tau^{2}$
statistic. One way of dealing with this is to assume a certain
fraction of the stars in a given pre-MS population (as defined by the
youth indicators) are actually non-members.

Starting from the original definition of $\tau^{2}$, as defined in
\cite{Naylor06},

\begin{equation}
\tau^{2} = -2 \sum_{i=1,\,N} \mathrm{ln}  \iint U_{i}(x-x_{i}, y-y_{i})\, \rho (x,y))\, dx\, dy.
\label{eqn:chp4:1}
\end{equation}

\noindent Now consider a model distribution that comprises two
components; one representing the cluster member model distribution
$(\rho_{\rm{c}})$ and the other signifying the non-member model
distribution $(\rho_{\rm{n}})$, such that
$\rho=\rho_{\rm{c}}+\rho_{n}$, then Eqn.~\ref{eqn:chp4:1} becomes

\begin{equation}
\tau^{2} = -2 \sum_{i=1,\,N} \mathrm{ln} \iint U_{i}(x-x_{i}, y-y_{i})\,
\{\rho_{\mathrm{c}}(x,y)+\rho_{\mathrm{n}}(x,y)\}\, dx\, dy.
\label{eqn:chp4:2}
\end{equation}

\noindent In regions away from the cluster sequence, it is assumed that
$\rho_{\rm{c}}=0$ and $\rho_{\rm{n}}$ is a constant i.e. the model is
simply a uniform distribution comprised of non-member stars with no
contribution from cluster members. Thus far from the sequence, for the
$i^{\rm{th}}$ point

\begin{eqnarray}
\tau^{2}_{i} & = & -2\, \mathrm{ln}\, \rho_{\mathrm{n}} \iint U_{i}(x-x_{i}, y-y_{i})\,
dx\, dy \\ \nonumber
\\
& = & -2\, \mathrm{ln}\, \rho_{\mathrm{n}}, \nonumber
\label{eqn:chp4:3}
\end{eqnarray}

\noindent due to the normalisation of the uncertainty function $U$
(see \citealp{Naylor09}). This therefore
defines a maximum value of $\tau^{2}_{i}$, which we shall call
$\tau^{2}_{\rm{c}}$, and the constant can be expressed as

\begin{equation}
\rho_{\mathrm{n}} = e^{-0.5 \tau^{2}_{\mathrm{c}}}.
\label{eqn:chp4:4}
\end{equation}

Defining the fraction of the stars in the model CMD which are cluster
members as $\mathfrak{F}$,
we can use the fact that the integral of the model distribution over
the entire CMD is also unity (see \citealp{Naylor09}), so
$\mathfrak{F}$ can be written as

\begin{eqnarray}
\mathfrak{F} & = & \frac{\iint \{\rho_{\mathrm{c}}(x,y)+\rho_{\mathrm{n}}(x,y)\}\, dx\, dy -
  \iint \rho_{\mathrm{n}}(x,y)\, dx\,
  dy}{\iint \{\rho_{\mathrm{c}}(x,y)+\rho_{\mathrm{n}}(x,y)\}\, dx\, dy} \\ \nonumber
\\
& = & 1 - A e^{-0.5 \tau^{2}_{\mathrm{c}}}, \nonumber
\label{eqn:chp4:5}
\end{eqnarray}

\noindent where $A$ is the area of the CMD. This can be re-arranged
for a specific $\tau^{2}_{\rm{c}}$, which represents the fraction
of stars in the model CMD which are members, such that

\begin{equation}
\tau^{2}_{\mathrm{c}} = -2\, \mathrm{ln} \left(\frac{1 -
      \mathfrak{F}}{A} \right).
\label{eqn:chp4:6}
\end{equation}

\noindent As way of an example, the model CMD for $\lambda$\,Ori
extends approximately $10\, \rm{mag}$ in $\gwfc$ and $3.5\,
\rm{mag}$ in $\giwfc$. If we then assume that
80 per cent of the stars in the sample of
pre-MS objects are members, then $\tau^{2}_{\rm{c}} \simeq
10$.

Practically, in the code, the integration from the brightest to the
faintest star in the model CMD implies $\iint \rho_{\rm{c}}\,
dx\, dy=1$. This therefore modifies Eqn.~\ref{eqn:chp4:6}, so that,

\begin{equation}
\tau^{2}_{\mathrm{c}} = -2\, \mathrm{ln}
\left(\frac{1-\mathfrak{F}}{\mathfrak{F} A}\right),
\label{eqn:chp4:7}
\end{equation}

\noindent however provided that $\mathfrak{F} \geq 0.5$, the value of
$\tau^{2}_{\rm{c}}$ only changes by $\simeq 1.5$.

There are two distinct points which need to be addressed concerning
the implementation of a uniform background contamination population;
i) is the uniform contamination model distribution sufficient to model the
non-uniform contaminating population and ii) does
the fraction of the contamination model distribution, relative to the member
stars, affect the derived age? To assess the effects of these points
we used
$\lambda$\,Ori as an example. Concerning the first
point, the fainter and bluer contamination of $\lambda$\,Ori was
isolated in CMD space. A random selection of this
contaminating population was added to the catalogue of
members (varying from $0-40$ per cent of the members) and the
$\tau^{2}$ fitting statistic used to derive the
pre-MS age for a given value of $\tau^{2}_{\rm{c}}$. The best-fit age
was affected by less than 10 per cent as a result of increasing the
contaminating population. For the second point, a fixed level of
contamination was adopted and the value of $\tau^{2}_{\rm{c}}$ varied
from $10-1$ in steps of 1. For each value of $\tau^{2}_{\rm{c}}$, the
pre-MS age was again derived and it was found that varying the level
of the uniform background contamination also affects the best-fit age
by less than 10 per cent. Hence, whilst adopting a uniform
background distribution of contaminating stars is unrealistic, the
model implementing this description is sufficiently robust to derive
reliable ages.

It is worth mentioning that the prescription presented here is 
mathematically equivalent to the soft-clipping
scheme discussed in the original description of $\tau^{2}$
by \cite{Naylor06}, as well as being analogous to the $n \sigma$
clipping scheme in the $\chi^{2}$ statistic. The subtle difference is
that whereas in the $\chi^{2}$ statistic, where data points that are clipped
are assigned a zero probability, in the $\tau^{2}$ regime, these points
are simply assigned a very low probability (equal to the constant
$\rho_{\rm{n}}$).

\label{lastpage}

\end{document}